\documentclass[preprint,3p,12pt]{elsarticle}

\usepackage{amssymb}

\usepackage{lineno,hyperref}
\modulolinenumbers[5]

\usepackage{subcaption}
\usepackage{xcolor}
\usepackage{soul}
\usepackage{url}
\usepackage{multirow}
\usepackage{enumitem}

\PassOptionsToPackage{normalem}{ulem}
\usepackage{ulem}

\providecolor{add_col}{rgb}{0,0,1}
\providecolor{del_col}{rgb}{1,0,0}
\providecolor{upd_col}{rgb}{0.01, 0.75, 0.24}

\newcommand{\deleted}[1]{}

\journal{Measurement}

\begin{document}
	
	\begin{frontmatter}
		
		
		
		\title{Validation and traceability of miniaturized multi-parameter cluster radiosondes used for atmospheric observations\\{\small Paper in preparation for submission to Measurement, Elsevier
		}}
		
		
		 \author[polito_disat,polito_det]{Shahbozbek Abdunabiev}
		 \author[inrim]{Chiara Musacchio}
		       \author[inrim]{Andrea Merlone}
		 \author[polito_disat]{Miryam Paredes}
		 \author[polito_det]{Eros Pasero}
		 \author[polito_disat]{Daniela Tordella}
		 \address[polito_disat]{Dipartimento di Scienza Applicata e Tecnologia, Politecnico di Torino, 10129 Torino, Italy}
		 \address[polito_det]{Dipartimento di Elettronica e Telecomunicazioni, Politecnico di Torino, 10129 Torino, Italy}
		 \address[inrim]{Istituto Nazionale di Ricerca Metrologica, 10135 Torino, Italy}
		
		\begin{abstract}
			In this work we designed and developed a cluster of light expendable radiosondes that can float passively inside warm clouds to study their micro-physical processes. This involves the tracking of both saturated and unsaturated turbulent air parcels. The aim of this new kind of observation system is to obtain Lagrangian statistics of the intense turbulence inside warm clouds and of the lower intensity turbulence that is typical of the air surrounding such clouds. Each radiosonde in a cluster includes an electronic board, which is mounted onto a small, biodegradable balloon filled with a mixture of helium and air. The cluster is able to float inside clouds for a few hours and to measure air temperature, pressure, humidity and the associated position, velocity, acceleration and magnetic field readings of each radiosonde along their trajectory.
			
		\end{abstract}
		
		
		
		
	\end{frontmatter}
	
	\clearpage
	\section{Introduction}
	\label{sec:intro}
	
	
	Clouds are the largest source of uncertainty in weather prediction and climate science. They continue to be a weak link in the modeling of atmospheric circulation. This uncertainty clouds depend on both physical and chemical processes that cover a huge range of scales, from the collisions of micron-sized droplets and particles to the airflow dynamics on a scale of some thousands of meters \cite{bodenschatz2010cloudturbulence}. Since, some ambiguities exist, related to the representation of clouds in climate models, more observations are needed \cite{onln:eucomplete}. \deleted{However, not all types of clouds are relevant, when discussing important issues, such as global warming}. {Clouds cool the Earth surface by reflecting sunlight back to the space by around 12 $^\circ$C, an effect that is basically caused by \textit{strato-cumulus (warm)} clouds. However, at the same time, the cooling effect of clouds is partially compensated by a 'blanketing' effect: cooler clouds reduce the amount of heat radiating into space by absorbing the heat coming from the Earth's surface and re-radiating some of it back downward. The blanketing effect warms Earth's surface of approximately 7$^\circ$C. These processes averages out to a net loss of 5 $^\circ$C\cite{onln:cloud_climatology}.}
	
	Radars are currently the main source of observational information about clouds in forecast models\cite{illingworth2007cloudnet}. They can provide information about the morphology of clouds, {humidity and} precipitation levels, and the liquid water content. In addition, dual-Doppler radar observations can also provide information on the flow three-dimensional mean velocity and vorticity fields \cite{xue2022three}. However, to understand how clouds evolve in space and time, it is necessary to know about the evolution of the internal fluctuations through direct measurements. In order to determine the fluctuations and forces that are relevant for cloud dynamics, we need to measure such quantities as temperature, pressure, moisture (humidity), velocity, acceleration, and the magnetic field inside clouds. This can be achieved by following flow parcels inside clouds in a Lagrangian manner to collect simultaneous multi-point observations in different parts of the trajectory. {Nevertheless}, these kinds of observations are still not directly available with the current instrumentation and measurement techniques that are available. The relative motion and relative measurements of the physical fluctuations are important to understand how turbulent dispersion and diffusion develop.	
	\deleted{This information, in turn, is crucial to comprehend the mechanisms that drive the transport and mixing processes of turbulent flows in nature and in industrial applications.} \citet{richardson1926atmospheric} was the first to examine the relative motion of a set of flow particles to establish the initial reasons for relative turbulent dispersion and diffusion. In addition to particle motion, measurements of physical quantities can also be conducted on each fluid particle along the trajectory as time passes.
	
	Direct numerical simulations (DNS) can provide insights into understanding the internal fluctuations and intermittency of clouds. However, DNS simulations can only resolve a small portion of clouds ($\sim$1-10 m) \cite{golshan2021,fossa2022,gallana2022}, and thus cannot provide a global picture on the scale of some tens of kilometers. A deterministic climate model usually has a grid size of 10 km, but state-of-the-art models can resolve smaller grids, e.g. 2.2 km \cite{onln:cl_sim_prace_2020, hentgen_2019_cl_sim}.
	Large amount of clouds at mid-level altitudes are due to strong and frequent updrafts, strong vertical mixing, and dynamical and microphysical conditions that are favorable for the formation of mixed-phase clouds\cite{hentgen_2019_cl_sim}. Convective parameterization schemes may be underestimated in climate simulations \cite{zang2022_gcm}, while convection-resolving simulations require huge amounts of computational resources (e.g., 90 million core hours \cite{onln:cl_sim_prace_2020}). With all this in mind, it is clear that more realistic numerical simulations and in-field experiments are needed. \deleted{: i) numerical simulations that can resolve small-scale dynamics of the clouds and their interaction with the surrounding clear air with a reasonable amount of computational resources; ii) in-field experiments that can provide small-scale variations of the physical quantities (velocity, acceleration, pressure, humidity, temperature, etc.) from direct measurements.}
	
	
	The present work discusses a new measurement system, based on a radiosonde cluster network, that is able to track fluctuations over a 10 km distance. \deleted{However, the application of the radiosonde network is not limited to cloud observations, and could also be extended to other contexts, such as environmental monitoring over urban and industrial areas}. This approach was inspired by the experimental method introduced by L. F. Richardson (1926) \cite{richardson1926atmospheric}. At the state of the art, balloon-borne radiosondes are used as Lagrangian markers in field observations for long periods of time \cite{businger1996balloons}, for example, for circumnavigations in the lower stratosphere around the earth, mainly along the southern and northern polar areas \cite{er1981relative, hertzog2004accuracy}.
	Some relevant instrumentation setups can be found to study tornadogenesis \cite{swenson2019, markowski2017drifter}, Lagrangian observations in the ocean \cite{lacasce2008statistics, novelli2017biodegradable} or very large-scale atmospheric observations at higher levels (200 hPa) of the atmosphere \cite{morel1974relative}. 
	
	The advantages of our proposed in-field measurement system {based on the use of a cluster of mini-expendable radiosondes passively transported by the carrier flow in the Atmospheric Boundary Layer (ABL)}  are threefold as the following can be obtained: (i) direct quantification of Lagrangian turbulent dispersion and diffusion from actual in-field measurement; (ii) the tracking \deleted{of small variations} {of the fluctuation of the physical} quantities inside warm clouds; (iii) a general understanding of the cloud dynamics, with simultaneous measurements in different parts of the cloud.
	
	\deleted{Our current in-field measurement system was designed and developed ab initio in the context of the H2020-COMPLETE project \cite{onln:eucomplete}. The project was an interdisciplinary attempt to decrease knowledge gaps in the understanding of cloud dynamics by combining skills from different areas: numerical simulations, laboratory experiments, and in-field experiments.} The in-field measurement system includes a network of ground stations and a cluster of radiosondes that were prototyped and implemented during the H2020-COMPLETE project \cite{onln:eucomplete}. Each miniaturized radiosonde consists of a 5 cm x 5 cm electronic board that weighs 7g, excluding the battery and the biodegradable balloon, which has a radius of 20 cm (much smaller than traditional weather balloons used for atmospheric sounding and circumnavigation around the earth).
	\deleted{A second version of the mini radiosonde has recently been developed; it has dimensions of 3.5 cm x 4 cm and weighs only 3g.}
	Each radiosonde includes various sets of sensors, that is, pressure, humidity, temperature, an IMU (Inertial Measurement Unit) and GNSS (Global Navigation Satellite System) sensors.

	Between June 2021 and November 2022, a series of preliminary experiments were conducted in the field using both single and multiple radiosondes under varying environmental conditions. The accuracy of sensor readings was verified by comparing them with reference values obtained from traceable instruments of {meteorological stations} provided by INRIM (Italian National Institute of Metrology Research, Turin), ARPA-Piemonte (Piemonte Regional Agency for Environmental Protection, Levaldigi Station, Piemonte, Italy), and OAVdA (Astronomical Observatory of the Autonomous Region of Aosta Valley, Saint-Barthelemy, Italy). \deleted{meteorological stations located in close proximity. Additionally, on September 29, 2021, a set of field tests was carried out at INRIM, involving a cluster of five radiosondes, and on November 3, 2022, another set of field tests took place at OAvdA, utilizing a cluster of ten radiosondes.}
	{Among these, we cite a field test carried out at INRIM, where a cluster of five tethered radiosondes  was involved (September 29th, 2021), and, on November 3rd, 2022, a field test carried out at OAVdA (1700 meters of altitude), where a cluster of ten freely floating radiosondes was launched}. These experiments explored the possibility of: i) performing the spectral analysis of fluctuations and ii) obtaining the distance neighbor graph statistics designed by L.F. Richardson (1926, \cite{richardson1926atmospheric}) for turbulent dispersion analysis in the atmosphere, a statistics which has yet to be realized in the context of in-field atmospheric observations. {However, the application of the radiosonde network is not limited to cloud observations, as it could also be extended to other contexts, such as environmental monitoring over urban and industrial areas}
	\deleted{In addition, the application scope of the sensor cluster could be extended to other contexts, such as environmental monitoring over urban and industrial areas.}

	
	
	The measurement system is described in Section \ref{sec:system_description}. The traceability of the system, the quality of the obtained dataset, and validation with reference systems are discussed in Section \ref{sec:result_discussion}. The results from the preliminary cluster in-field experiment are provided in Section \ref{sec:field_tests}.  \deleted{, and we conclude our discussion in} {In Section \ref{sec:conclusion}, the main conclusions are discussed.}
	
	
	\section{Description of the measurement system}\label{sec:system_description}
	
\begin{figure*}[h!]
		\centering
		\begin{subfigure}[b]{0.8\textwidth}
			\includegraphics[width=\linewidth]{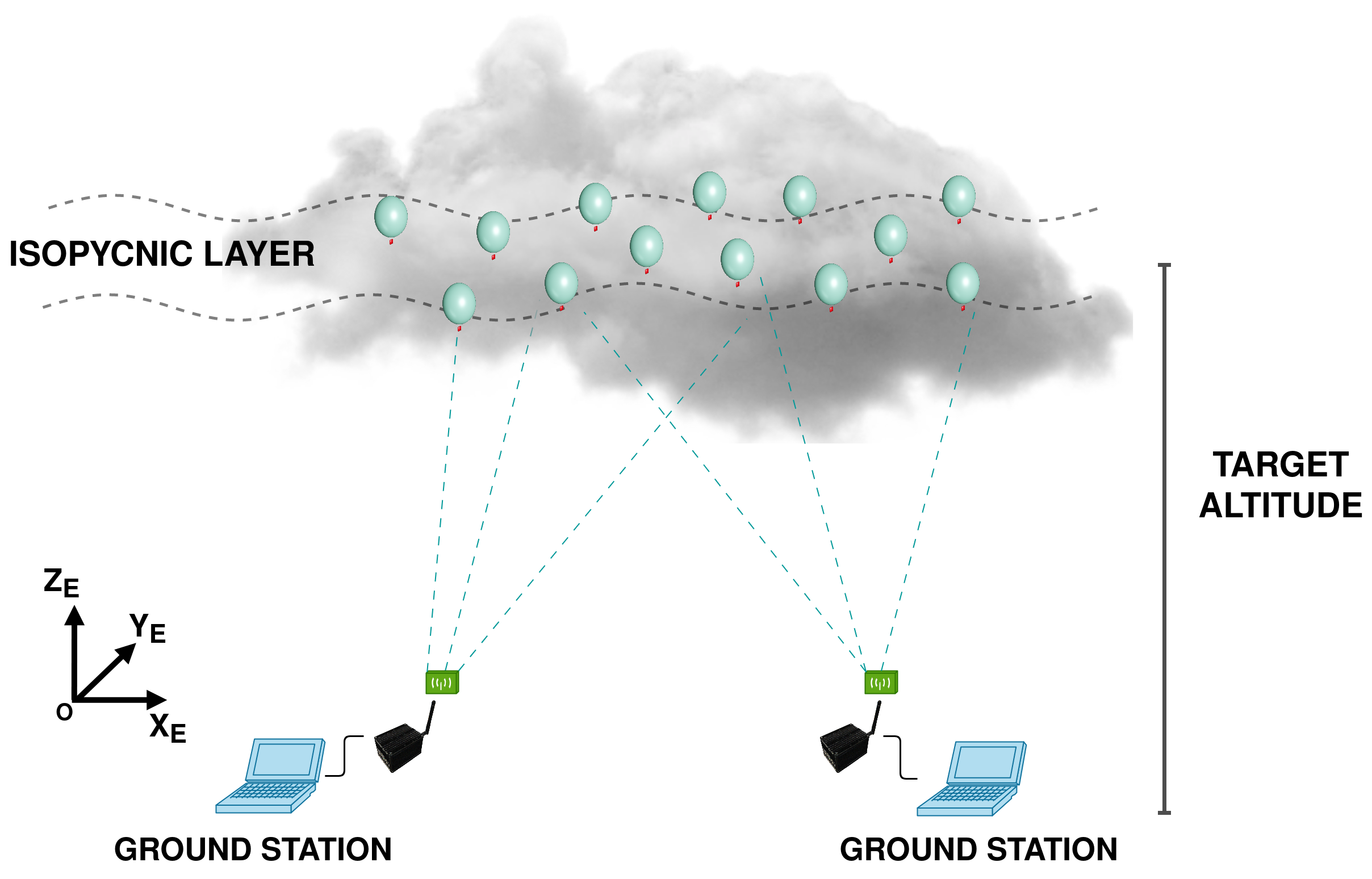}
		\end{subfigure}
		\caption{Illustration of the in-field experiment with a cluster of radiosondes and a set of receiver stations. The cluster of radiosondes is floating across the isopycnic layer at the pre-configured target altitude (1-2 km). The launching point of the cluster is considered as the origin of the experiment observation frame, X$_E$,Y$_E$,Z$_E$.}
		\label{fig:scheme}
	\end{figure*}

	The making the probes to observe relevant parts of the cloud over a different range of scales during a cloud's lifetime is challenging in terms of instrumentation setup. The following measurement system was suggested to accomplish this challenging task for in-field experiments as shown in Figure \ref{fig:scheme}. The measurement system {consists of three main building blocks}: a {cluster} of radiosondes, a set of receiver stations, and a post-processing machine. The aim was to place a set of radiosondes inside warm clouds (or any other atmospheric environment), where each radiosonde could passively follow the fluid flow across isopycnic layers at the target altitude. Thus, it would be possible to obtain information about the real dynamics of the surrounding fluid, which is a cloud, when a balloon is inside, and clear air when the balloon is outside the cloud.
	
	Each radiosonde transmits sensor readings to ground stations through the Lora radio transmission protocol. LoRa is a relatively new proprietary communication technology that allows long distance communications while consuming relatively low power. It utilizes license-free industrial, scientific, and medical (ISM) frequency bands to exchange information at low data rates. Ground stations receive data from radiosondes and are connected to post-processing machines. All the data are stored in a post-processing machine. The same data transmitted by a radiosonde can be received by different ground stations to reduce data losses. The design of the radiosonde electronic board, the tests conducted inside the environmental chamber, and the initial performance evolution of the radiosonde in field experiments were described in a previous work by Paredes et al. \cite{miryam_sensors2021}.
	
	\subsection{Radioprobes}
\begin{figure*}
		\centering
		\begin{subfigure}[b]{0.205\textwidth}
			\includegraphics[width=\linewidth]{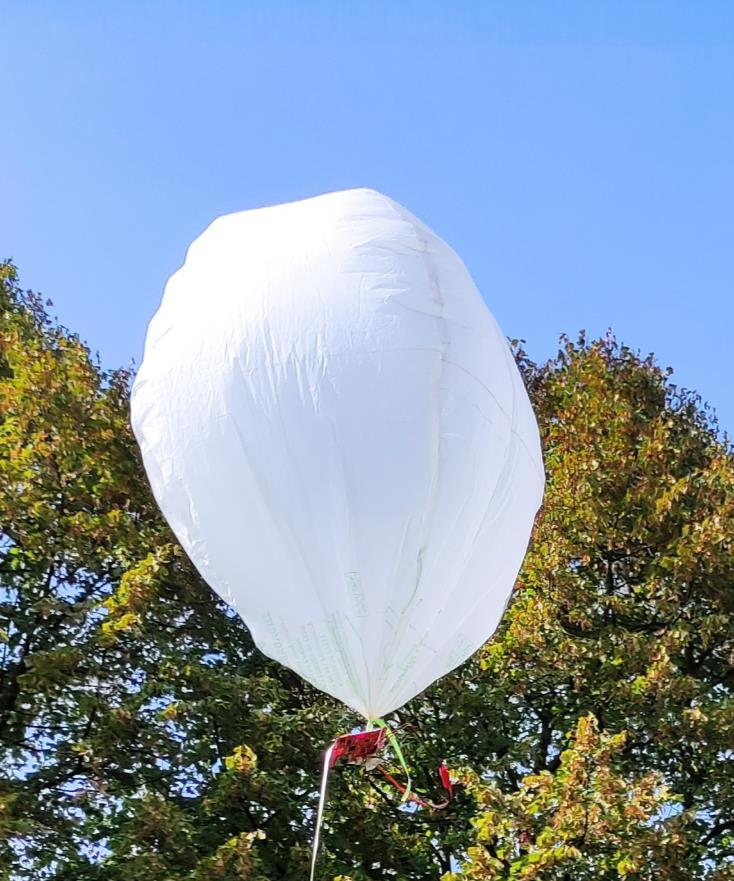}
			\caption{}
		\end{subfigure}
		\hspace{0.02\textwidth}
		\begin{subfigure}[b]{0.22\textwidth}
			\includegraphics[width=\linewidth]{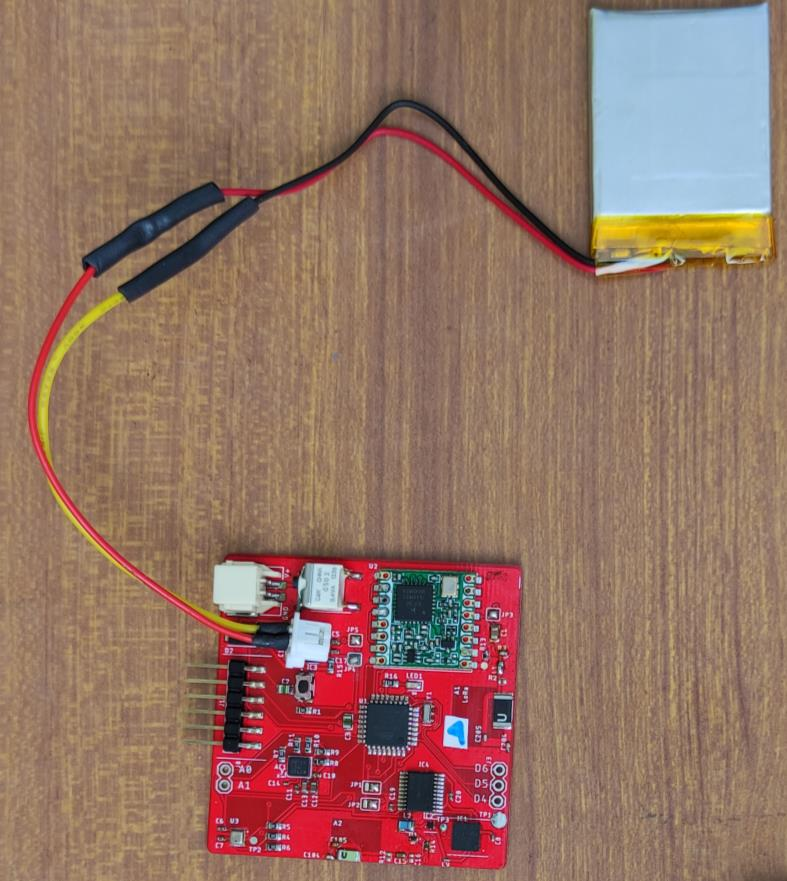}
			\caption{}
		\end{subfigure}
		\hspace{0.02\textwidth}
		\begin{subfigure}[b]{0.48\textwidth}
			\includegraphics[width=\linewidth]{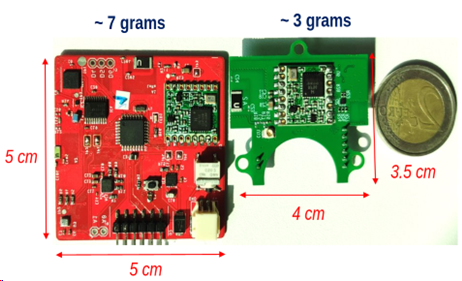}
			\caption{}
		\end{subfigure}
		\caption{(a) Radiosonde attached to the ground with a thread during an in-field test. (b) The current version of the radioprobe electronic board with a battery. (c) {The current prototype (red) is presented together with a prospective smaller two-layer design (green) and a two-euro coin for size comparison.}}
		\label{fig:radioprobeversions}
	\end{figure*}

	The assembled radiosonde can be seen in Figure \ref{fig:radioprobeversions}(a), which includes a biodegradable balloon and a radioprobe electronic board. The radiosonde can float (stay on air) with the helium filled balloon, and can transmit the sensor measurements using a battery-powered radioprobe as shown in Figure \ref{fig:radioprobeversions}(b) for several hours. The embedded electronics \deleted{(microprocessor, radio module, and sensors)} can measure velocity, acceleration, pressure, temperature, and humidity fluctuations in the surrounding environment. \deleted{This configuration was selected on the basis of the results of in-field experiment tests (see Sections \ref{sec:result_discussion} and \ref{sec:field_tests})}. {Figure \ref{fig:radioprobeversions}(c) shows both the current prototype (red) and a prospective two-layered smaller design (green) of the radioprobe.} \deleted{The new version of the radioprobe prototype is currently undergoing hardware and software tests.} This paper focuses on the current working prototype. {The radioprobe electronic board comprises several essential components, such as a microcontroller, a power module, a radio transmission module, a PHT (pressure, humidity and temperature) module, GNSS and IMU sensors. Additionally, it is equipped with two ceramic chip antennas: one for the GNSS sensor and another for the radio module.}
	
	The \textit{microcontroller} is a data-processing and control unit that allows to control other {components, the acquisition of sensor readings, and the automated execution of function calls within the device}. The \textit{radio transmission module} of the radioprobe enables one-way wireless communication with ground stations using radio frequency signals. \textit{PHT} \deleted{(Pressure, Humidity, and Temperature)}, \textit{IMU} \deleted{(Inertial Measurement Unit)}, and \textit{GNSS} \deleted{(Global Navigation Satellite System)} sensors provide readings of physical quantities. {The list of measured} quantities are described in Table \ref{tbl:sensor_ranges}, together with the operating ranges of the sensors, {declared accuracies and the corresponding sensor devices}\deleted{provider functional units}.
	\begin{table}[h!]
		\centering
		\caption{{List of physical quantities measured during the experiments. The measurement ranges, accuracy and the corresponding sensor component names are reported. Accuracy values are reported as given in sensor datasheets. Note that in the 'Acceleration' row, \textbf{g} represents the standard gravity acceleration of 9.81 m/s$^2$.}}
		\resizebox{\textwidth}{!}{%
			\begin{tabular}{p{3cm} p{3cm} p{3.2cm} p{3cm}} 
				\hline
				Physical quantity & Range & Declared accuracies & Device Name\\ [0.5ex] 
				\hline\hline
				Pressure & [300, 1100] mbar & ± 1 hPa & \multirow{3}{3cm}{Bosch BME280 \cite{docs:pht_datasheet}} \\ 
				Humidity & [0, 100] \%  & ± 3 \% &  \\
				Temperature & [-40, 85] $^{\circ}$C & ± 1-1.5 $^{\circ}$C  &  \\ \hline
				Longitude & [-180, 180] degrees  & \multirow{2}{3cm}{Horizontal accuracy = ±3.5m} & \multirow{4}{*}{UBLOX ZOE-M8B \cite{docs:ubx_datasheet}} \\
				Latitude & [-90, 90] degrees  & &  \\ 
				Altitude & $<$ 50000 m & ± 7.0 m & \\
				Speed & $<$ 500 m/s & ± 0.4 m/s & \\ \hline
				Acceleration & [-16, 16] g  & ± 90 mg & \multirow{2}{*}{STM LSM9DS1 \cite{docs:imu_datasheet}} \\
				Magnetic field & [-16, 16] gauss  & ± 1 gauss & \\ [1ex] 
				\hline
			\end{tabular}%
		}
		\label{tbl:sensor_ranges}
	\end{table}
	
	The sensors were chosen on the basis of their compact size and low-power consumption. Furthermore, they were configured to work in energy-efficient mode {during the experiments}. In fact, GNSS, by U-blox, has a compact size and can be configured to operate in {super power efficient mode (Super-E mode \cite{docs:ubx_datasheet})}\deleted{e-mode}. In other studies, researchers exploited high precision GNSS sensors \cite{swenson2019}, but such sensors consume more power, an aspect that is crucial for our application context. Moreover, the current GNSS sensor can provide compact PVT (position, velocity, and time) navigation information by using the proprietary {UBX-PVT} protocol \cite{docs:ubx_rec_desc}, which cannot be provided in a single sensor reading when using the traditional NMEA {(National Marine Electronics Association) packets}\deleted{protocol}\cite{onln:nmea}.
	
	{Besides its compact and lightweight design, the radioprobe electronic board offers lower unit cost than the commercially available radiosondes for sounding experiments. The cost of each individual radiosonde was reported to be around 200 US\$ for high altitude soundings\cite{TheLifeCycleofaRadiosonde2013} and 155 US\$ for low altitude soundings\cite{viablealternativers2014} without considering the balloon cost. The unit cost of the current radioprobe electronic board is 90 US\$ for an order amount of 20 pieces. The cost goes down to 57 US\$ and 54 US\$,  for an order of 1 000 and an order of 10 000 pieces, respectively. Here, the unit cost consists of the costs of the sensors, PCB substrate, assembly, battery and other electronic components.}
	
	\subsection{Isopycnic floating and biodegradable balloon}
	The radiosonde system needed to float at an almost constant altitude during the experiments. {To achieve this, we designed balloons using non-elastic, biodegradable material, Mater-Bi. The balloon material keeps its quasi-spherical shape without expansion while floating at a constant altitude}\cite{basso2020,Yajima2009}.
	{Furthermore, the use of biodegradable materials for both balloons and possibly electronic boards serves to minimize the environmental impact of the entire radiosonde system.} The characteristics of the balloon material, the processing methods, and the polymer coatings were studied in the COMPLETE project by Basso et al. \cite{basso2020}. During this study, green polymers, such as Mater Bi and PLA were examined and compared with materials used for traditional weather balloon production, such as latex and mylar. The properties of the above-mentioned materials were analyzed in laboratory experiments in collaboration with IIT (Italian Institute of Technology) Genoa. The main properties of interest were the \textit{tensile strength, hydrophobicity, helium permeability, and resistance to variations} of the surrounding temperature and humidity. As a result of these experiments, it was concluded that Mater-Bi with applied coatings {(such as a mixed solution of carnauba wax, pine resin, and acetone or hydrophobic nano-silica combined with dimethyl silicone oil)} performed the best to meet the predefined requirements \cite{basso2020}.
	
	Spherical balloons ($R_b$ = 20 cm) were made for the recent in-field experiments from store-bought Mater-Bi bags. The selected Mater-Bi material was 20 $\mu$m thick and had a density of 1.24 g cm$^{-3}$, and was thus thinner than that used in the previous studies (30 $\mu$m) carried out by Basso et al.\cite{basso2020}. Therefore, the balloon mass was reduced by a factor of 1.5, which in turn reduced the overall payload \deleted{budget} (Eq. \ref{eq:volballoon}). The balloon dimensions were identified considering the weight of the radiosonde electronic board with a battery and standard atmospheric parameters \cite{atmosphere1975international} at a target floating altitude. The volume of the balloon has to satisfy the following equation for stable floating at a fixed altitude:
	\begin{equation}
		V_b = \frac{m_r + m_b}{\rho_a - \rho_g} = \frac{m_r + m_b}{\rho_a (1- M_g/M_a)},
		\label{eq:volballoon} 
	\end{equation}
	{where subscript $b$ refers to the balloon, $r$ represents the radioprobe, $a$ stands for air and $g$ for gas, in this case, helium. $m_r$ denotes the mass of the radioprobe with the battery and the connections, while $m_b$ is the mass of the balloon. $\rho_a$ and $\rho_g$ stand for air and gas densities at a given altitude, and $M_a$ and $M_g$ represent the molar masses of air and gas inside the balloon, $V_b = 4/3\pi R_b^3$ is the volume of the balloon and $m_b = S\Delta \rho_m = 4 \pi R_b^2\ \Delta\, \rho_m$, where $S$ is the surface area of the  balloon with radius $R_b$, while $\Delta$ and $\rho_m$ refer to the sheet thickness and density of the Mater-Bi material}.
	
	
	\begin{figure}[h!]
		\centering
		\includegraphics[width=0.8\linewidth]{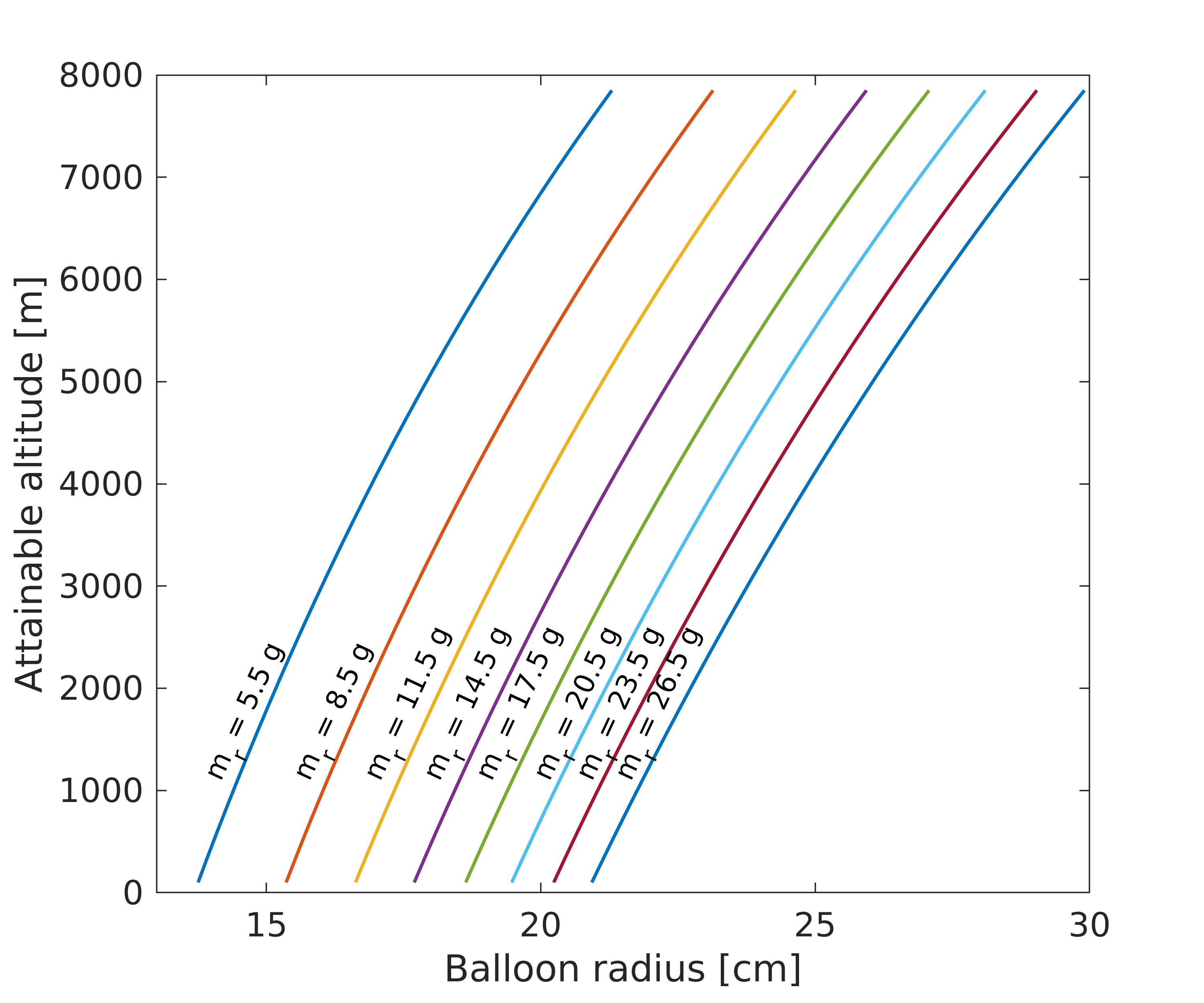}
		\caption{{The attainable altitude is computed as a function of the balloon dimensions by considering the properties of the selected material and type of gas used (Helium). In our case we selected biodegradable Mater-Bi material by Novamont®, which has a density of 1.240 kg/m3 and a thickness of 20 microns. Here, different solid lines represent the different values of possible radiosonde weights (radioprobe, battery and connections) between 5.5 grams and 26.5 grams. The green line corresponds to the weight of the current prototype, which is 17.5 grams.} A detailed breakdown of the weights of the radiosonde and the balloon is given in Table \ref{tbl:probeweight}.}
		\label{fig:weightvsvol}
	\end{figure}
	
	Figure \ref{fig:weightvsvol} highlights the relationship between {attainable altitude and balloon radius}\deleted{total liftable payload (excluding the balloon weight, $m_b$) and the floating altitude for the chosen balloon dimensions}. In the current design, the weight of the radioprobe with the battery and the connections, $m_r$, is 17.5 g as specified in Table \ref{tbl:probeweight}. It can be seen from Figure \ref{fig:weightvsvol} that a 20 cm radius balloon can lift the radiosonde to approximately 1700 m above sea level, while a 21 cm radius balloon can lift it to around 2600 m, and so forth. \deleted{The second version of the prototype (8.5 g) instead allows even smaller balloons and smaller amounts of gas (e.g. Helium) to be used to reach the same floating altitude.}
	
	\begin{table}[h!]
		\centering
		\caption{Distribution and comparison of the total payload of the radiosonde. The balloon weight was computed for a spherical balloon with the radius value indicated inside brackets.}
		\begin{tabular}{l  l}
			\hline \rule{0pt}{2.6ex}
			Part & Mass [grams] \\ \hline 
			Radioprobe & 7 \\ 
			Battery & 8  \\
			Connections & 2.5 \\
			Balloon & 12.5 (R=20 cm)\\ \hline \rule{0pt}{2.6ex}
			Total & 30\\
			\hline
		\end{tabular}
		\label{tbl:probeweight}
	\end{table}
	
	\subsection{Architecture of the radiosonde cluster and data processing}
	The LoRa-based wireless sensor network (WSN) concept was adopted for the radiosonde network. A star architecture, in which, each radiosonde is connected to the ground receiver station with a point-to-point link, was adopted. A feasibility analysis of the selected network architecture was carried out for different application scenarios \cite{bertoldo2018feasibility, bertoldo2018urbannoisy, paredes2019propagation}. The results of the first in-field tests of the network architecture in the current application context were presented in the previous work of Paredes et al.\cite{miryam_sensors2021}. LoRa protocol-based WSN networks are generally used within the LoRaWAN infrastructure. However, in this work, the LoRa protocol has been used to create an ad hoc private network and to adapt the technology to the working scenario. Therefore, RFM95, the commercial off-the-shelf LoRa-based transceiver module from HopeRF was used. This module features long-range spread spectrum communication links and high immunity to interference and it optimizes the power use \cite{miryam_sensors2021}. The ground stations and the transmitters were equipped with the same radio module, RFM95.
	
	\begin{figure}[h!]
		\centering
		\includegraphics[width=0.7\linewidth]{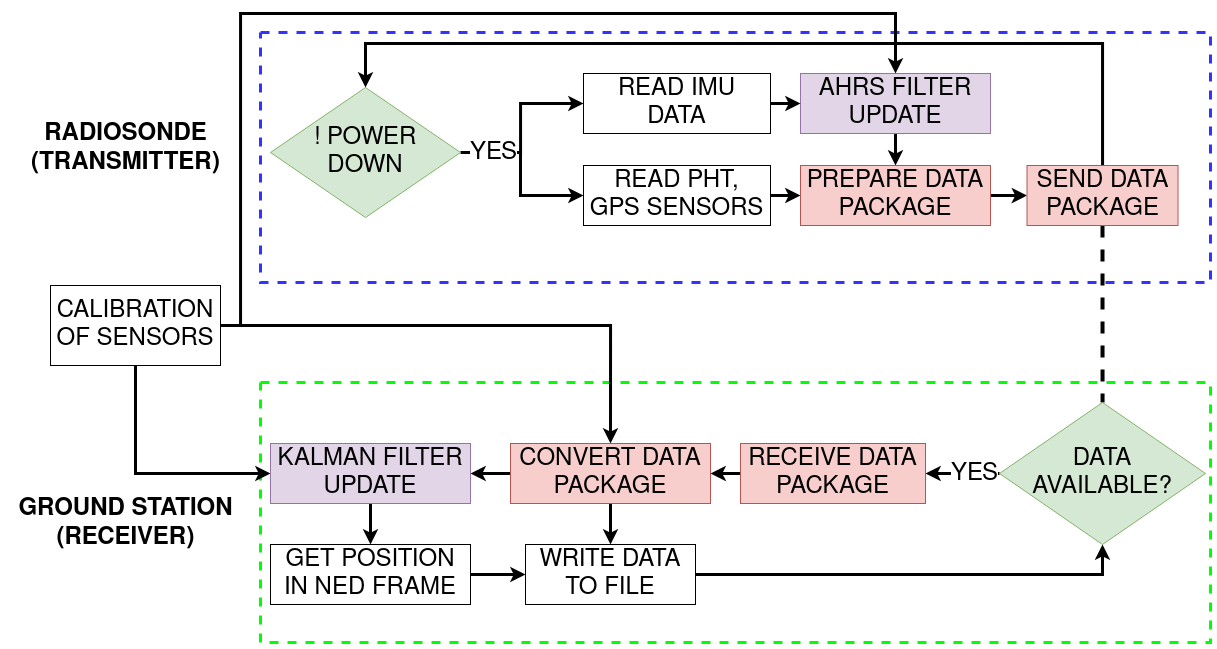}
		\caption{The data processing flow of the radiosonde network is shown as the communication between a single radiosonde (dashed blue rectangle) and a ground station (dashed green rectangle). Calibration values of the sensors were used for both the radiosonde and the ground station. Specific pre-launch calibrations were carried out to identify the possible bias offset values of the accelerometer, the magnetometer and the  pressure humidity and temperature \deleted{(PHT)} MEMS {(Micro-electromechanical systems)} sensors.}
		\label{fig:probeworkingprinciple}
	\end{figure}
	
	The data-processing flow of the radioprobe can be seen in Figure \ref{fig:probeworkingprinciple}. The flow consists of the steps that have to be performed by the radioprobe (transmitter) and ground station (receiver). Some of the processing is performed directly by the transmitter, and more power- and time-consuming parts are performed by the receiver, albeit with the help of the post-processing machine. As can be observed in Figure \ref{fig:probeworkingprinciple}, the sensor data is processed by the AHRS (Attitude and Heading Reference System) filter before being sent to the ground station. The AHRS filter acquires readings from a 9-DOF IMU sensor (3x accelerometers, 3x gyroscopes, and 3x magnetometers) and provides the course (orientation) of the radioprobe as output. In order to remove any possible errors introduced by the sensor readings, the AHRS filter also uses sensor calibration data \cite{madgwick2011}. IMU sensor readings of the radioprobe are provided in the body frame (xyz) of the IMU sensor. These readings can be translated into local experiment frames (X$_e$Y$_e$Z$_e$) by using orientation data from the AHRS filter. The acceleration data provided in the local experiment frame, can also serve as positioning information during GNSS outages. This can be achieved by integrating the acceleration data with GNSS sensor data using a Kalman filter, which operates in two different operating modes: \textit{predict} and \textit{update} \cite{Kalman1960ANA}. In the predict mode, IMU data can used to provide information on the position realtive to the previous reference position. As soon as GNSS data are available, the reference position can be updated. In this way, position information could be available during GNSS outages. Since the GNSS sensor consumes much more power than the IMU and other sensors, this approach may help to reduce power consumption.
	
	\subsection{Metrological traceability}
	
	The validation process of the mini-radiosondes is intended to have a robust metrological foundation, with the aim of ensuring comparability of the readings of the radiosondes and of obtaining a link with the absolute value of the temperature. INRIM was involved in this validation process because of its well-established expertise in the metrology of meteorology and the climate \cite{meteomet2015, meteomet2018} and because of its previous experiments on radiosondes \cite{rosoldi2022vaisala, musacchio2015arctic}. {The preliminary calibration and characterization of the sensors were performed through climate chamber experiments in INRIM's Applied Thermodynamics Laboratory. These experiments employed the Kambic KK190 CHLT climate chamber, platinum resistance thermometers (PT100), and a Delta Ohm humidity probe to evaluate the sensitivity and accuracy of the selected temperature and humidity sensors (BOSCH BME280) for the current radiosonde system.} {The uncertainty of the Pt100 ranges from 0.011 $^\circ$C for positive temperatures and 0.020 $^\circ$C for negative temperatures. The total uncertainty of the Delta Ohm probe declared is ±3\% RH}\cite{miryam_sensors2021}.
	
	In-field test procedures were defined for the experiments prior to the launch of \textit{tethered} and \textit{freely floating} balloons equipped with the radiosondes. {The first stage of experiments was carried out at INRIM campus, and radiosonde sensor readings were compared with reference sensors calibrated at the INRIM laboratory.} {Additionally, the test procedures allowed us to choose a proper radiosonde layout configuration between two proposed configurations (see Figure \ref{fig:inr1_conf}) by quantifying radiosonde sensor accuracies with respect to the reference sensors}.
	
	\deleted{The first experiment was carried out at INRIM campus, and the results were compared with those of a Vaisala WXT510 automatic weather station. The Vaisala station is located at the height of 2 m on natural grass in an open area on the campus. The temperature and pressure sensors of the station were calibrated in the INRIM laboratory and thus provide accurate measurements. The different configurations of the radiosonde sensors were tested while lying on the Vaisala station measurement area to compare their readings with reference ones.}
	
	At a second stage, a transportable system was assembled for the purpose, to act as on site calibration device for pre-launch checks. The system was installed for the in-field experiment with a cluster of freely floating radiosondes (OAVdA, St. Bathelemy, Italy, Nov 3, 2022). The system was equipped with a PT100 CalPower platinum resistance thermometer, calibrated in the INRIM laboratory, and used as a reference sensor. PT100 was installed in a Barani helical passive solar shield. A further PRT temperature sensor was added without a shield, to reproduce unprotected radiation conditions in order to estimate the magnitude of the influence of solar radiation. A Fluke DAQ 1586A multimeter was used for resistance data acquisition and the calculation of corrected temperature values by implementing a calibration curve. The system was adapted to the location by selecting an appropriate, obstacle-free area in the launching zone to perform the pre-launch check of the radiosondes. Mini-radiosondes were checked against INRIM facility measurements during the in-field experiment, by coupling them to the system in an open area for 10 minutes before the balloons were launched. Sensor readings were acquired to evaluate their comparability in terms of ${\Delta}$T$_{rel}$ (difference between the mean temperature and temperature reading of each sonde) and the correct temperature value by comparing mini-radiosonde readings and reference sensor readings (${\Delta}$T$_{abs}$). This procedure allows both the spread of readings between the tested radiosondes to be evaluated and the correction, ${\Delta}$T, at the pre-launch condition to be calculated.
	
	{In addition to above calibration and pre-launch test procedures, dual-sounding experiments were performed to quantify uncertainties of radiosonde sensor measurements. Temperature, humidity, pressure and positioning dataset were compared with the reference radiosonde measurements during three dual-sounding experiments from October 2020 to July 2023.} {For temperature measurements, the mean differences ($<T-T_{ref}>$) and the normalized mean differences relative to the reference sensor readings ($<(T-T_{ref})/T_{ref}>$) were examined.  This process was carried out along the altitude in different experiment sites (see Section \ref{results_vertical_launch} for results and discussion). In Table \ref{tbl:sensor_ranges} above, accuracy values of the sensors were reported from manufacturer datasheets. These values were considered as barely preliminary, and they should be revisited as shown by the observed accuracies deduced  from the comparison tests. Observed accuracy values of the current radiosonde sensors are presented in the following Sections 3 and 4. }
	
	
	\section{Measurements and validation}\label{sec:result_discussion}
	As mentioned in the previous sections, the preliminary results were presented in the works of Basso et al.\cite{basso2020} and Paredes et al.\cite{miryam_sensors2021}. However, not all the components of the measurement system were fully field-tested or confirmed in the earlier works. Furthermore, some post-processing techniques can only be used after appropriate in-field tests have been conducted. {In this section, the proposed measuring system is compared and validated relative to established measurement methods and instrumentation during fixed point measurements at the ground level and vertical profiling observations of the atmosphere.} {Table \ref{tbl:experiment_list} gives a list of experiments carried out during the development of the radiosonde cluster system. The list includes experiments mentioned inside this section and the following Section \ref{sec:field_tests}.} \deleted{The initial experiments were conducted at the INRIM campus in Turin and at the Levaldigi airport, Cuneo, Italy, to evaluate the radiosonde setup and to compare the sensor readings with reference readings. These experiments were necessary steps to implement Lagrangian tracking and correlation measurements on a cluster of radiosondes. The preliminary in-field experiments with a cluster of radiosondes are discussed in Section \ref{sec:field_tests}.}
	
	\begin{table}
		\caption{{In-field measurement campaigns during the development of the radiosonde cluster and measurement system.}}
		\resizebox{0.95\textwidth}{!}{%
			\begin{tabular}{p{2.3cm}|p{4.5cm}|p{2cm}|p{4cm}} 
				\hline
				\textbf{Date} & \textbf{Description} & \textbf{Place} & \textbf{Coverage} \\ [0.5ex] 
				\hline\hline
				\textbf{Oct 28, 2020} \newline \textbf{June 9, 2021} & Two dual launch experiments with Vaisala RS-41 SG probe in collaboration with ARPA-Piemonte & Levaldigi Airport, Cuneo, Italy  & Vertical atmospheric profiling, up to 14 km in distance and 9 km in altitude. \\ \hline
				
				\textbf{July 20, 2021} & Radiosonde configuration testing; 2 configurations & \multirow{2}{2cm}{INRIM, Turin, Italy}  & \multirow{2}{4cm}{short-range, controlled setup, up to 100 m.} \\ \cline{1-2}
				
				\textbf{Sep 29, 2021} & Testing the cluster of tethered radiosondes (5 sondes) &   & \\ \hline
				
				\textbf{Feb 10, 2022} & -Testing the cluster of tethered radiosondes (5 sondes) in approximately operational environment; -Radiosonde balloon tracking with stereo vision analysis. & \multirow{2}{2cm}{OAVdA, St. Barthelemy, Aosta, Italy}  & short-range, controlled setup, up to 100 - 150 m. \\ \cline{1-2} \cline{4-4}
				
				\textbf{Nov 3, 2022} & The first experiment with the free flying cluster of 10 radiosondes &   & long-range, freely floating setup, up to 9 km \\ \hline
				
				\textbf{July 6, 2023} & Dual launch experiment with Vaisala RS-41 SGP probe in collaboration with MET OFFICE (UK)  & Chilbolton Observatory, Chilbolton, UK &  long-range, vertical profiling, up to 34 km horizontally within 100 – 12000m altitude range during the ascent  \\ [1ex] 
				\hline
			\end{tabular}%
		}
		\label{tbl:experiment_list}
	\end{table}
	
	\subsection{Pre-launch calibration and fixed-point measurements}
	\begin{figure}[h!]
		\centering
		\begin{subfigure}[b]{0.25\textwidth}
			\includegraphics[width=\linewidth]{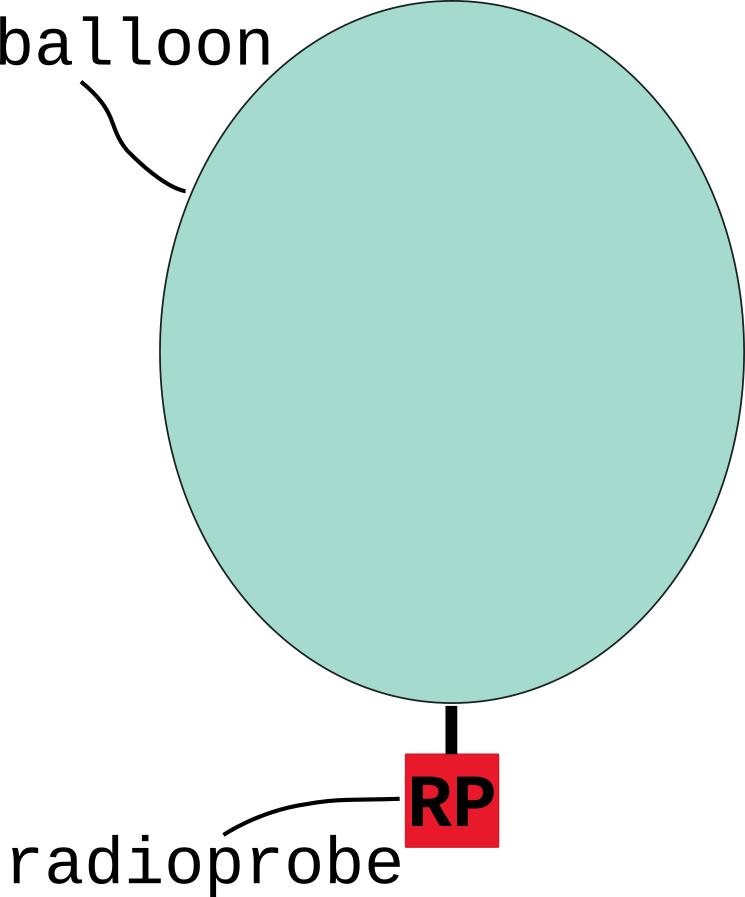}
			\caption{}
		\end{subfigure}
		\hspace{0.05\textwidth}
		\begin{subfigure}[b]{0.25\textwidth}
			\includegraphics[width=\linewidth]{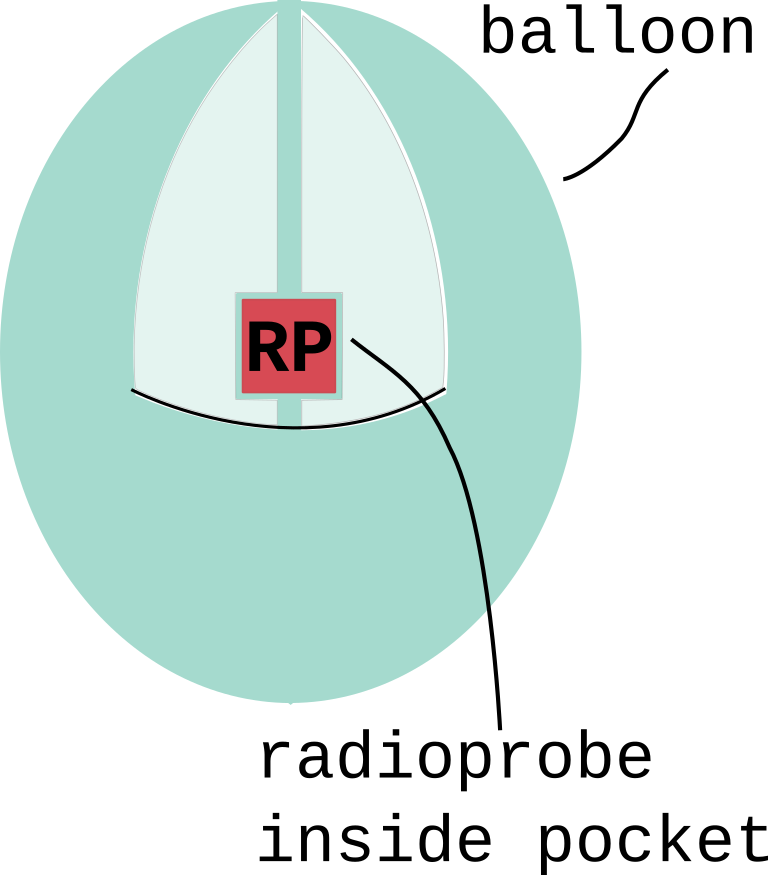}
			\caption{}
		\end{subfigure}
		
		\caption{The radioprobe electronic board was tested in two different radiosonde configurations. (a) Configuration A: the radioprobe board is \textbf{outside} the balloon. (b) Configuration B: the radioprobe board is in a pocket \textbf{inside} the balloon.}
		\label{fig:inr1_conf}
	\end{figure}
\begin{figure}[ht!]
		\centering
		\begin{subfigure}{0.4\textwidth}
			\caption*{Before, without balloons}
			\includegraphics[width=\linewidth]{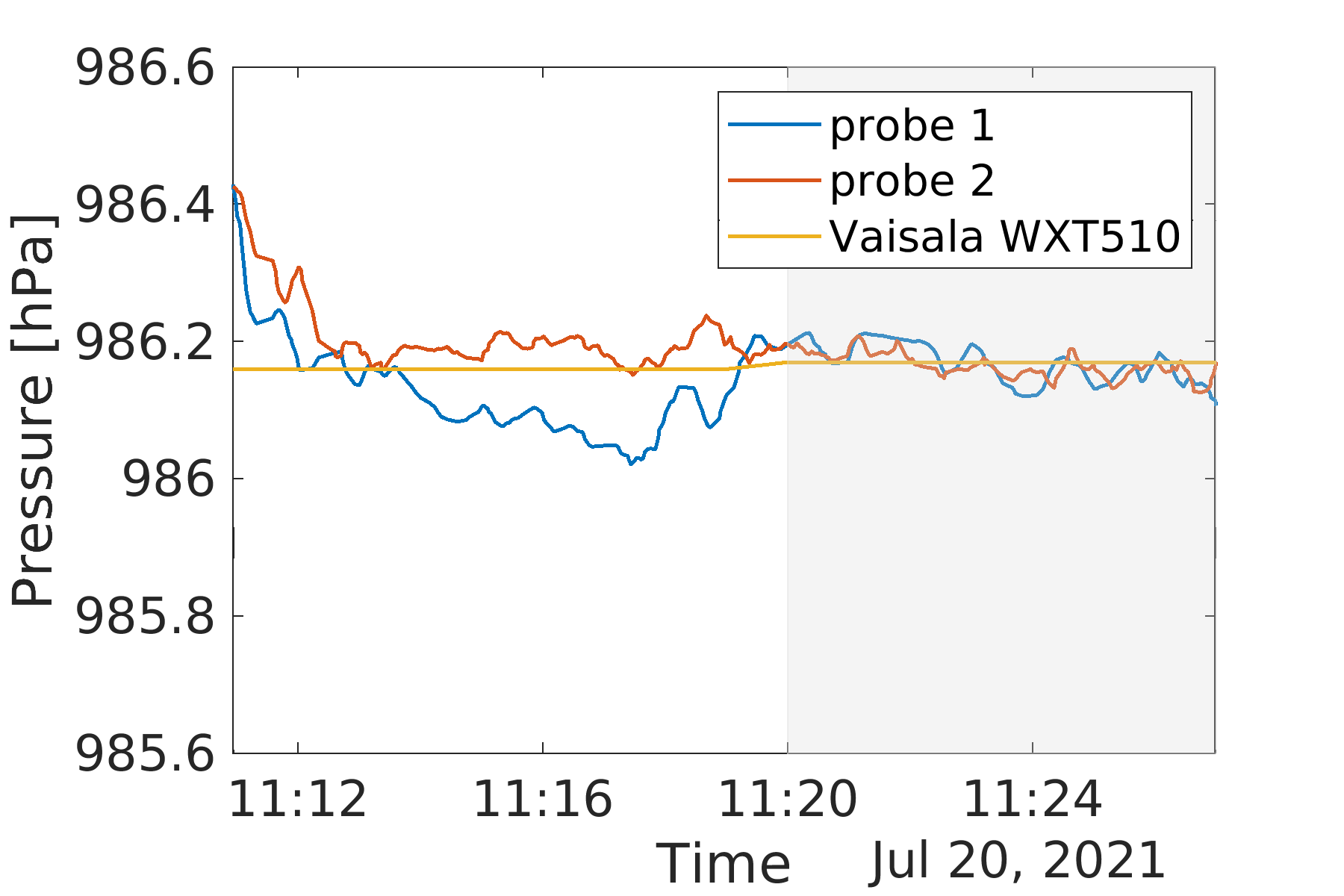}
			\caption{}
		\end{subfigure}
		\hspace{0.01\textwidth}
		\begin{subfigure}{0.4\textwidth}
			\caption*{After, with balloons}
			\includegraphics[width=\linewidth]{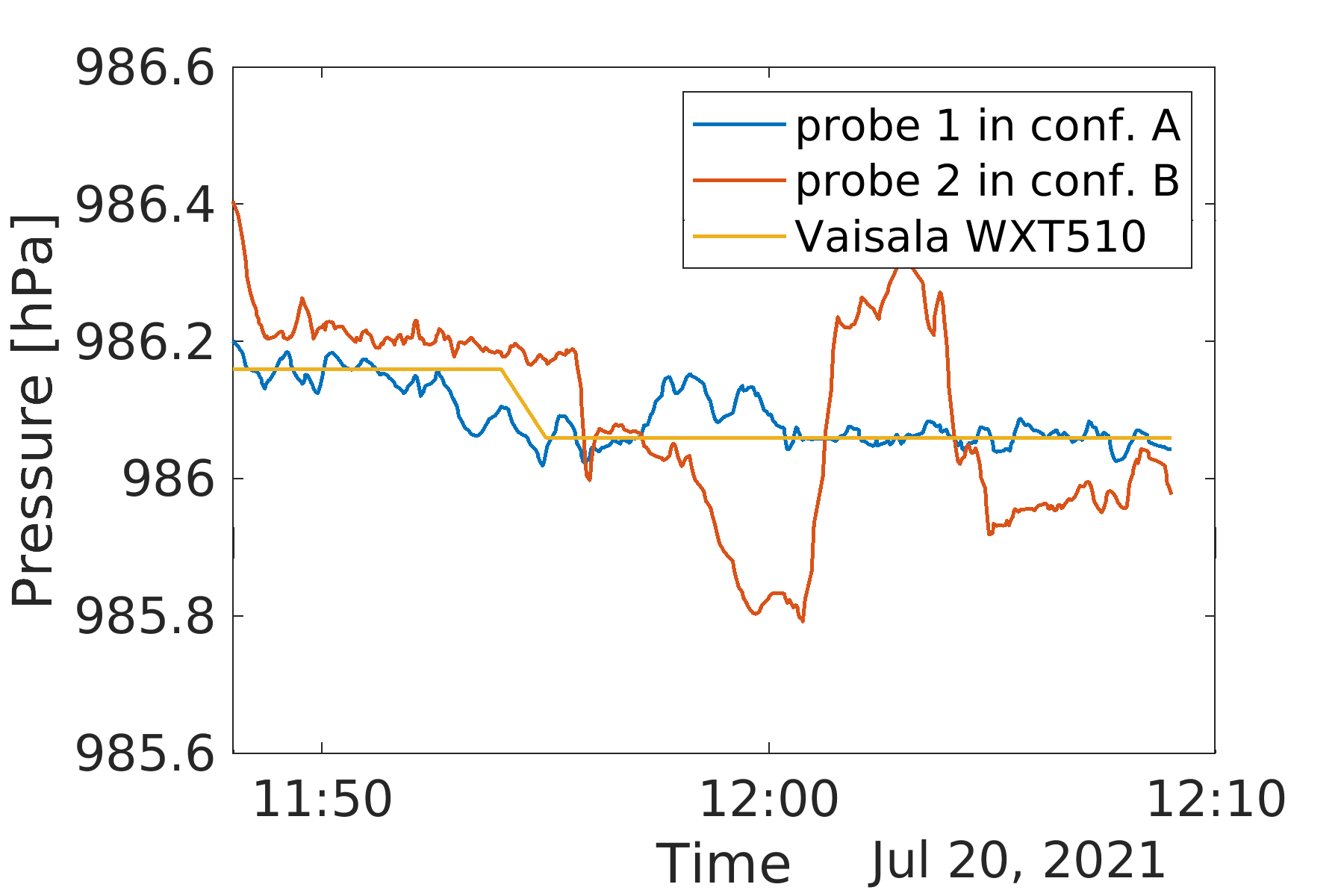}
			\caption{}
		\end{subfigure}
		\vspace{0.001\textheight}
		\begin{subfigure}{0.4\textwidth}
			\includegraphics[width=\linewidth]{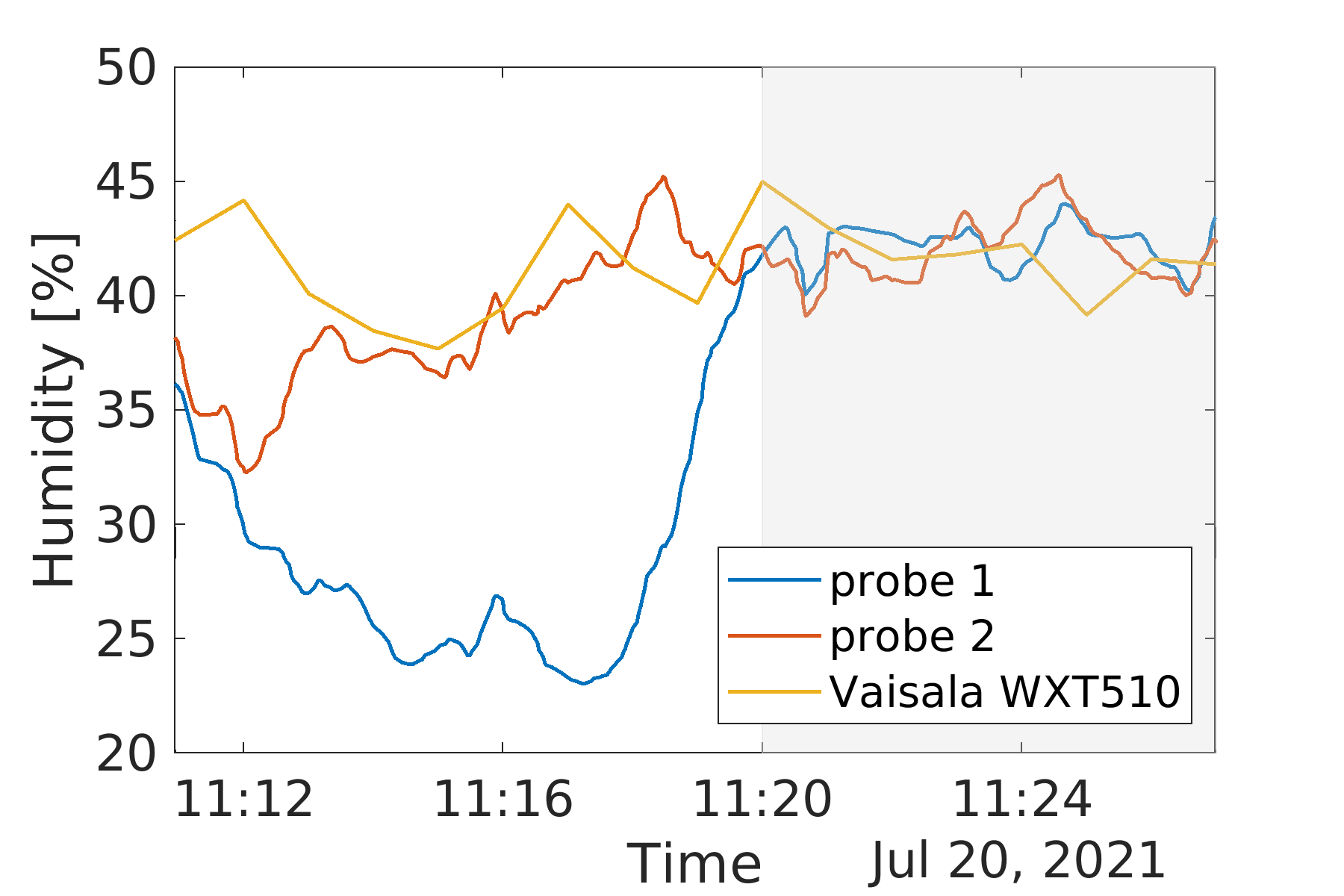}
			\caption{}
		\end{subfigure}
		\hspace{0.01\textwidth}
		\begin{subfigure}{0.4\textwidth}
			\includegraphics[width=\linewidth]{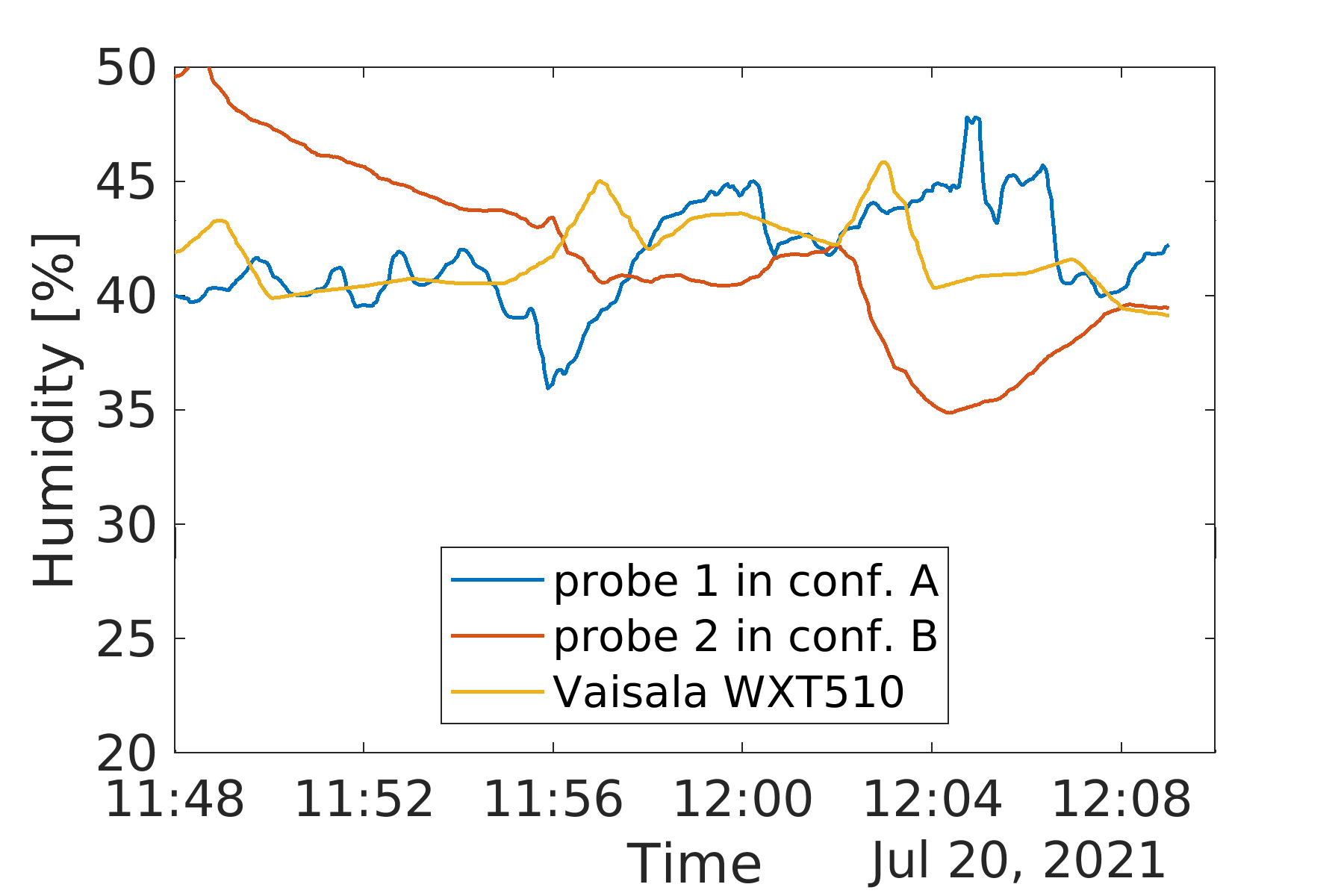}
			\caption{}
		\end{subfigure}
		\vspace{0.001\textheight}
		\begin{subfigure}{0.4\textwidth}
			\includegraphics[width=\linewidth]{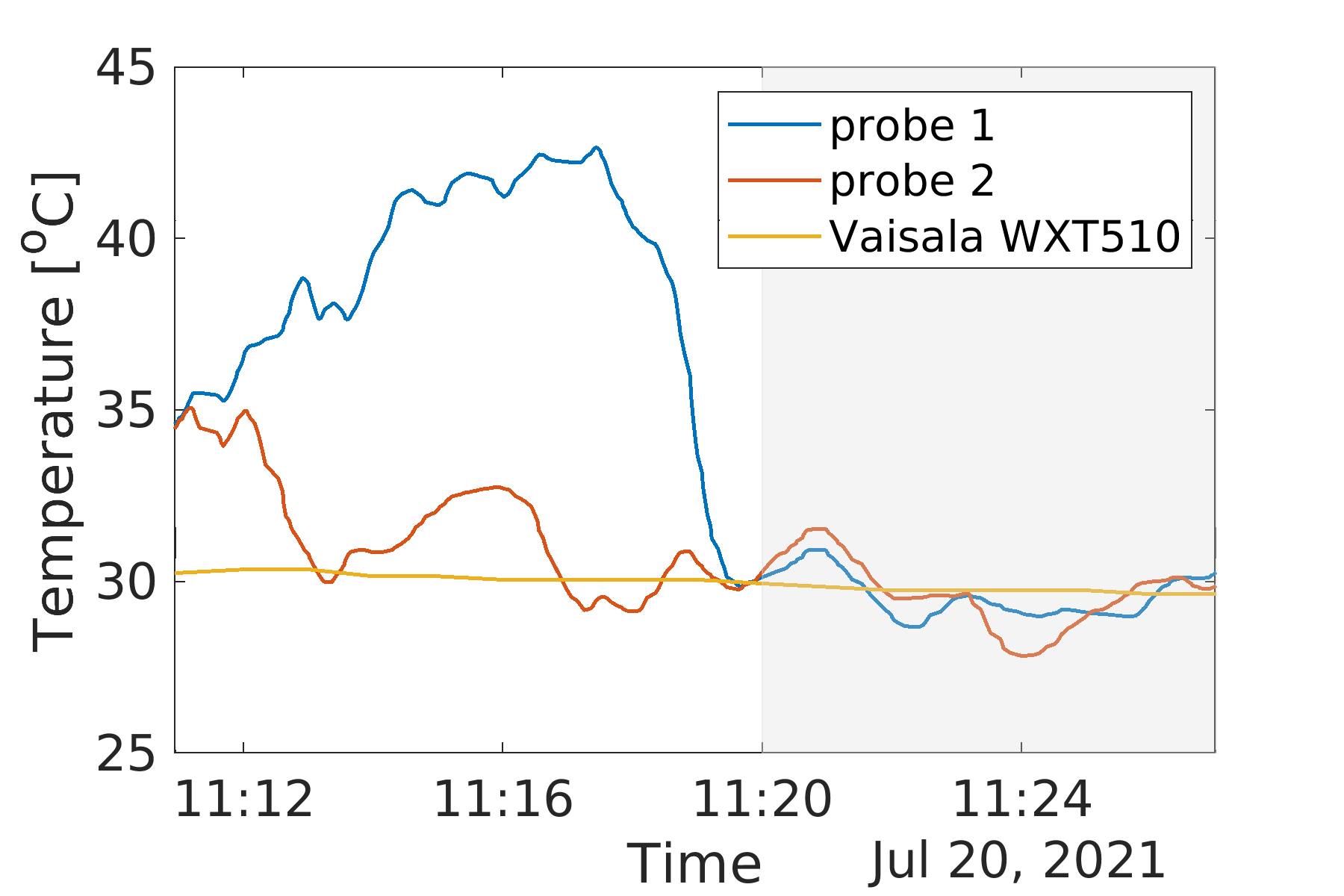}
			\caption{}
		\end{subfigure}
		\hspace{0.01\textwidth}
		\begin{subfigure}{0.4\textwidth}
			\includegraphics[width=\linewidth]{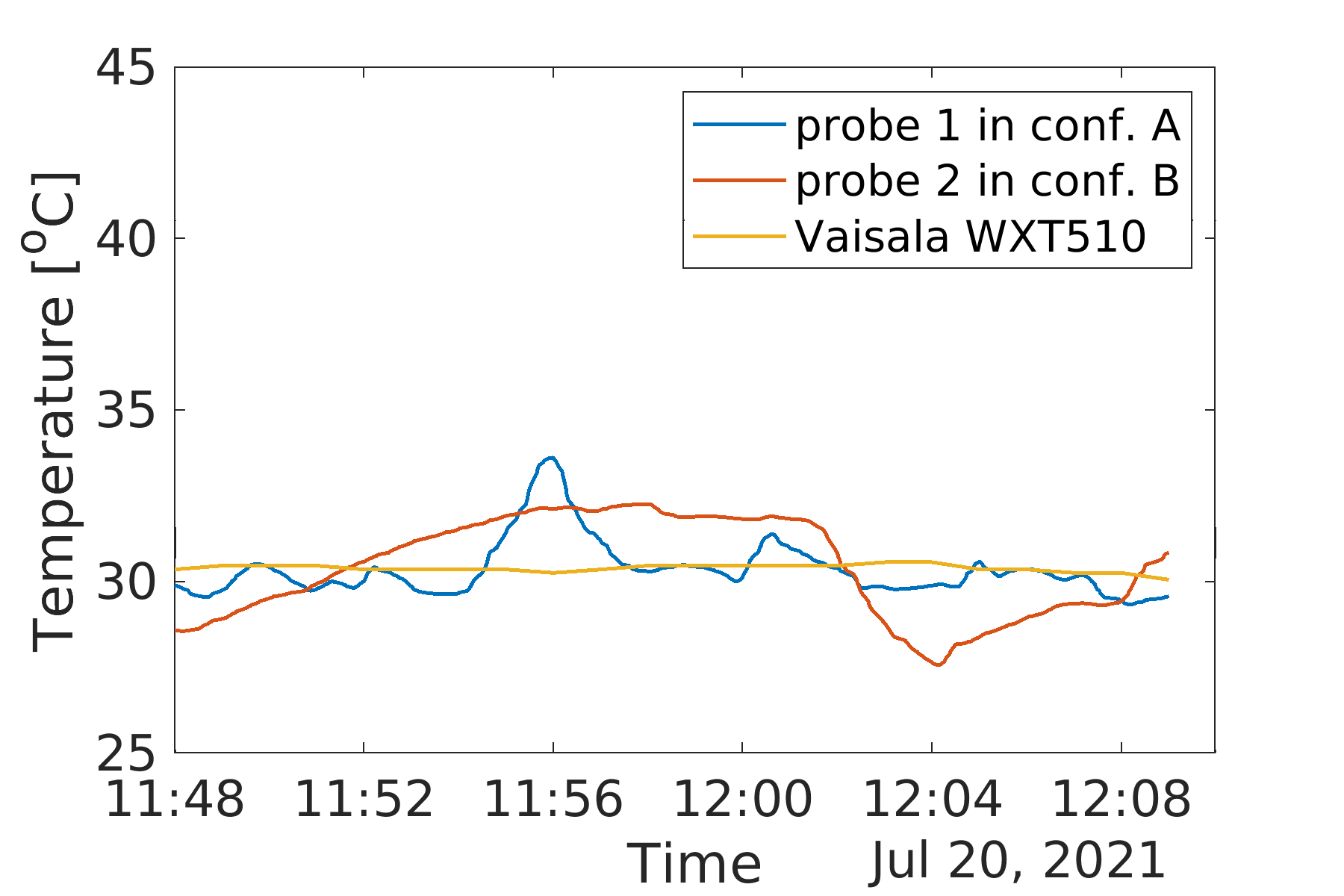}
			\caption{}
		\end{subfigure}
		\caption{Pressure, humidity, and temperature readings from the radiosondes in configurations A and B, see Fig. \ref{fig:inr1_conf}. First column, panels a, c, and e: sensor readings when the probe was still on the ground, but was not attached to the balloon. The gray shadowed regions represent the probe readings after the warm-up transient period has passed. It should be noted that, being continuously operational, the WXT510 station does not show a warm-up transient. Second column, panels b, d, and f: comparison of the sensor readings when the radiosondes were assembled to the balloon; in configuration A: the radioprobe is \textbf{outside} the balloon; in configuration B: the radioprobe is \textbf{inside} the balloon.}
		\label{fig:inr1_pht}
	\end{figure}

	{In-field tests began with the evaluation of various configurations of the radioprobe, aiming to validate sensor measurements in each configuration against the fixed-point ground station. For this test, two radiosondes were assembled in two different configuration layouts as shown in panels a and b of Figure \ref{fig:inr1_conf}. The validation tests were located around the Vaisala WXT510 station inside the INRIM campus, Turin, on July 20th, 2021. In order to choose the proper configuration, the radioprobe sensor readings were analyzed in these two setups. Figure \ref{fig:inr1_pht} shows the comparison of the pressure, humidity, and temperature readings from two radiosonde configurations. Pressure, humidity and temperature measurements were provided before (panels a, c and e) and after (panels b, d and f) attaching the radioprobe electronic board to a balloon.} The sensor readings were also compared with measurements taken by the WXT510 station. The pre-calibrated sensors of the station provide accurate temperature, pressure, and humidity readings throughout the day at 1-minute time intervals.
	
	In the first period of the experiment, the radiosonde sensors took a short period of time (from 11.10 to 11.20) to warm up and catch up with the WXT510 station readings. This is typical behavior of MEMS sensors, and especially of humidity and temperature sensors used for atmospheric measurements. After attaching the radioprobes to the balloons, the readings from configuration B started to show mismatches with the reference station measurements as shown in the rightmost panels of Figure \ref{fig:inr1_pht}, while the readings from the probe in configuration A were better aligned with the reference station measurements, particularly those of pressure and temperature. However, some small fluctuations in the temperature and humidity readings, with respect to the fixed station could have been due to the positioning and movement of the probes around the station.
	
	{Configuration B is preferable from the floating dynamical point of view and provides a better protection of the electronic board (e.g., water-resistance).} {However, in this configuration, the temperature and humidity readings were substantially biased, mainly related to the insulation effect of the balloon, which slows the effective time response of sensors. This led us to adopt configuration A for the succeeding in-field experiments.} \deleted{This experiment helped to identify an appropriate configuration of the radiosonde and quantify biases and the warm-up times of the sensors.}
	
	\subsection{{Dual-sounding experiments}}\label{results_vertical_launch}
	{The first two dual-sounding experiments were} conducted in collaboration with the \deleted{Regional Agency for the Protection of the Environment of the Piedmont region (ARPA)}{ARPA-Piemonte} on October 28 and on June 9, 2021, at Levaldigi Airport, Cuneo, Italy. The experiment site was equipped with an automatic sounding system, where ARPA-Piemonte launches radiosondes twice a day for atmospheric profiling measurements. We observed interference problems with the GNSS sensor during the first dual-sounding experiment. At that time, the radioprobe board was attached directly to the Vaisala RS41-SG probe. In order to resolve this issue the radioprobe was attached to the reference RS41-SG probe during the second launch with an 80 cm offset. {The last dual-sounding experiment was conducted in the context of Wessex Convection Campaign, at Chilbolton Observatory, Chilboton, UK on July 6, 2023.}
	
	{In this section, we primarily presented the results from the second experiment conducted in collaboration ARPA-Piemonte. In the following figures, when comparing our sensor readings with the reference data, we denoted our radiosonde as COMPLETE.}
	
	\subsubsection{{Data transmission}}
	\begin{figure}[bht!]
		\centering
		\centering
		\begin{subfigure}{0.4\textwidth}
			\includegraphics[width=\linewidth]{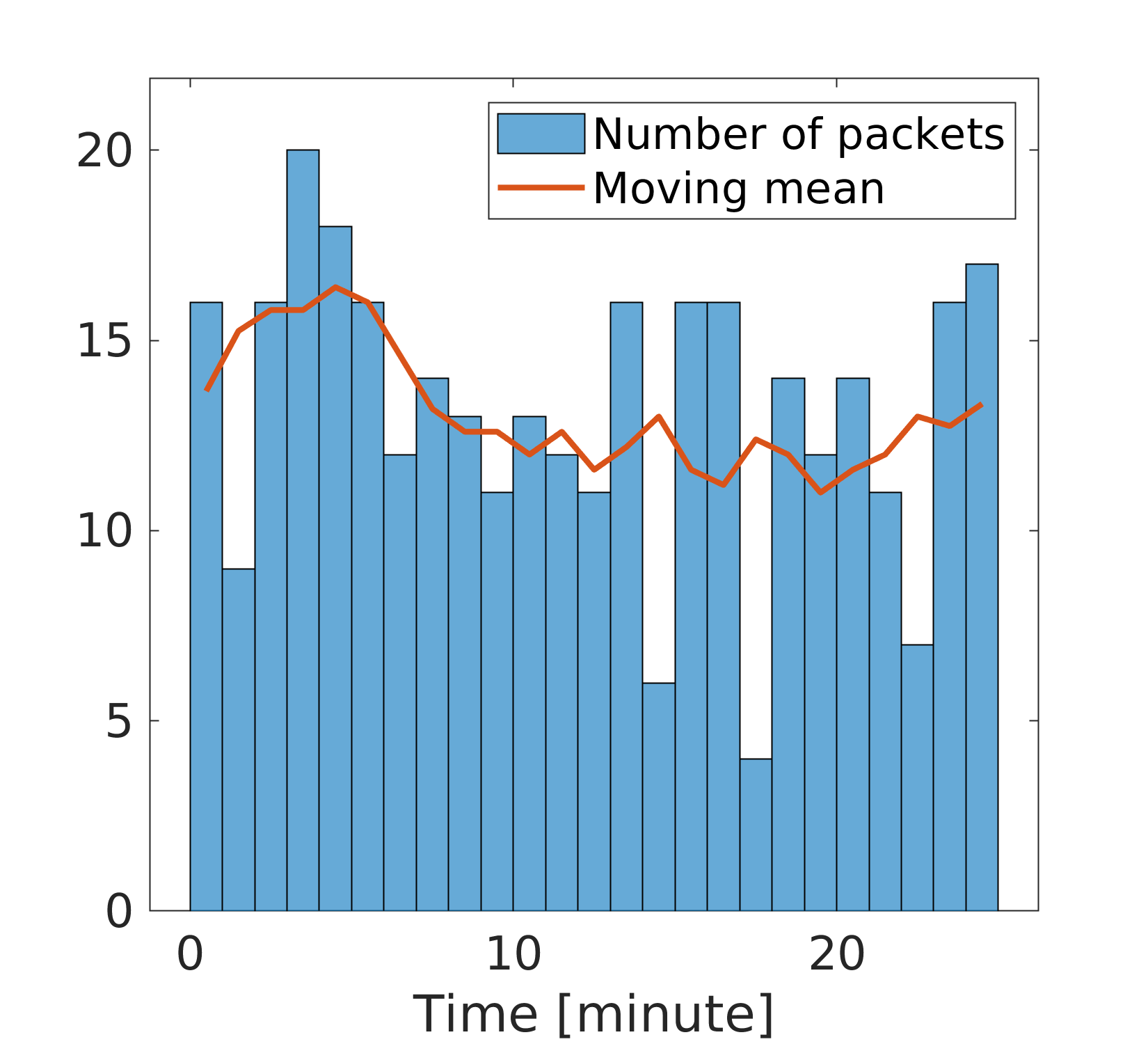}
			\caption{}
		\end{subfigure}
		\hspace{0.001\textwidth}
		\begin{subfigure}{0.4\textwidth}
			\includegraphics[width=\linewidth]{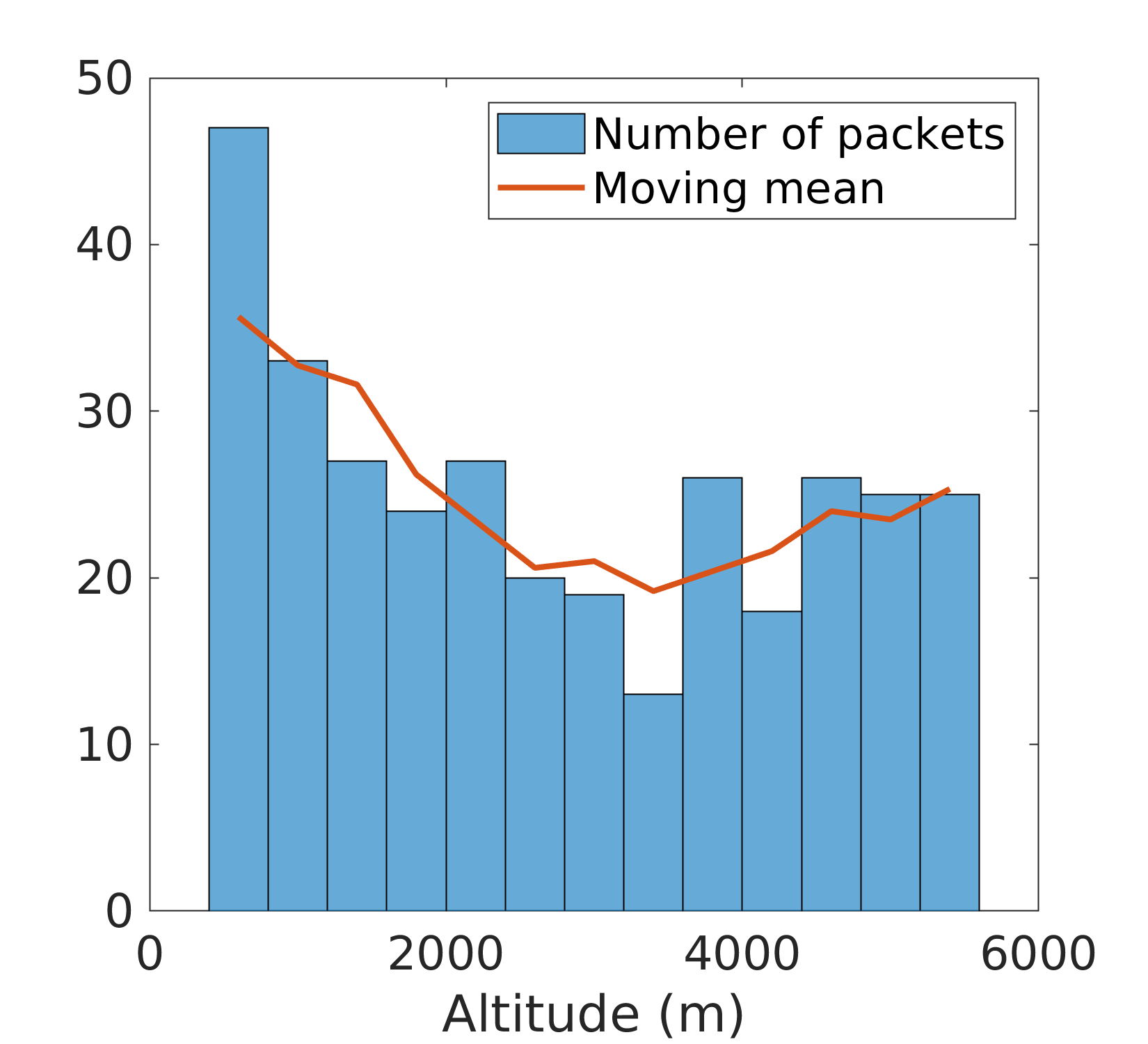}
			\caption{}
		\end{subfigure}
		\caption{Transmission tests for long distances (up to 12-13 km) and for a packet size of the order of 100 bytes during the dual-sounding experiment at Levaldigi Airport. The tests were carried out in collaboration with ARPA-Piemonte, on June 9, 2021 \cite{golshan2023cloud, paredes2021electronic}. (a) Number of packets received in each minute. (b) Number of packets along the altitude levels (bin size = 400 meters). The red lines represent the average transmission trends over time and altitude.}
		\label{fig:arpa_transmission}
	\end{figure}
	During the second dual-sounding experiment on June 9, 2021, data transmission continued for about 1 hour until the radioprobe reached almost $\sim$ 9 km in altitude and 13 km in distance. {Panel a of Figure \ref{fig:arpa_transmission} illustrates the average number of packets received in each minute for the first 25 minutes of the launch, while panel b shows the average number of packets received at a given altitude  for altitude range between 400 m and 6000 m. The radioprobe transmitted packets once in each 3–4 seconds during the flight. These performances were promising and within our target altitude range (1–3 km).} The original idea was to reach a 1 Hz transmission rate. However, this is difficult to achieve with the current radioprobe computational parameters and data packet size. Furthermore, delays and packet losses could be introduced due to congestion in the receiver. The current prototype of the receiver station is based on Adafruit Feather 32u4, which was designed for direct P2P LoRa communication. It has been agreed that the design of a more powerful multi-channel receiver station could alleviate this congestion and reduce receiver delays. For this reason, we are currently developing a new receiver station that can simultaneously receive data packets from 10–20 radiosondes without incurring packet losses due to packet collisions. Additionally, the effect of packet losses can be reduced with the help of appropriate post-processing, re-sampling, and filtering operations. Post-processing should be conducted while taking into account the application context. {For instance, in atmospheric measurements, factors like the atmospheric lapse rate, temperature gradient, and complementary information from sensors can be employed in conjunction with others, such as pressure and GNSS altitude, or acceleration and GNSS velocity, to enhance the accuracy and/or the completeness of the results.}
	
	\subsubsection{{Position and velocity measurements}}
	\begin{figure}[bht!]
		\centering
		\begin{subfigure}{0.49\textwidth}
			\includegraphics[width=\linewidth]{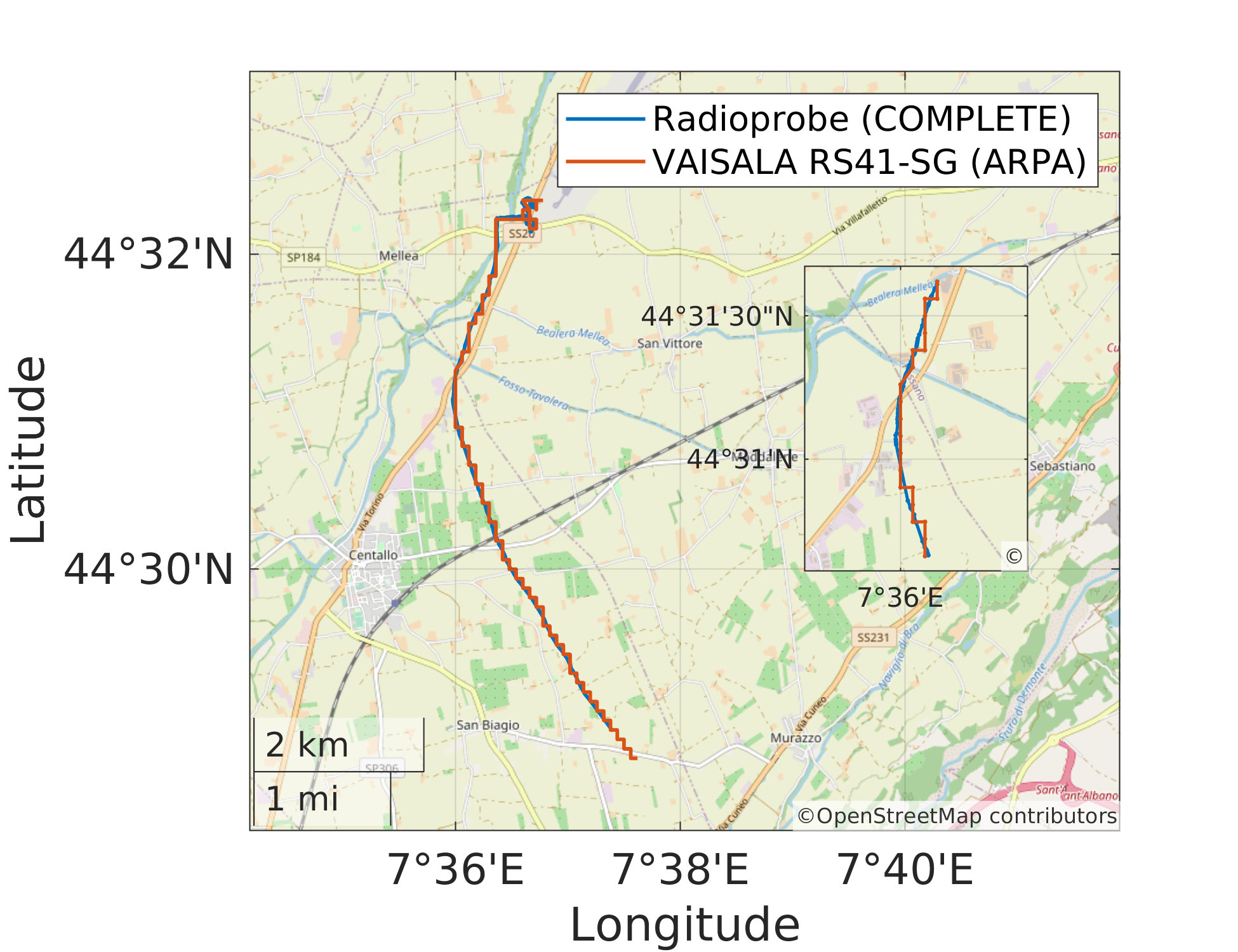}
			\caption{}
		\end{subfigure}
		\hspace{0.001\textwidth}
		\begin{subfigure}{0.49\textwidth}
			\includegraphics[width=\linewidth]{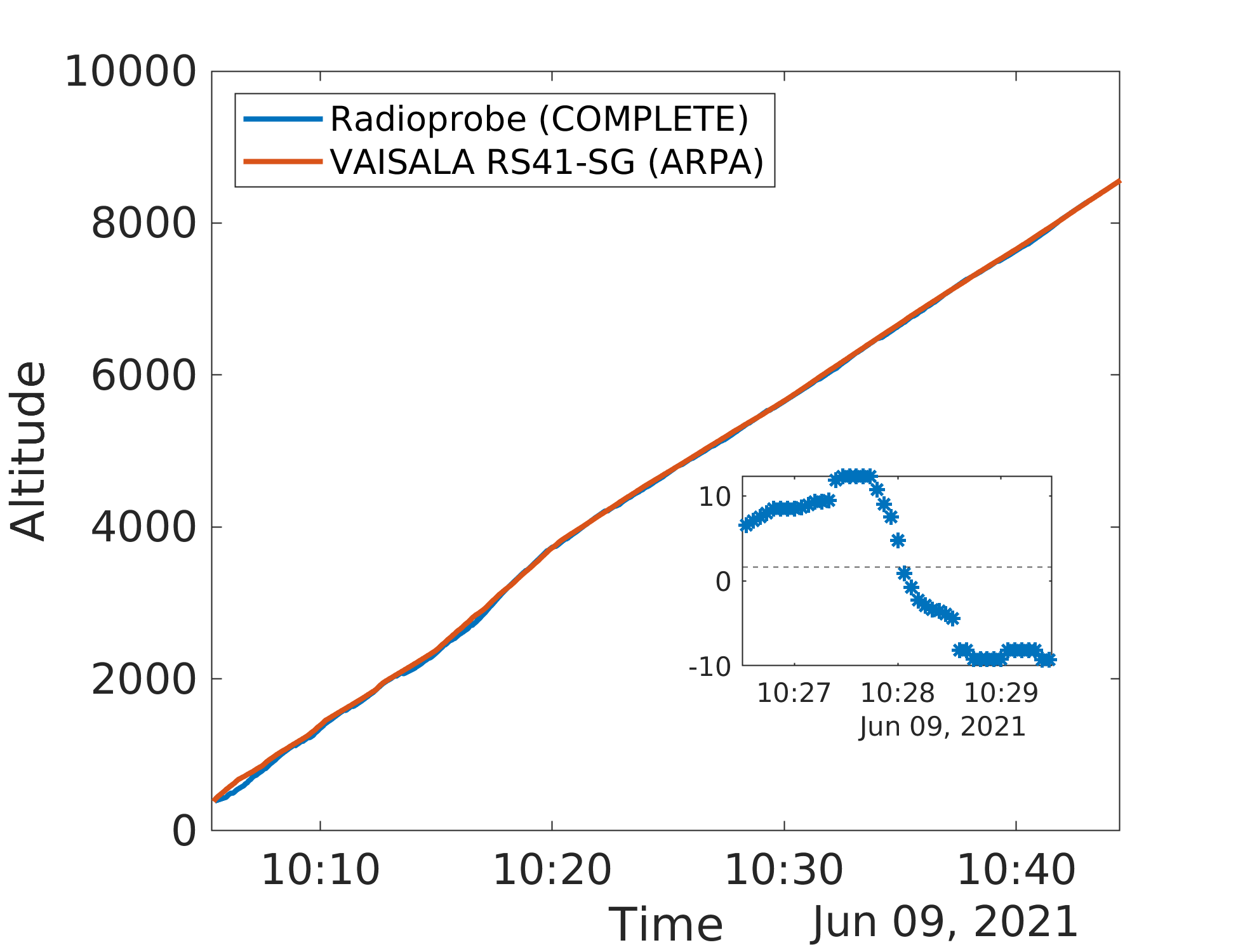}
			\caption{}
		\end{subfigure}
		\caption{GNSS positioning measurements of the COMPLETE radioprobe and comparison with the Vaisala RS41-SG probe during the dual-sounding experiment on June 9, 2021, at Levaldigi Airport, Cuneo, Italy. {(a) Trajectories of the radiosondes on the map. For longitude readings: RMSE = 2.97e-4 and MBE = 2.95e-5 degrees, for latitude readings: RMSE = 3.7e-4 and MBE = -2.41e-4 degrees. (b) Altitude readings from radiosondes over the time axis. In this case, RMSE = 22.3 m and MBE = 11.05 m. The inset plot highlights the altitude difference, $\Delta Z = Z_{complete} – Z_{RS41}$, between the COMPLETE and reference RS41-SG probe for a shorter time interval.}}
		\label{fig:arpa2_gnss_lla}
	\end{figure}
	
	Figure \ref{fig:arpa2_gnss_lla} shows a comparison of the GNSS sensor measurements of the radiosonde with the reference Vaisala RS41-SG radiosonde. It can be seen from panel a and b that raw longitude, latitude, and altitude readings from GNSS sensor {exhibit a good agreement with the reference RS41-SG dataset}. {The inset plot on panel a shows magnified view for a shorter longitude and latitude ranges, while the inset plot of panel b highlights the altitude difference between radiosondes, $\Delta Z = Z_{complete} – Z_{RS41}$. The comparison yielded the following estimations of RMSE (Root Mean Square Error) and MBE (Mean Bias Error) values:}
	\begin{enumerate}[label=(\roman*),wide, labelwidth=!]
		\item {For longitude readings : RMSE = 2.97e-4 and MBE = 2.95e-5 degrees.}
		\item {For latitude readings: RMSE = 3.7e-4 and MBE = -2.41e-4 degrees.}
		\item {For altitude readings: RMSE = 22.3 m and MBE = 11.05 m.}
	\end{enumerate}
	{The longitude and latitude readings provided by ARPA-Piemonte displayed lower resolution during the comparison, consequently impacting the computation of MBE and RMSE values. We consistently observed constant values in ARPA's RS41-SG longitude and latitude readings during time intervals in between 20 to 40 seconds. In fact, the panel a of Figure \ref{fig:arpa2_gnss_lla} displays stair-like pattern in the reference readings. This pattern results from the ARPA dataset providing longitude and latitude readings accurate to the 3rd decimal place (e.g., 7.612 and 44.538), whereas our probe provides readings up to the 6th decimal place (e.g., 7.611639 and 44.538215).}
	
	\begin{figure}[bht!]
		\centering
		\begin{subfigure}{0.49\textwidth}
			\includegraphics[width=\linewidth]{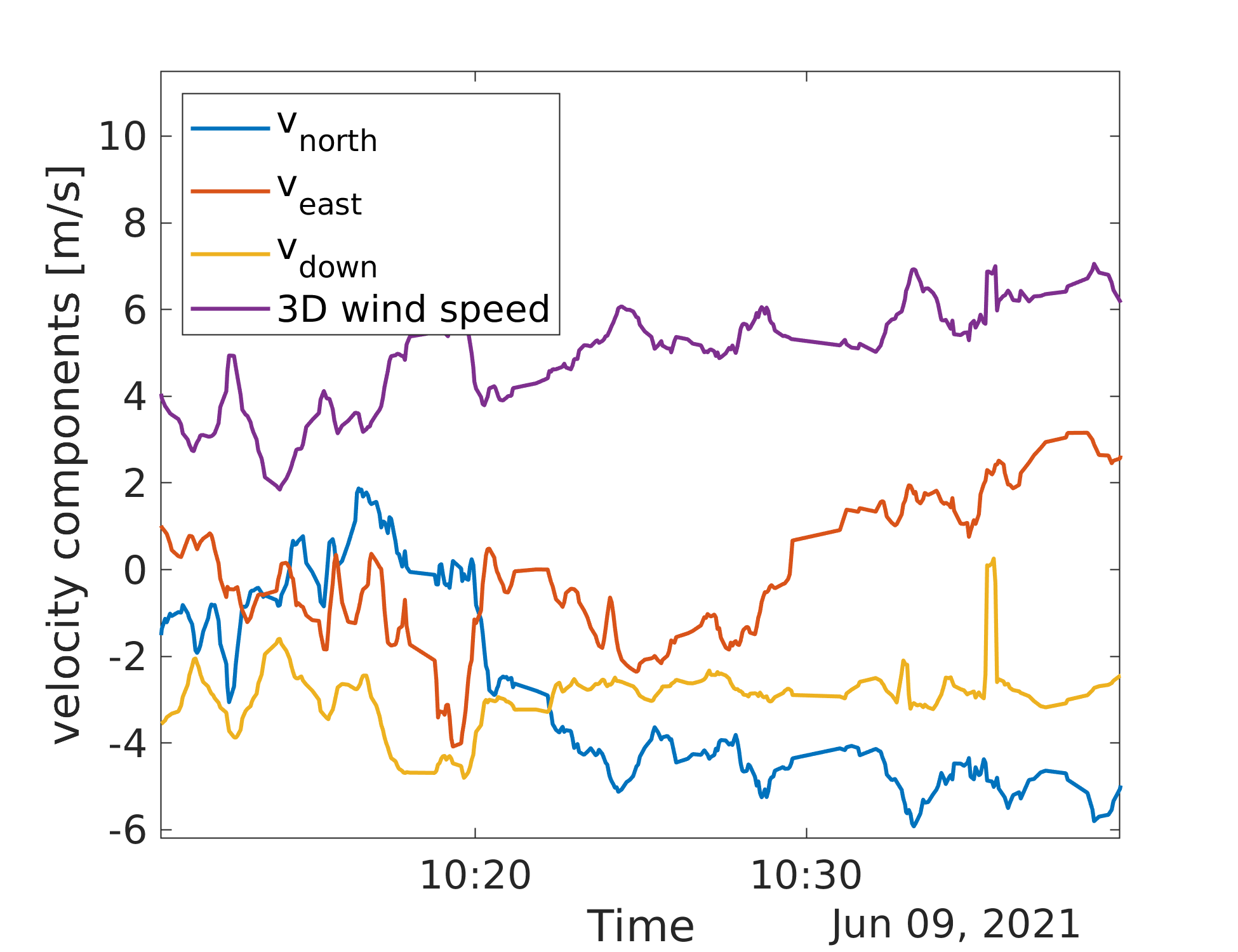}
			\caption{}
		\end{subfigure}
		\hspace{0.001\textwidth}
		%
		\begin{subfigure}{0.49\textwidth}
			\includegraphics[width=\linewidth]{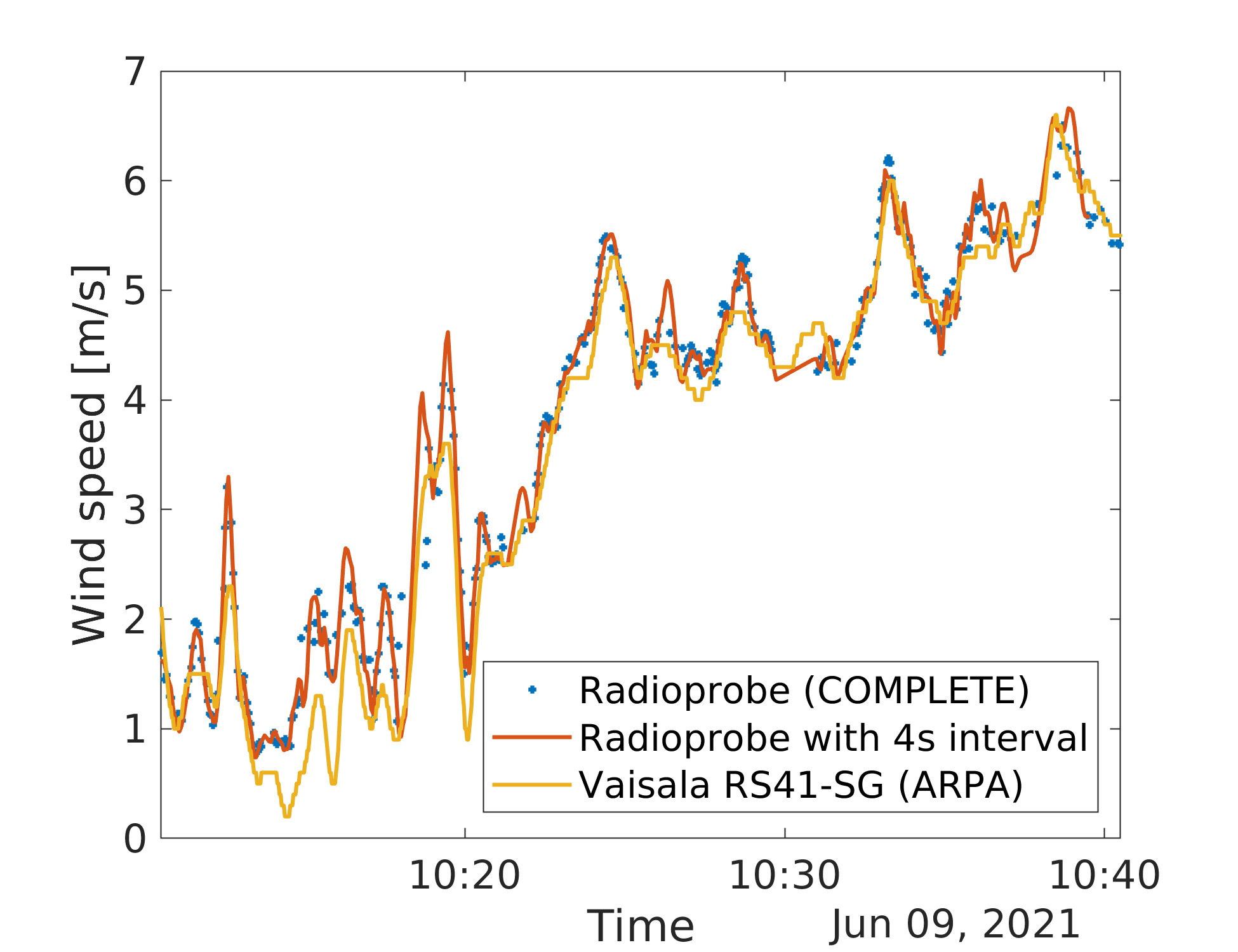}
			\caption{}
		\end{subfigure}
		\hspace{0.01\textwidth}
		\begin{subfigure}{0.49\textwidth}
			\includegraphics[width=\linewidth]{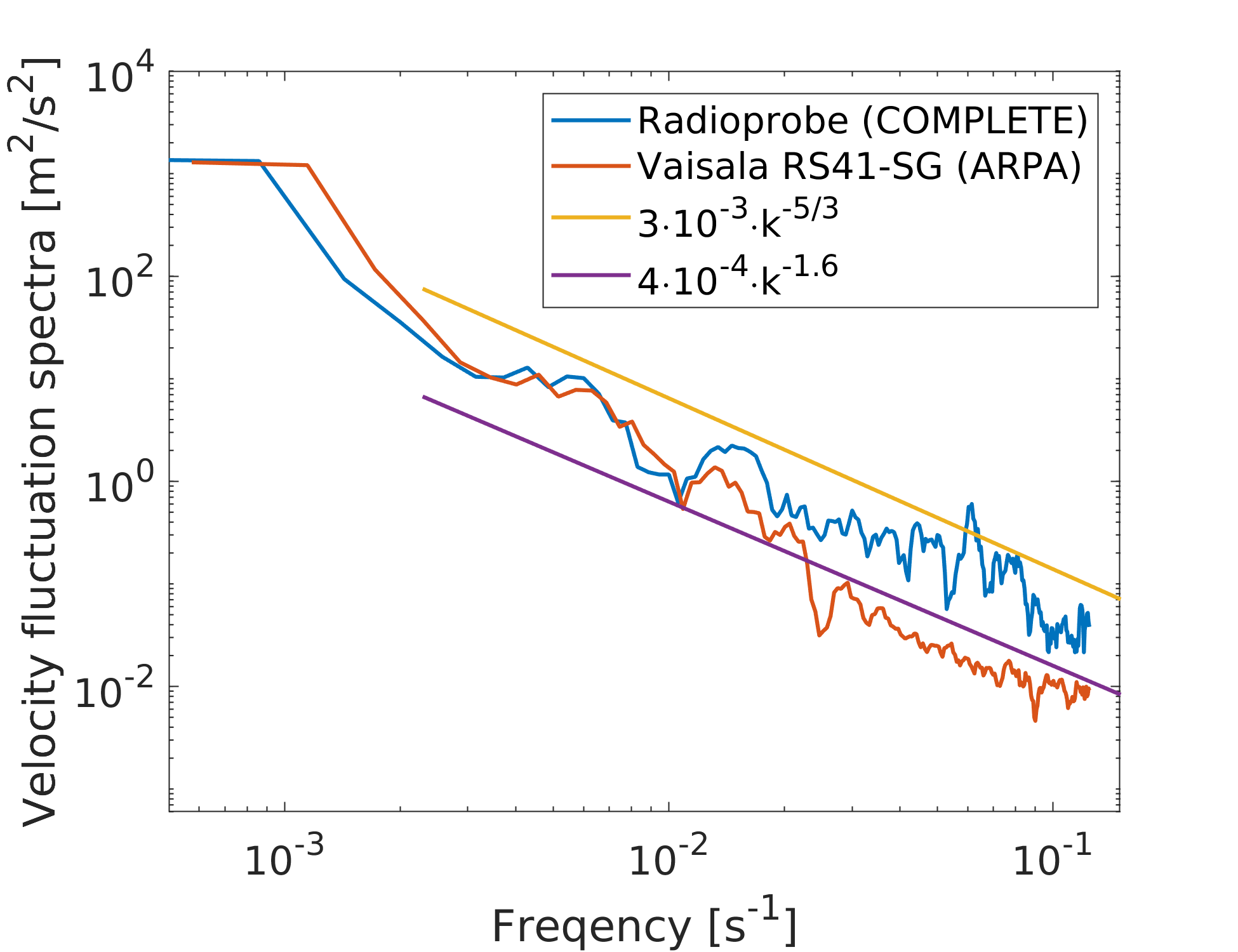}
			\caption{}
		\end{subfigure}
		\caption{Velocity measurements of the radioprobe during the dual-sounding experiment on June 9, 2021, at Levaldigi Airport, Cuneo, Italy. (a) Velocity components and the magnitude of the 3D wind speed derived from GNSS sensor of COMPLETE radioprobe. {(b) Comparison between the horizontal wind speed measurement provided by the COMPLETE radioprobe and the RS41-SG radiosonde of ARPA-Piemonte. Both raw (in blue) and resampled datasets with 4-second regular intervals (in orange) are included for the COMPLETE radioprobe. (c) A comparison of the power spectrum of wind speed fluctuations is shown between the two probes. Alongside the raw spectrum dataset, two trend lines (in yellow and violet) are presented for reference.} The frequency range is based on the Nyquist theory and represents half of the sampling frequency, denoted as $f_s/2 = 0.125 s^{-1}$.}
		\label{fig:arpa2_gnss_vel}
	\end{figure}
	
	A {UBX-PVT packet from GNSS sensor provides} velocity readings in the north, east, and down directions, as shown in the plots in Figure \ref{fig:arpa2_gnss_vel}a. The horizontal wind speed was computed from the north and east velocity components and was compared with the horizontal wind speed readings of the RS41 probe, see Fig. \ref{fig:arpa2_gnss_vel}b. The wind speed was further analyzed with the FFT (Fast Fourier Transform) to obtain preliminary results of the power spectra of the fluctuations. In order to compute the spectra, a 30-minute wind speed dataset with a 4s time step was sampled (see Fig. \ref{fig:arpa2_gnss_vel}c), which gives a frequency range from 5$\cdot 10^{-4}$s$^{-1}$ to 0.25 s$^{-1}$ and a Nyquist frequency of 0.12 s$^{-1}$ (2$\pi$/8 = $\pi$/4 rad/s).
	
	The same kind of analysis can be performed with vertical velocity and temperature datasets. The power spectra analysis of the vertical velocity can be used to identify a cutoff point of the Brunt-Vaisala frequency, while the vertical temperature profile (Fig. \ref{fig:arpa_pht}c) can be used to derive a complete profile of the Brunt-Vaisala frequency along the altitude \cite{nath2010turbulencecharacteristics, jaiswal2020gpsradiosonde}.
	
	\subsubsection{{Pressure, humidity and temperature measurements}}
	
	\begin{figure}[ht!]
		\centering
		\begin{subfigure}{0.48\textwidth}
			\includegraphics[width=\linewidth]{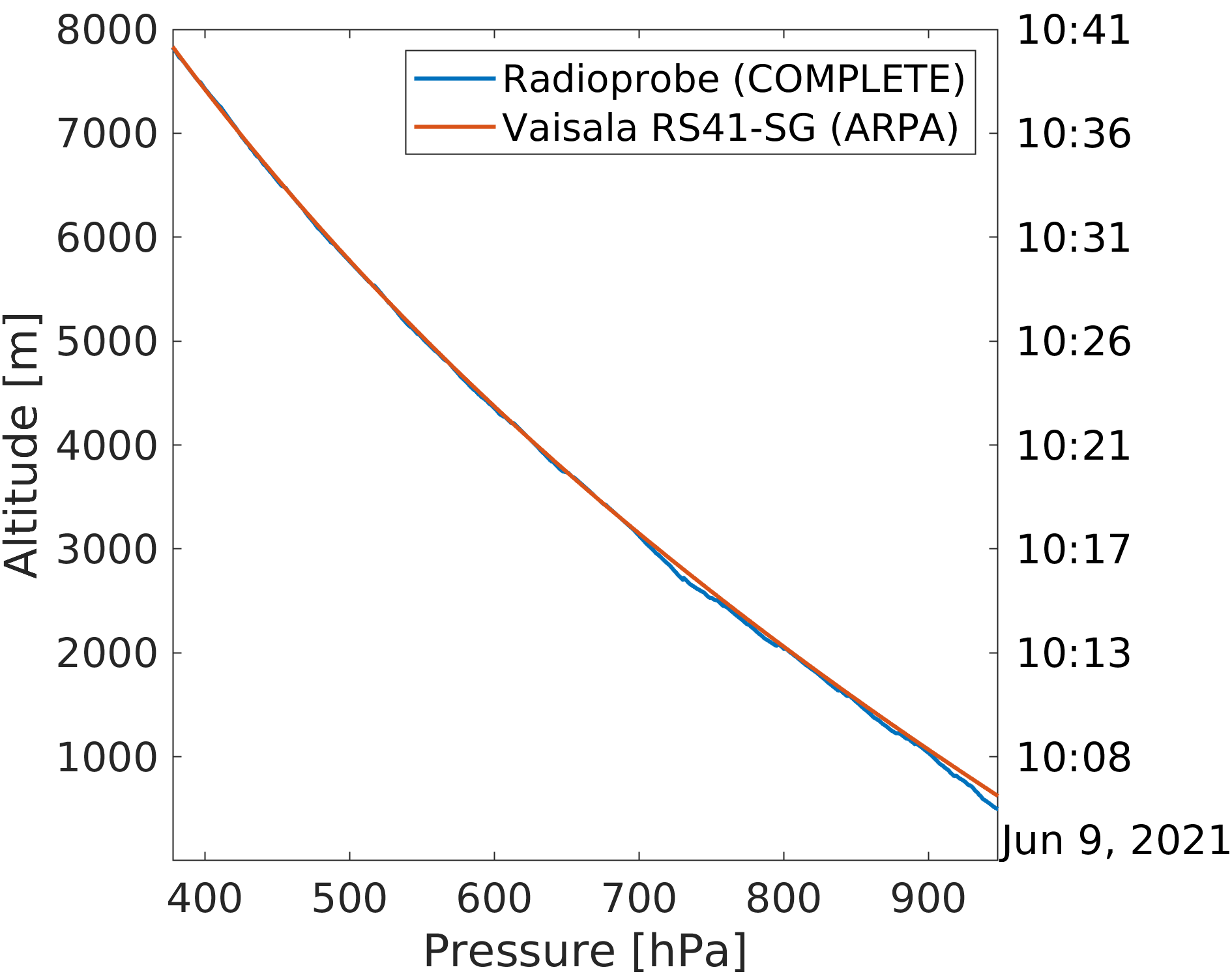}
			\caption{}
		\end{subfigure}
		\hspace{0.01\textwidth}
		\begin{subfigure}{0.48\textwidth}
			\includegraphics[width=\linewidth]{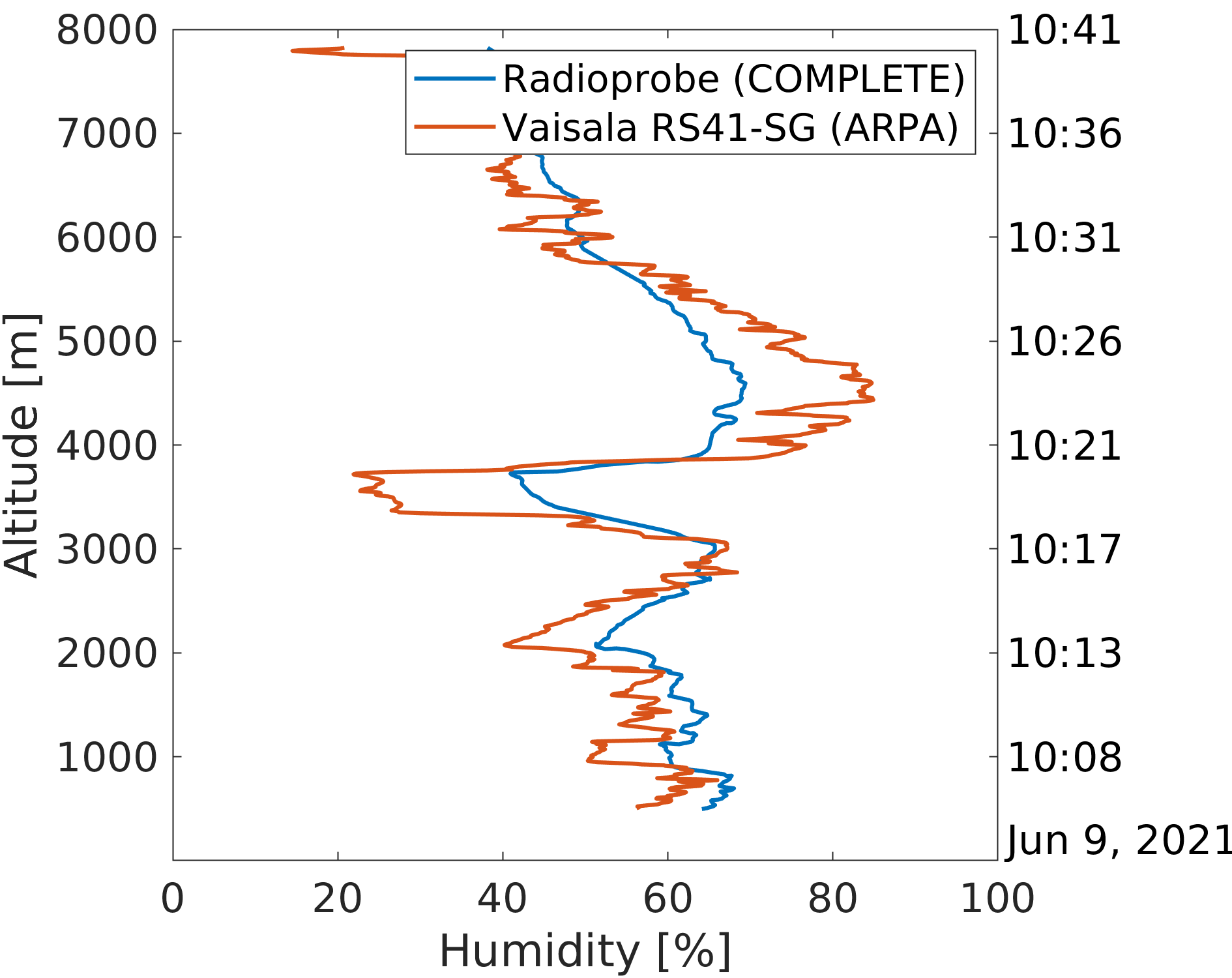}
			\caption{}
		\end{subfigure}
		\vspace{0.001\textheight}
		\begin{subfigure}{0.48\textwidth}
			\includegraphics[width=\linewidth]{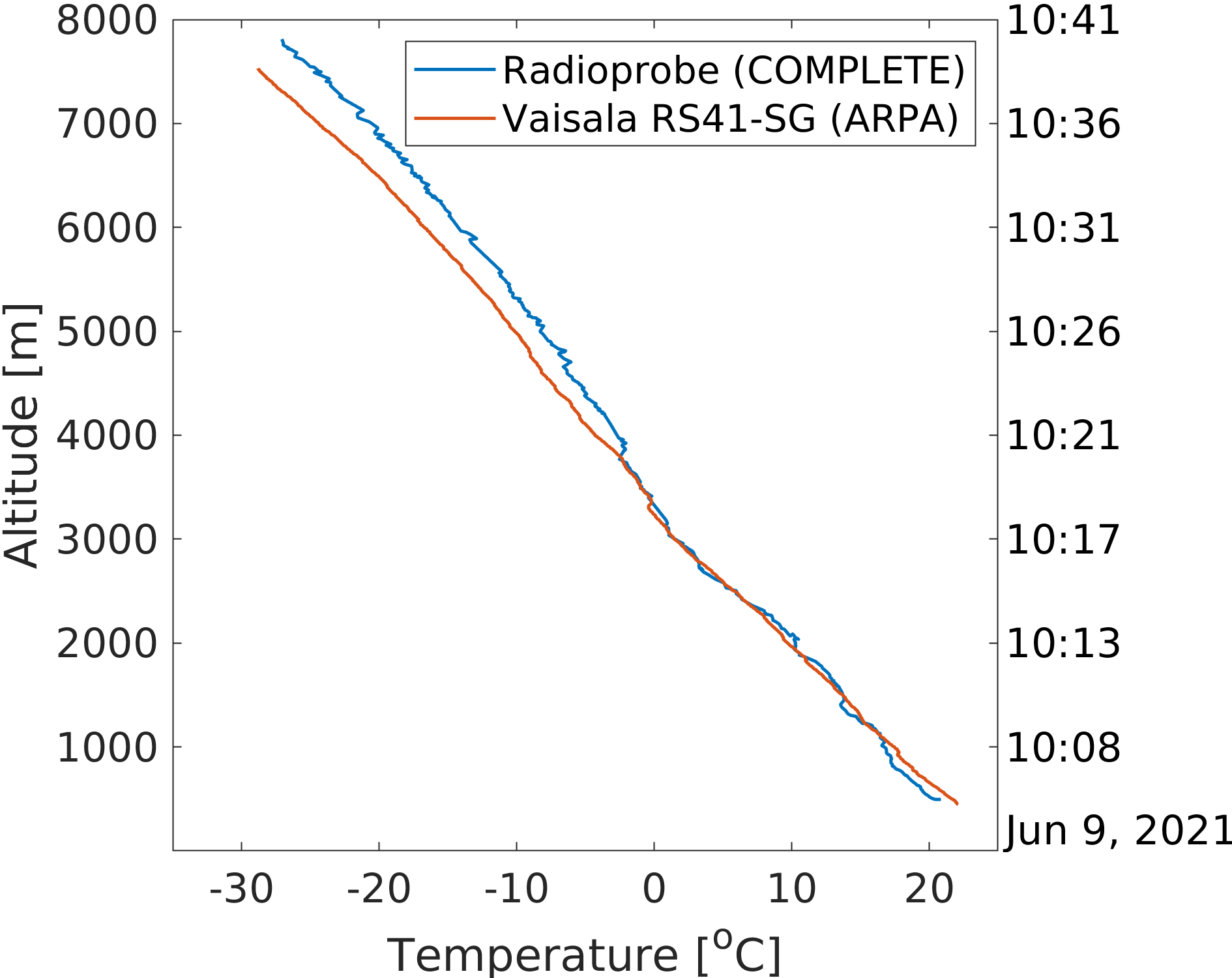}
			\caption{}
		\end{subfigure}
		\hspace{0.01\textwidth}
		\begin{subfigure}{0.48\textwidth}
			\includegraphics[width=\linewidth]{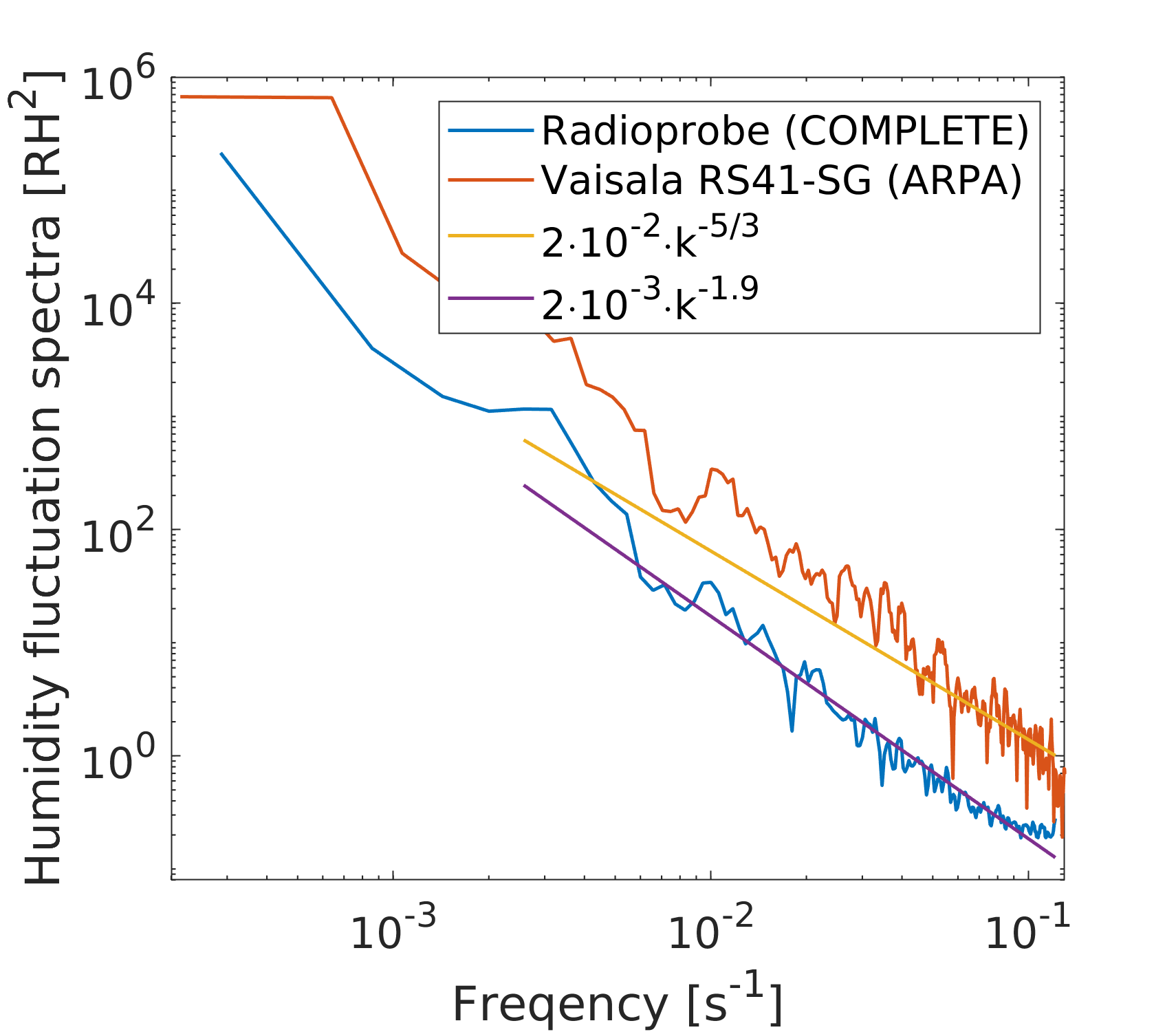}
			\caption{}
		\end{subfigure}
		\caption{Pressure, humidity, and temperature readings from radiosondes during the dual-sounding experiment on June 9, 2021, at Levaldigi Airport, Cuneo, Italy. Panels a, b, and c present a comparison of the measurements between our radioprobe (COMPLETE) and the Vaisala RS41-SG radiosonde of ARPA-Piemonte. (d) Comparison of the humidity fluctuation spectra between two probes. Two trend lines (yellow and violet) are provided for comparison purposes.}
		\label{fig:arpa_pht}
	\end{figure}
	
	{A comparison between PHT readings of COMPLETE probe and the reference RS41-SG probe is illustrated in Figure \ref{fig:arpa_pht}. We found a good agreement for pressure measurements, while temperature and humidity measurements showed mismatches and deviations from the reference readings. For the temperature, the COMPLETE probe readings are better aligned with the reference dataset below 4000 m of altitude. Above this altitude, the readings showed a nearly linear deviation trend. For the humidity measurements, the COMPLETE probe mounting the BME280 MEMS underestimated the readings in the higher RH ranges (65 - 85\%) and overestimated them in lower RH ranges (20 - 40 \%). However, BME280 humidity sensor was able to follow the variation trend of the reference readings and to track the humidity fluctuations as shown by the power spectra shown in panel d of Figure \ref{fig:arpa_pht}. As expected from the humidity profile in panel b, the magnitude of the fluctuation spectra was lower than the reference spectra computed from the RS41-SG readings.}
	
	{The intended operational altitude range for our radiosonde network typically spans from surface to 2500 m, with occasional extensions to a maximum of 3000-4000 m. We examined temperature measurements obtained from radiosondes during dual-launch experiments and compared them to reference radiosonde data across an altitude range of 400 to 3600 m. To facilitate this comparison, we divided the altitude into 400-meter intervals, where variations in air density are relatively modest. Subsequently, we calculated the mean temperature difference ($<$T-T$_{ref}>$) and the normalized mean temperature difference ($<$T-T$_{ref}$/T$_{ref}>$) relative to the reference radiosonde data (T$_{ref}$) for each altitude interval. Table \ref{tbl:dual_soundings} provides a comprehensive summary of the comparative analysis for three dual-sounding experiments: ARPA-2020 (conducted on October 28, 2020, at Levaldigi Airport in Cuneo, Italy), ARPA-2021 (performed on June 9, 2021, at Levaldigi Airport in Cuneo, Italy), and MET-OFFICE-2023 (carried out on July 6, 2023, at Chilbolton Observatory in Chilbolton, United Kingdom).}
	
	\begin{table}[ht!]
		\caption{{Comparative analysis of temperature measurements across the altitude ranges with respect to the reference radiosonde data in dual-launch experiments conducted from October 2020 to July 2023. The table includes mean differences ($<$T-T$_{ref}>$), normalized mean differences relative to the reference sensor readings ($<$T-T$_{ref}$/T$_{ref}>$) and the temperature measurement range (T$_{ref}$) for each experiment site.}}
		\resizebox{\textwidth}{!}{%
			\begin{tabular}
				{|p{2.2cm}|p{1.5cm}|p{1.8cm}|p{2.7cm}|p{1.5cm}|p{1.8cm}|p{2.7cm}|p{1.5cm}|p{1.8cm}|p{2.7cm}|}
				\hline
				\multirow{3}{2.3cm}{Altitude interval} & \multicolumn{9}{c|}{\textbf{Experiment sites}} \\ \cline{2-10}
				& \multicolumn{3}{p{6.5cm}|}{ARPA-2020, Levaldigi, Italy(Ref: Vaisala RS41-SG; Oct 28, at 12:05, local CET time; mild wind, clear-sunny day), COMPLETE probe fastened to the Vaisala sonde case} & \multicolumn{3}{p{6.5cm}|}{ARPA-2021, Levaldigi, Italy (Ref: Vaisala RS41-SG; June 9, at 12:05, local CET time; mild wind, clear-sunny day), COMPLETE probe hanged with a 0.8m long wire to the Vaisala sonde case} & \multicolumn{3}{p{6.5cm}|}{MET-OFFICE-2023, Chilbolton, UK(Ref:   Vaisala RS41-SGP; July 6, at 09:04, local UK time; strong wind, partially cloudy, partially rainy), COMPLETE probe fastened to the RS41-SGP sonde case} \\ \cline{2-10}
				& Temp. Range (T$_{ref}$) & $<$T-T$_{ref}>$ & $<$T-T$_{ref}$/T$_{ref}>$ & Temp. Range (T$_{ref}$) & $<$T-T$_{ref}>$ & $<$T-T$_{ref}$/T$_{ref}>$ & Temp. Range (T$_{ref}$) & $<$T-T$_{ref}>$ & $<$T-T$_{ref}$/T$_{ref}>$ \\ \hline
				400 – 800 m & 282.4 ÷ 284.0 K & 0.14 K & 0.05 \% & 291.9 ÷ 295.0 K & -1.26 K & -0.4 \% & 283.1 ÷ 285.3 K & -0.25 K & -0.08 \% \\ \hline
				800 – 1200 m & 280.5 ÷ 282.2 K & -0.16 K & -0.06 \% & 288.7 ÷ 291.9 K & -0.57 K & -0.19 \% & 279.5 ÷ 283.1 K & 0.20 K & 0.07 \% \\ \hline
				1200 –1600 m & 279.9 ÷ 282.4 K & -0.24 K & -0.08 \% & 286.1 ÷ 288.7 K & -0.36 K & -0.13 \% & 275.8 ÷ 279.5 K & 0.20 K & 0.07 \% \\ \hline
				1600 – 2000 m & 280.9 ÷ 283.2 K & -0.62 K & -0.21 \% & 282.9 ÷ 286.1 K & 0.27 K & 0.09 \% & 272.5 ÷ 275.8 K & 0.58 K & 0.21 \% \\ \hline
				2000 – 2400 m & 277.6 ÷ 280.7 K & 0.78 K & 0.28 \% & 279.5 ÷ 282.9 K & +0.49 K & 0.17 \% & 272.5 ÷ 276.9 K & 0.69 K & 0.25 \% \\ \hline
				2400 – 2800 m & 275.0 ÷ 277.4 K & 1.40 K & 0.51 \% & 276.1 ÷ 279.5 K & -0.37 K & -0.13 \% & 274.6 ÷ 276.9 K & 1.53 K & 0.56 \% \\ \hline
				2800 – 3200 m & 273.6 ÷ 275.0 K & 1.11 K & 0.41 \% & 273.5 ÷ 276.1 K & 0.03 K & 0.01 \% & 271.5 ÷ 274.6 K& 2.81 K & 1.03 \% \\ \hline
				3200 – 3600 m & 270.3 ÷ 273.5 K & 1.29 K & 0.47 \% & 271.9 ÷ 273.5 K & -0.06 K & -0.02 \% & 269.3 ÷ 271.5 K & 3.31 K & 1.22 \%  \\ \hline
			\end{tabular}%
		}
		\label{tbl:dual_soundings}
	\end{table}
	
\begin{figure}[h!]
		\centering
		\caption*{\scriptsize Chilbolton, UK. July 6th, 2023. COMPLETE probe was directly fastened to Vaisala RS-41 SGP probe with a scotch tape, \textbf{possible heating from RS41-SGP}.}
		\begin{subfigure}[b]{0.3\textwidth}
			\includegraphics[width=\linewidth]{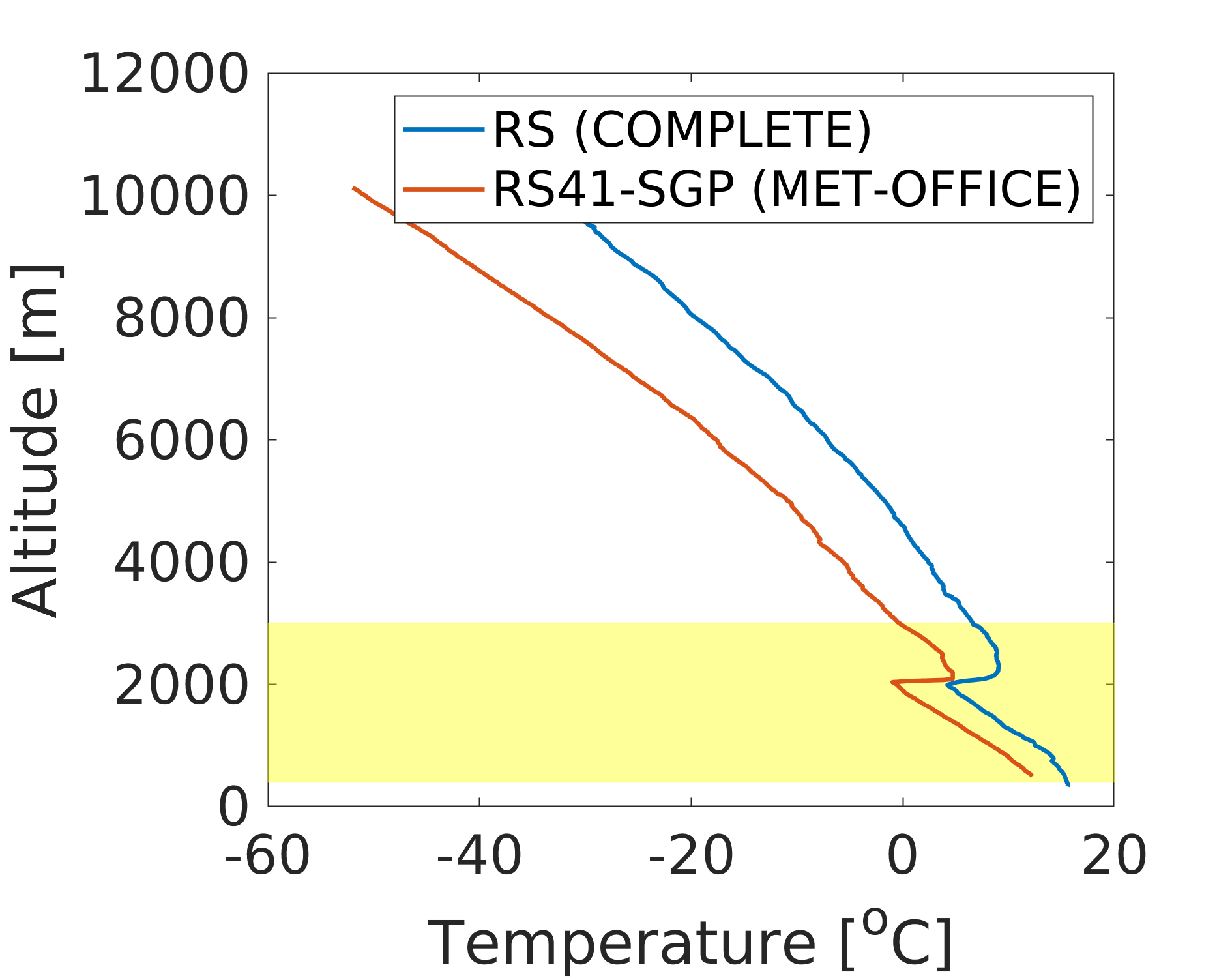}
			\caption{}
		\end{subfigure}
		\hspace{0.01\textwidth}
		\begin{subfigure}[b]{0.3\textwidth}
			\includegraphics[width=\linewidth]{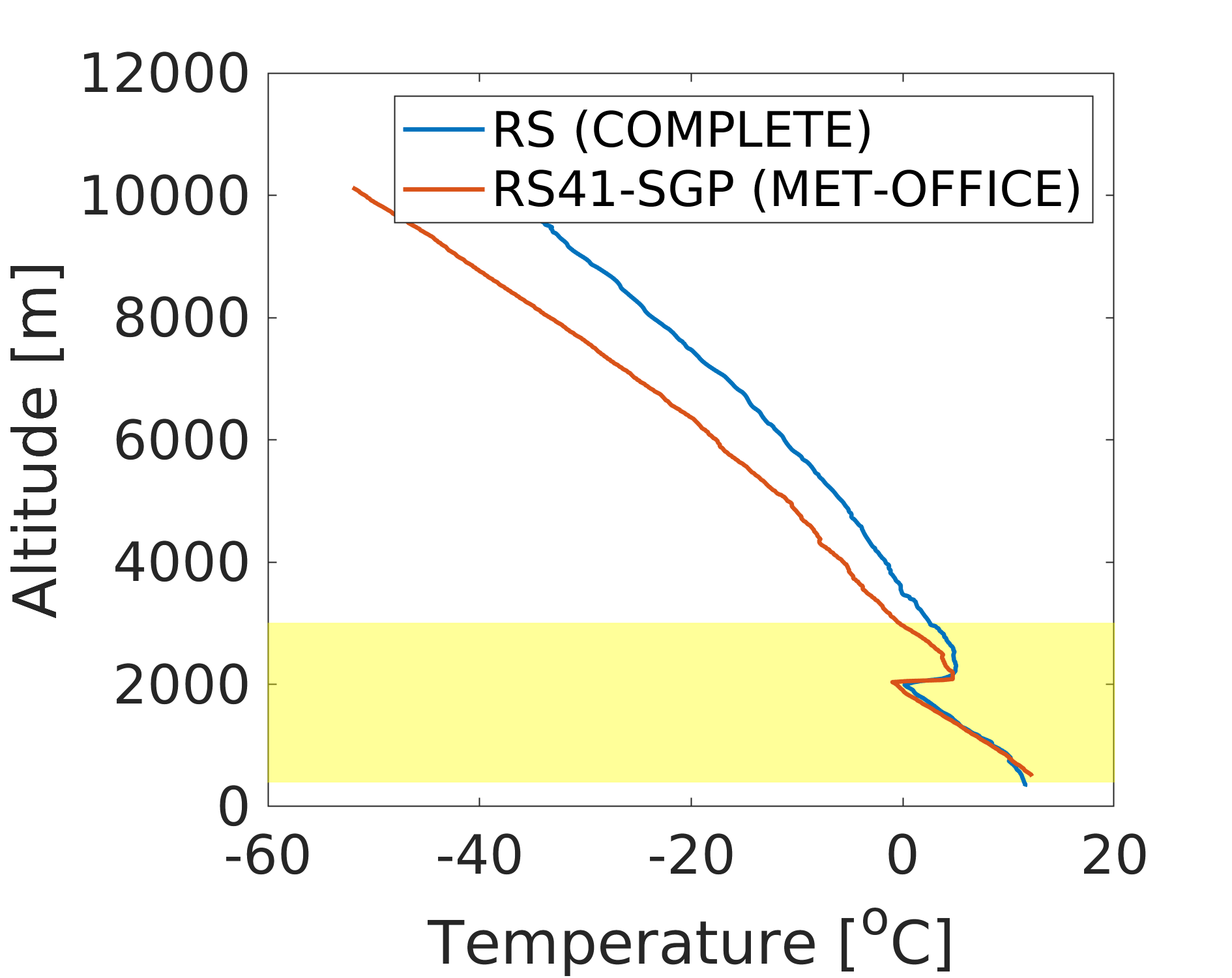}
			\caption{}
		\end{subfigure}
		\hspace{0.01\textwidth}
		\begin{subfigure}[b]{0.3\textwidth}
			\includegraphics[width=\linewidth]{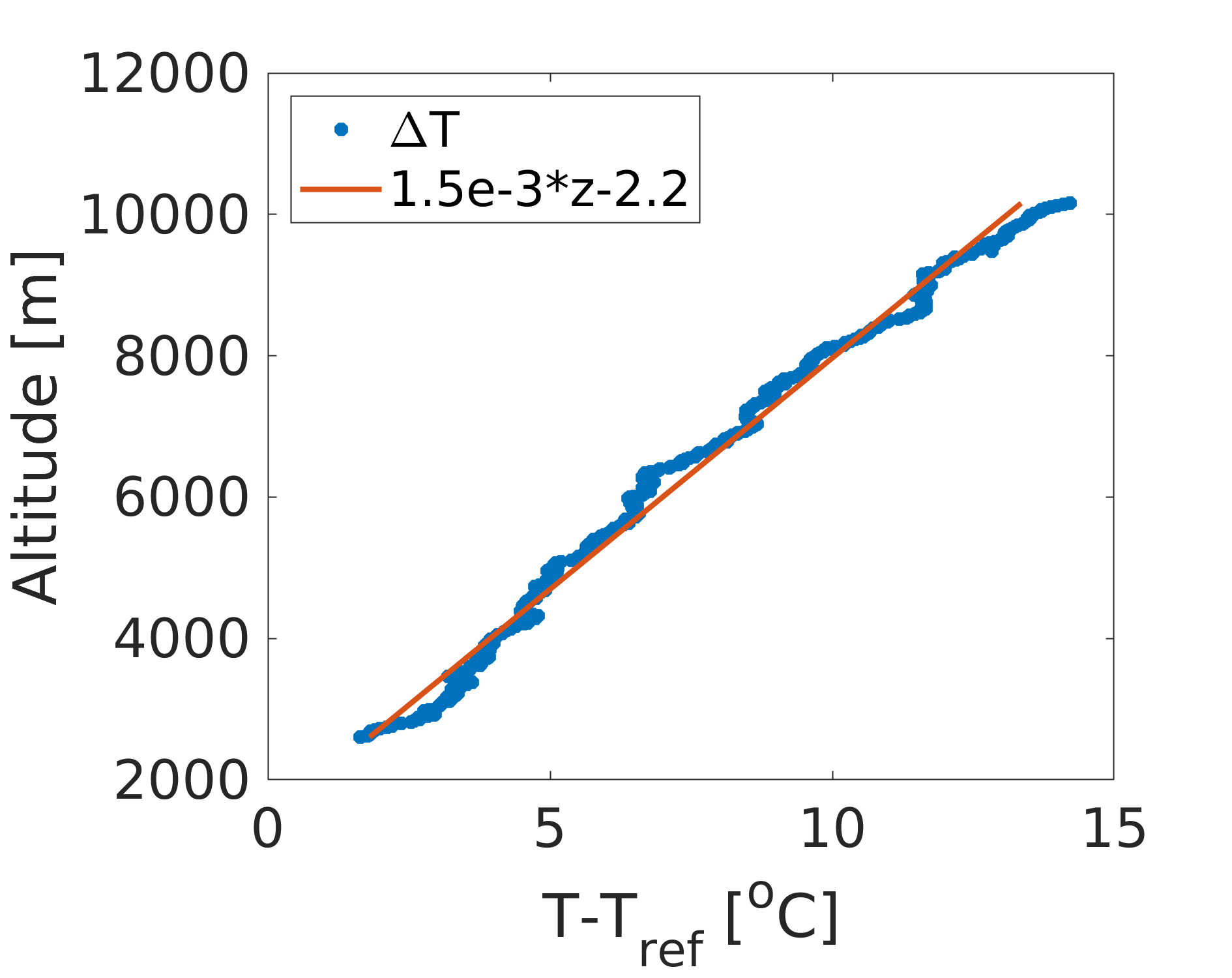}
			\caption{}
		\end{subfigure}
		\caption*{\scriptsize Levaldigi Airport, Cuneo, Italy. June 9th, 2021. COMPLETE probe attached with a 80-cm long wire to RS41-SG. \textbf{No heating from RS41-SG}.}
		\begin{subfigure}[b]{0.3\textwidth}
			\includegraphics[width=\linewidth]{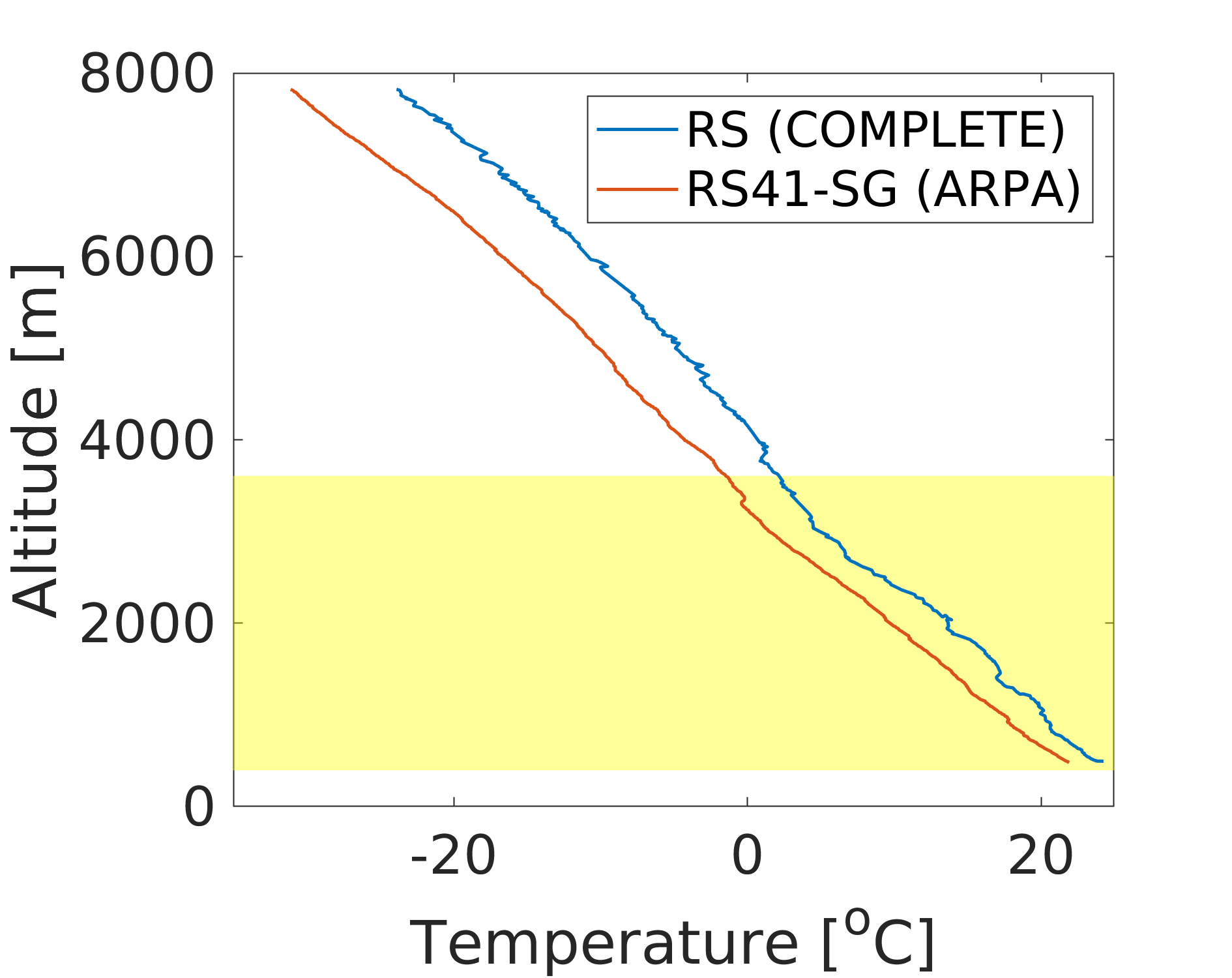}
			\caption{}
		\end{subfigure}
		\hspace{0.01\textwidth}
		\begin{subfigure}[b]{0.3\textwidth}
			\includegraphics[width=\linewidth]{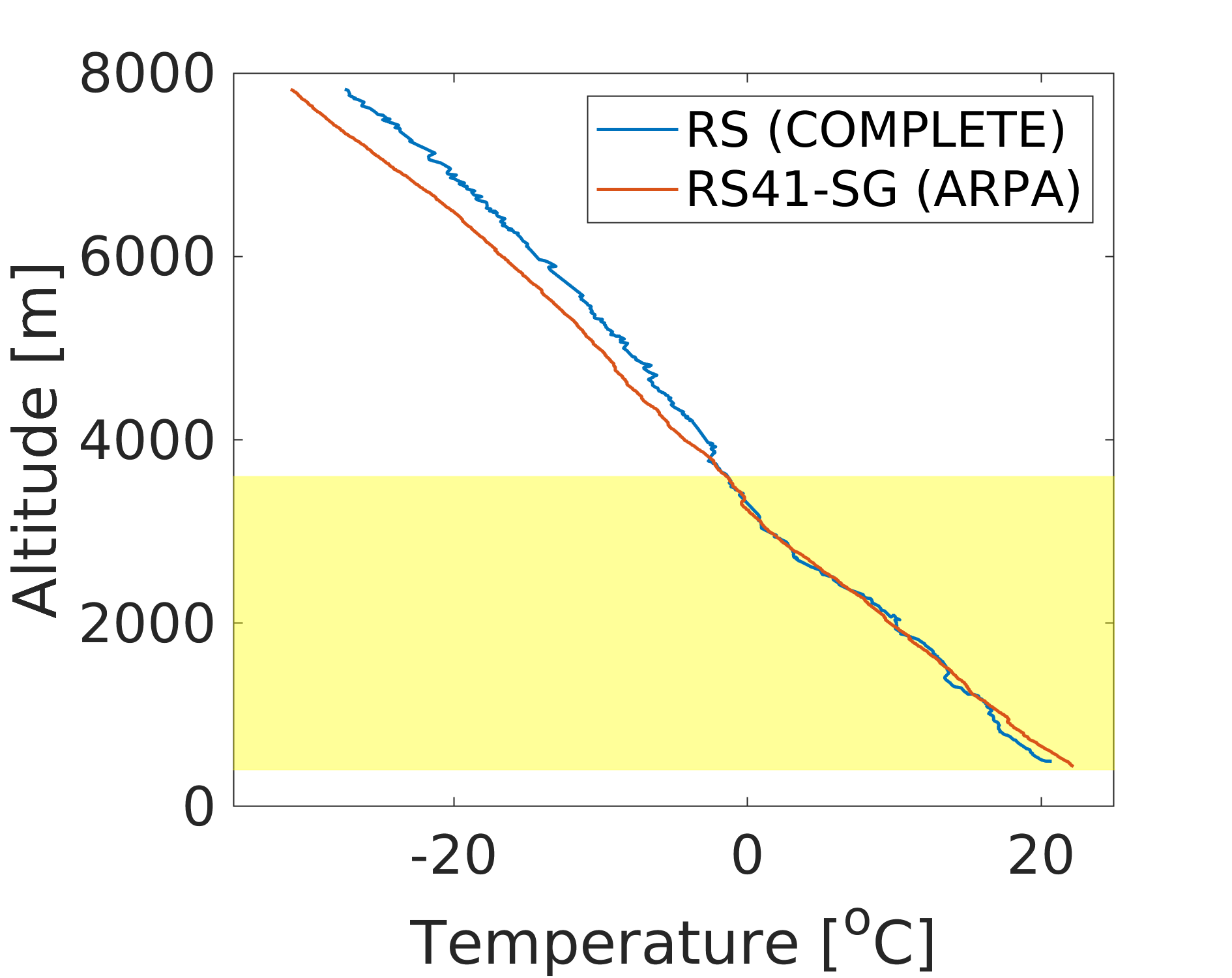}
			\caption{}
		\end{subfigure}
		\hspace{0.01\textwidth}
		\begin{subfigure}[b]{0.3\textwidth}
			\includegraphics[width=\linewidth]{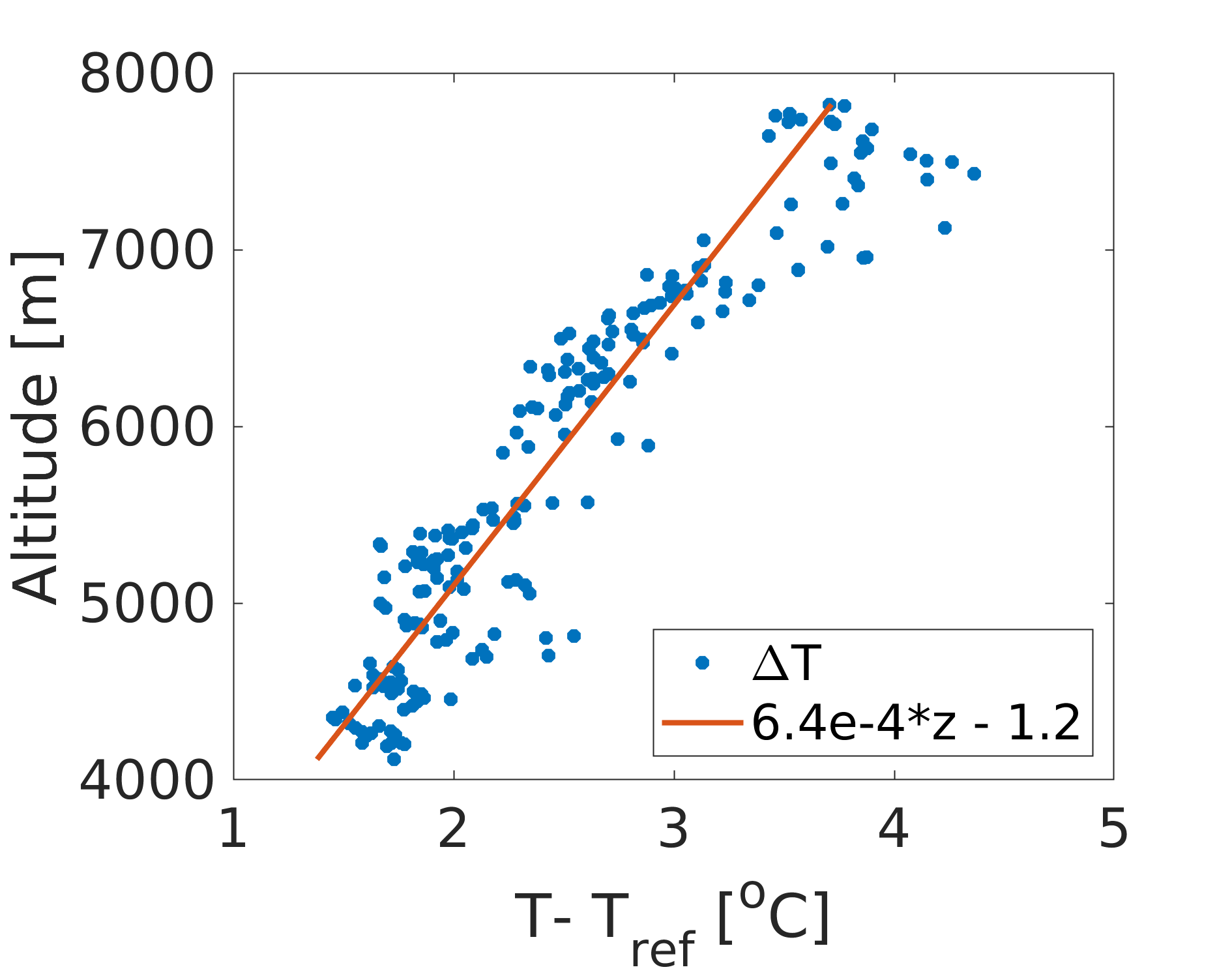}
			\caption{}
		\end{subfigure}
		\caption*{\scriptsize Levaldigi Airport, Cuneo, Italy. October 28th, 2020. COMPLETE probe fastened to Vaisala RS-41 SG probe with a scotch tape, \textbf{possible heating from RS41-SGP}.}
		\begin{subfigure}[b]{0.3\textwidth}
			\includegraphics[width=\linewidth]{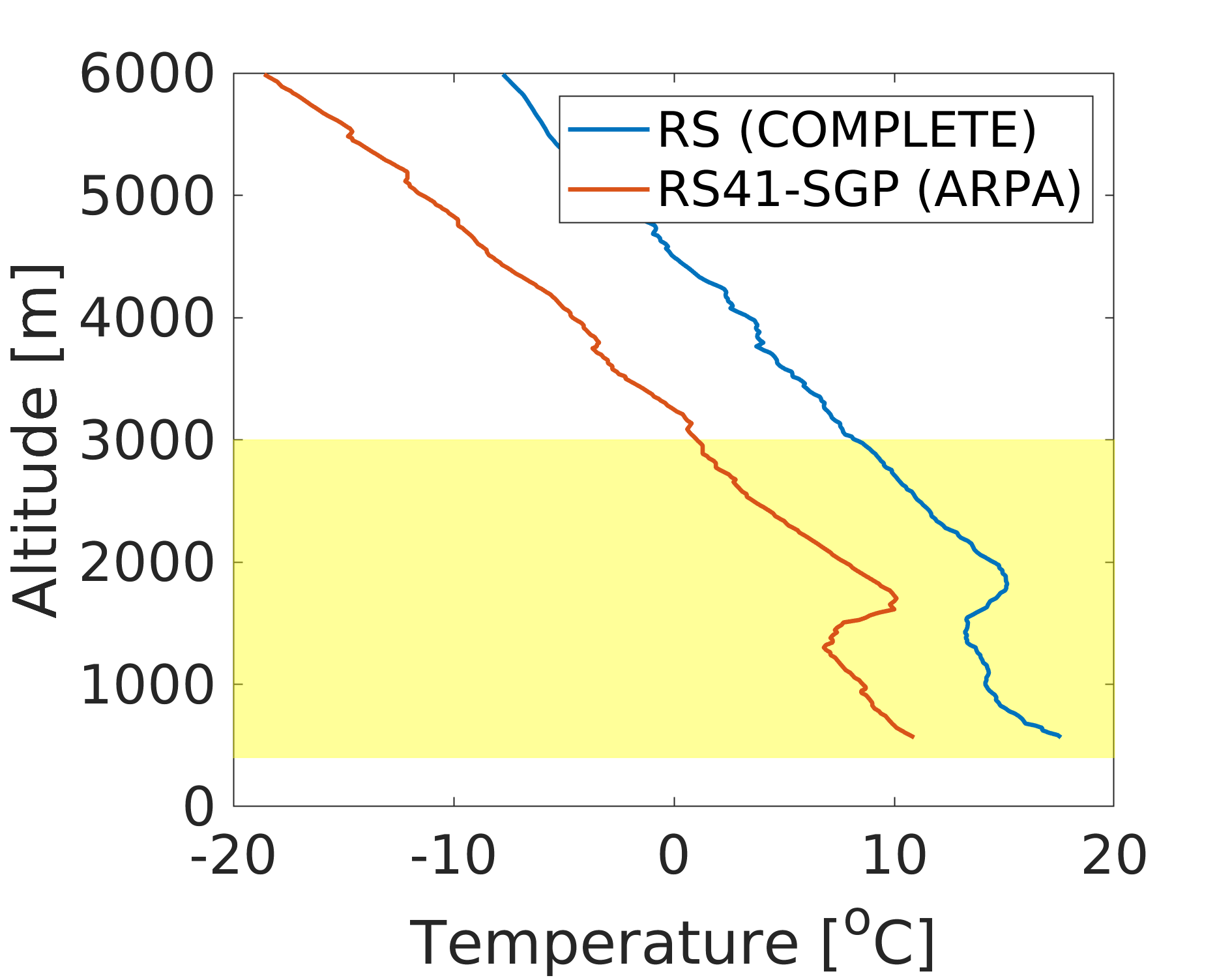}
			\caption{}
		\end{subfigure}
		\hspace{0.01\textwidth}
		\begin{subfigure}[b]{0.3\textwidth}
			\includegraphics[width=\linewidth]{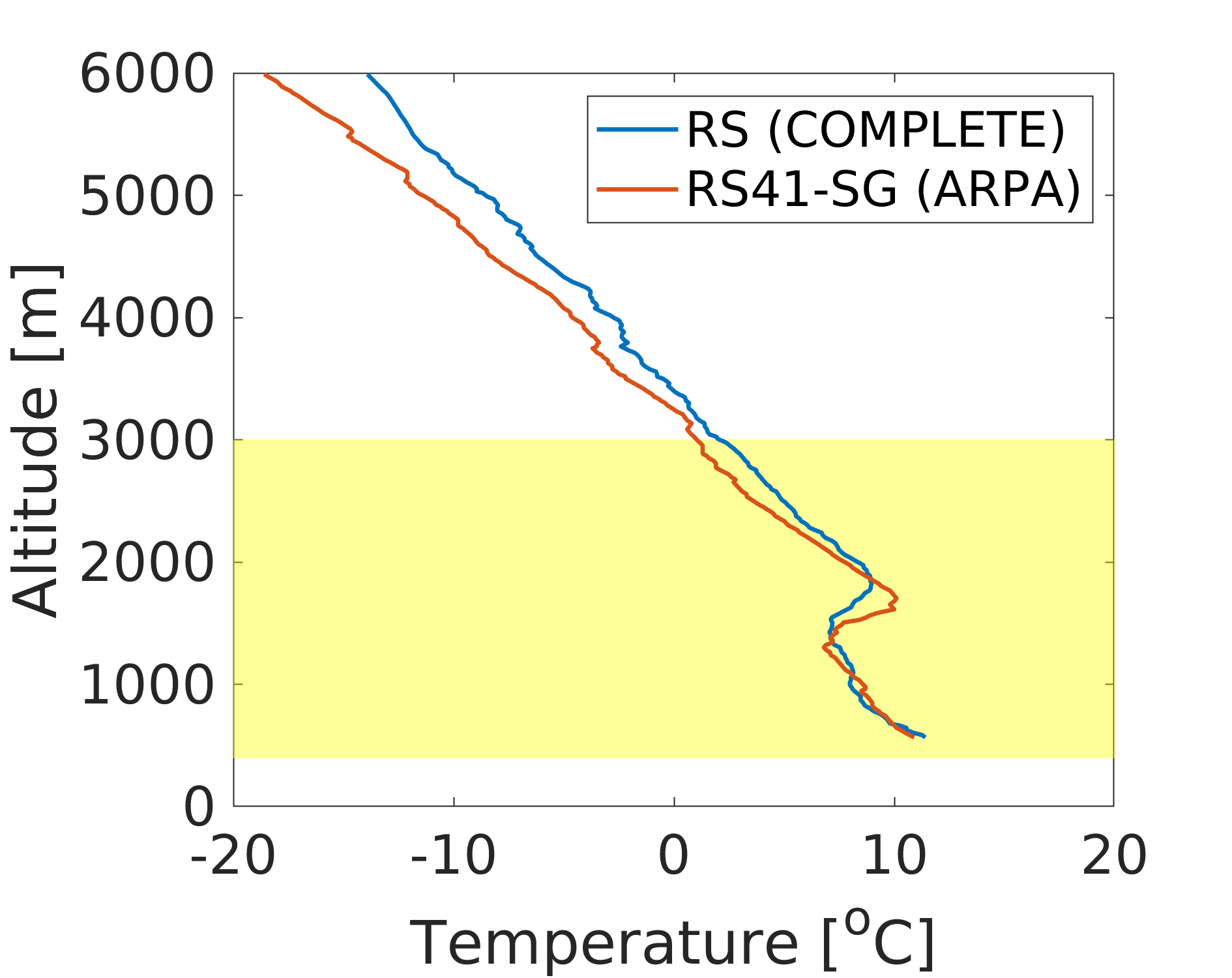}
			\caption{}
		\end{subfigure}
		\hspace{0.01\textwidth}
		\begin{subfigure}[b]{0.3\textwidth}
			\includegraphics[width=\linewidth]{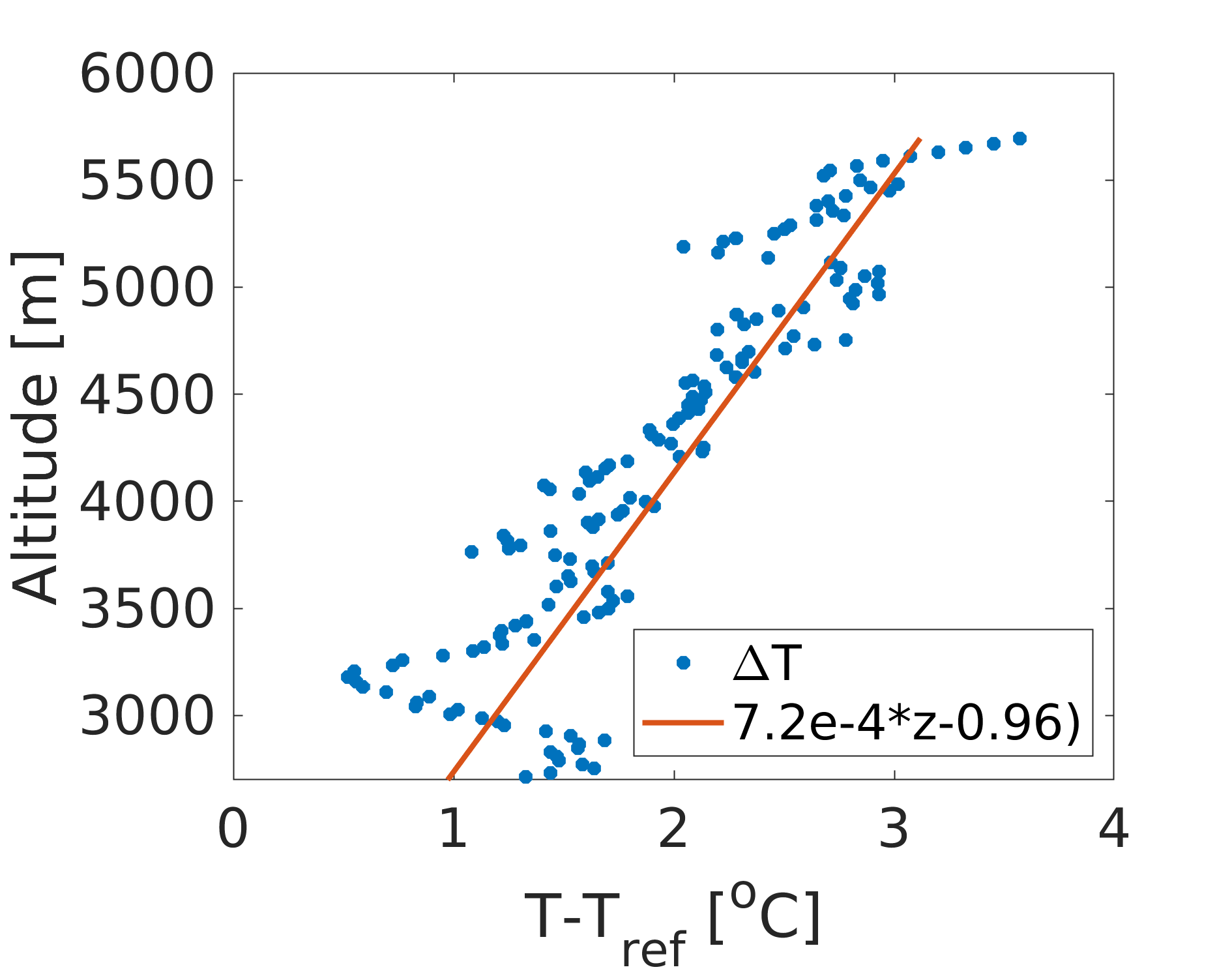}
			\caption{}
		\end{subfigure}
		\caption{{Dual-soundings of COMPLETE probe linked to Vaisala RS41-SG/SGP radiosonde. (a,d,g) Raw temperature readings along the altitude. (b,e,h) Temperature readings along the altitude after removing an initial bias. (c,f,i) Linear trends for the temperature measurement drift above the empirically observed thresholds due to radiation issues.}}
		\label{fig:dual_soundings}
	\end{figure}

	{As previously mentioned, temperature readings exhibit linear drift at higher altitudes, above ABL. To analyze this linear deviation trend, we considered data from all three dual-sounding tests.  In the case without an inversion cap (ARPA-2021), the linear temperature deviation begins above 4000 m. When an inversion cap is present (ARPA-2020 and MET-OFFICE-2023), the deviation starts at lower altitudes, around 3000 m. In Figure \ref{fig:dual_soundings}, we present both the raw data (panels a, d, g) and data compensated using the ground fixed-point bias obtained from Vaisala weather stations at each sounding launch location (panels a, d, g). The legends in the three rightmost panels of Figure \ref{fig:dual_soundings} illustrate the linear scaling of the temperature measurement drift at an altitudes above the empirically observed thresholds whihc is anyway higher than the design operative layer of the mini radioprobe cluster.}
	{The temperature drift observed above the ABL is surely caused, at least in part, by the solar irradiance. The effect of the radiation was already compensated for the reference RS-41SG radiosondes as reported by Vaisala \cite{onln:rs41preformance}. At the moment, the present realization of the mini-radiosondes does not yet apply any correction for radiation effects on the sensing element.  In the future, the link between above mentioned deviation trends and the solar irradiance can be determined within an upper-air simulator (UAS) as done by Lee et al.\cite{lee2022radiationrs41} and therefore proper corrections can be applied. Additionally, we were able to study a radiation effect during the pre-launch test of the recent experiment with a cluster of radiosondes at OAVdA, Saint-Barthelemy, Aosta, Italy, on February 10, 2022 (see Section \ref{sec:cluster_launch})}
	
	\subsubsection{{Accuracy of humidity measurements}}
	
	The plots in Figure \ref{fig:arpa_pht} show that the radioprobe suffered from some biases during the launching, with respect to the Vaisala RS-41 probe. The biases are evident, especially for the humidity and temperature readings, and are mainly due to heating and radiation on a sunny day. 
	The sensor datasheet suggests that the air flow in the direction of the vent hole of the PHT sensor should be engineered to allow a sufficient air exchange from the inside to the outside. This aspect was already considered during the design of the PCB board in tests in an environmental chamber, and in field experiments \cite{miryam_sensors2021}. However, it is believed that an improved board design will improve the humidity measurements. Furthermore, due to the slow response time, it has also been suggested to use low data rates for atmospheric observation applications. An air-flow velocity of approximately 1 m/s is needed to observe the effects on the response time of the device to the humidity measurements, which requires 1 second to reach 63\% of the step change \cite{docs:pht_datasheet}. For this purpose, response of the sensor to environmental changes was tested and validated inside the Kambic KK190 CHLT climatic chamber, which is located in the Applied Thermodynamics Laboratory of the \deleted{Italian National Metrology Institute(INRiM)}{INRiM}. {The climate chamber allows to control relative humidity within the range of 10\% to 98\%. During the tests, RH levels were maintained at constant values of 10\%, 20\%, 40\%, and 60\% within the climate chamber for a duration of 30 minutes each.} The obtained results matched well with the specifications provided by the manufacturer (see Figure 7 and Tables 4-5 of Paredes et al.\cite{miryam_sensors2021}).
	
	It is also worth mentioning that the PHT sensor, like the other sensor components, was selected because of its compact size, and its low-power and low-cost characteristics. {Our observations fully confirm that the use of low-cost RH sensors can lead to underestimation of high humidity values as reported in the study by Wilson et al.\cite{wilson2013effect}, even though low-cost sensors are becoming an integral part of the current IOT (Internet-Of-things) systems \cite{tagle2020field}. One of the potential solutions is to develop a compensation technique similar to the one developed for Vaisala RS-41SG probes \cite{onln:rs41preformance}. Another solution can be the adoption of an alternative sensor specifically designed for atmospheric monitoring applications, such as the KFS140-FA \cite{docs:kfs140fa_datasheet} or the P14 4051 Rapid Thermo \cite{docs:rapidp14_datasheet}.}
	
	\deleted{The observed biases of the temperature readings were mainly related to a radiation effect on the sensors, although the radiation effect is corrected in the Vaisala RS-41 radiosondes \cite{onln:rs41preformance}. The issue of the effect of radiation and the related corrections were discussed by \cite{lee2022radiationrs41}. The present realization of the mini-radiosondes does not apply any correction for radiation effects on the sensing element. Since the mini-radiosonde has to be used inside clouds, the radiation effect should be minimized. Moreover, the data analysis refers to relative measurements, that is, differences in the readings between the radiosondes when the balloons are fluctuating in the same area under similar radiation conditions. However, we were able to study a radiation effect during the recent experiment with a cluster of radiosondes at OAVdA, Saint-Barthelemy, Aosta, Italy, on February 10, 2022 (see Section \ref{sec:cluster_launch}).}
	
	
	\section{Preliminary in-field measurements with a cluster of radiosondes}\label{sec:field_tests}
	\deleted{The tests and validation results of the instrumentation that is currently in use, which includes fixed ground stations and vertical profiling radiosondes, are reported in the previous section.}
	{In the previous sections, the focus was primarily the discussion on measurements obtained from a single radiosonde. Here, we present the preliminary results obtained through simultaneous measurements from a cluster of radiosondes. An experiment involving a radiosonde cluster can allow us an endoscopic  observation of the atmospheric flow by concurrently collecting data from various regions of the flow domain.
		
		The trajectory data of each radiosonde can be effectively integrated with the recorded measurements of physical quantities along their respective trajectories. This integration process leads to the generation of a multi-Lagrangian dataset. The resulting Lagrangian dataset is valuable for conducting turbulent diffusion analysis. Furthermore, using clusters and tracking multiple physical quantities greatly increases the number of combinations of the Lagrangian cross-correlations.}

\subsection{{Pre-launch tests}}
\begin{figure}[ht!]
	\centering
	\begin{subfigure}{0.75\textwidth}
		\includegraphics[width=\linewidth]{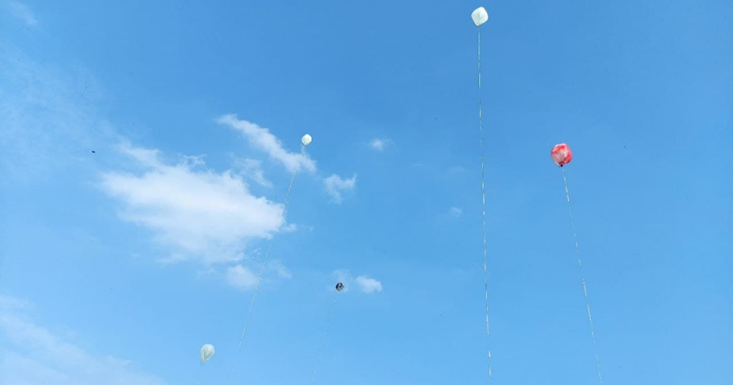}
	\end{subfigure}
	\caption{{Experiment with five tethered radiosondes at INRIM on September 29, 2021. Two radiosondes tracked using video camera were marked in red and black.}}
	\label{fig:inr2_setup}
\end{figure}
{Experimental tests were conducted with radiosondes in a tethered mode, under controlled conditions, before proceeding with a free-flying experiment involving a radiosonde cluster. Two such tests were performed with a cluster of five tethered radiosondes at the INRIM campus on September 29, 2021, and at OAVdA, St. Barthelemy, on February 10, 2022. During these tests, we assessed the overall performance of the cluster setup with five radiosondes as transmitters and two ground stations as receivers. Figure \ref{fig:inr2_setup} displays the configuration of the tethered radiosondes during the INRIM test.} \deleted{We employed a Sony HDV camera for the position validation to assess the relative position of the radiosondes from the camera viewpoint. During the experiment, two radiosondes were marked, one in red and one in black to help them to be seen by the camera. However, one camera was insufficient to precisely reconstruct the relative positions and separations between the radiosondes. Therefore, it was necessary to use a set-up that, included at least two cameras in order to carry out a reliable stereo vision analysis. This arrangement was also implemented in the subsequent in-field tests in the Aosta Valley.}

\begin{figure}[ht!]
	\centering
	\begin{subfigure}{0.45\textwidth}
		\includegraphics[width=\linewidth]{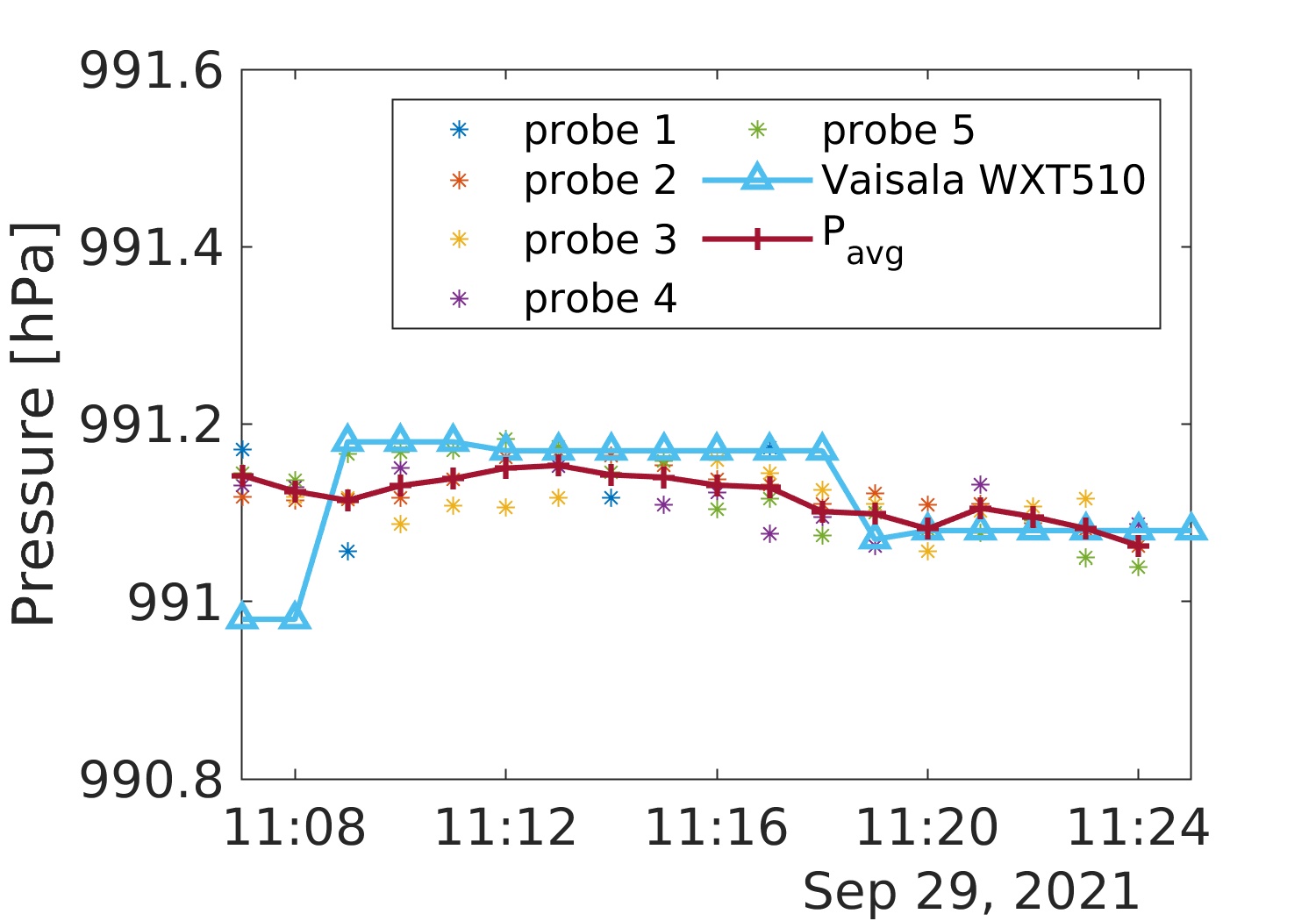}
		\caption{}
	\end{subfigure}
	\hspace{0.01\textwidth}
	\begin{subfigure}{0.45\textwidth}
		\includegraphics[width=\linewidth]{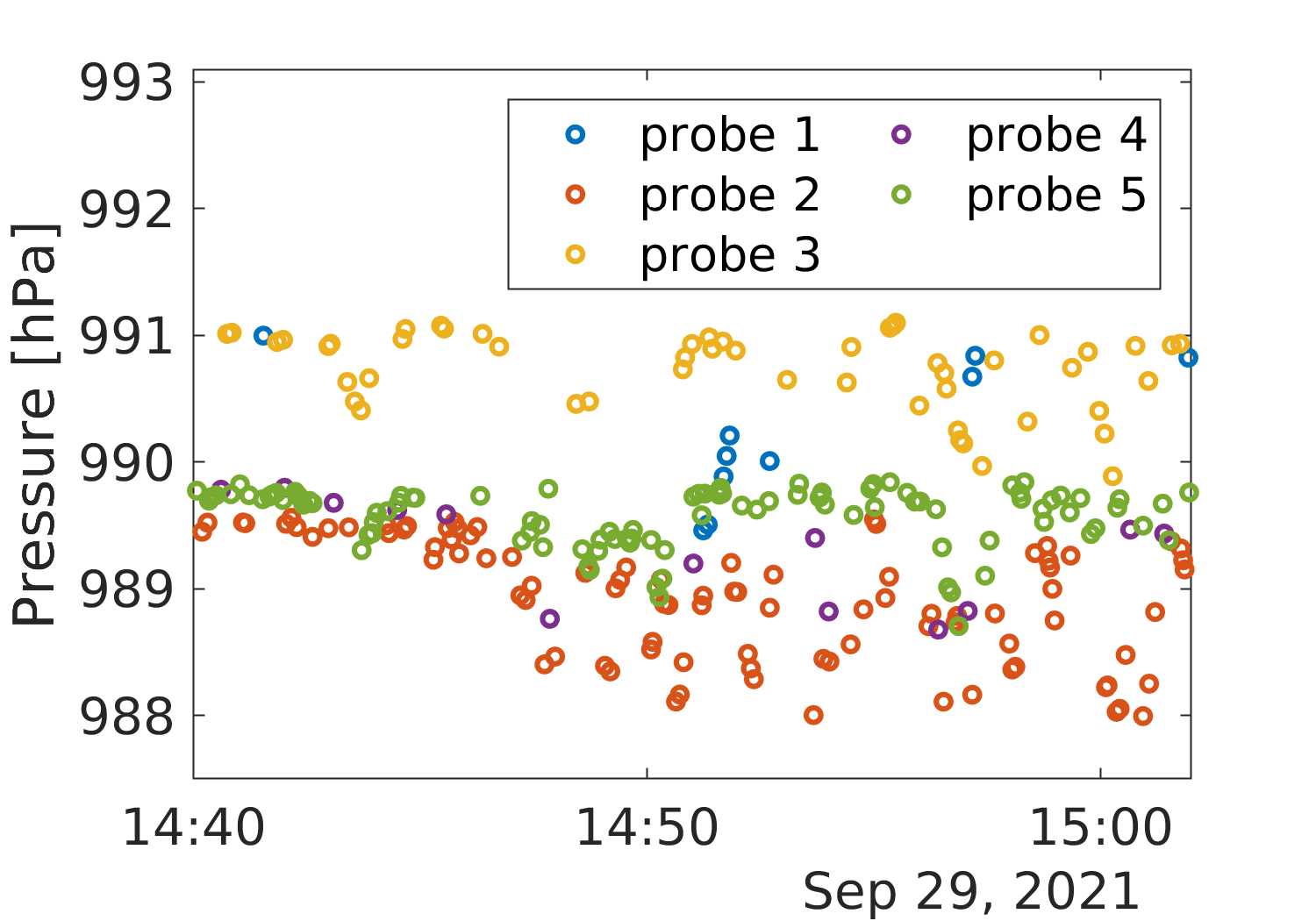}
		\caption{}
	\end{subfigure}
	\vspace{0.01\textwidth}
	\begin{subfigure}{0.45\textwidth}
		\includegraphics[width=\linewidth]{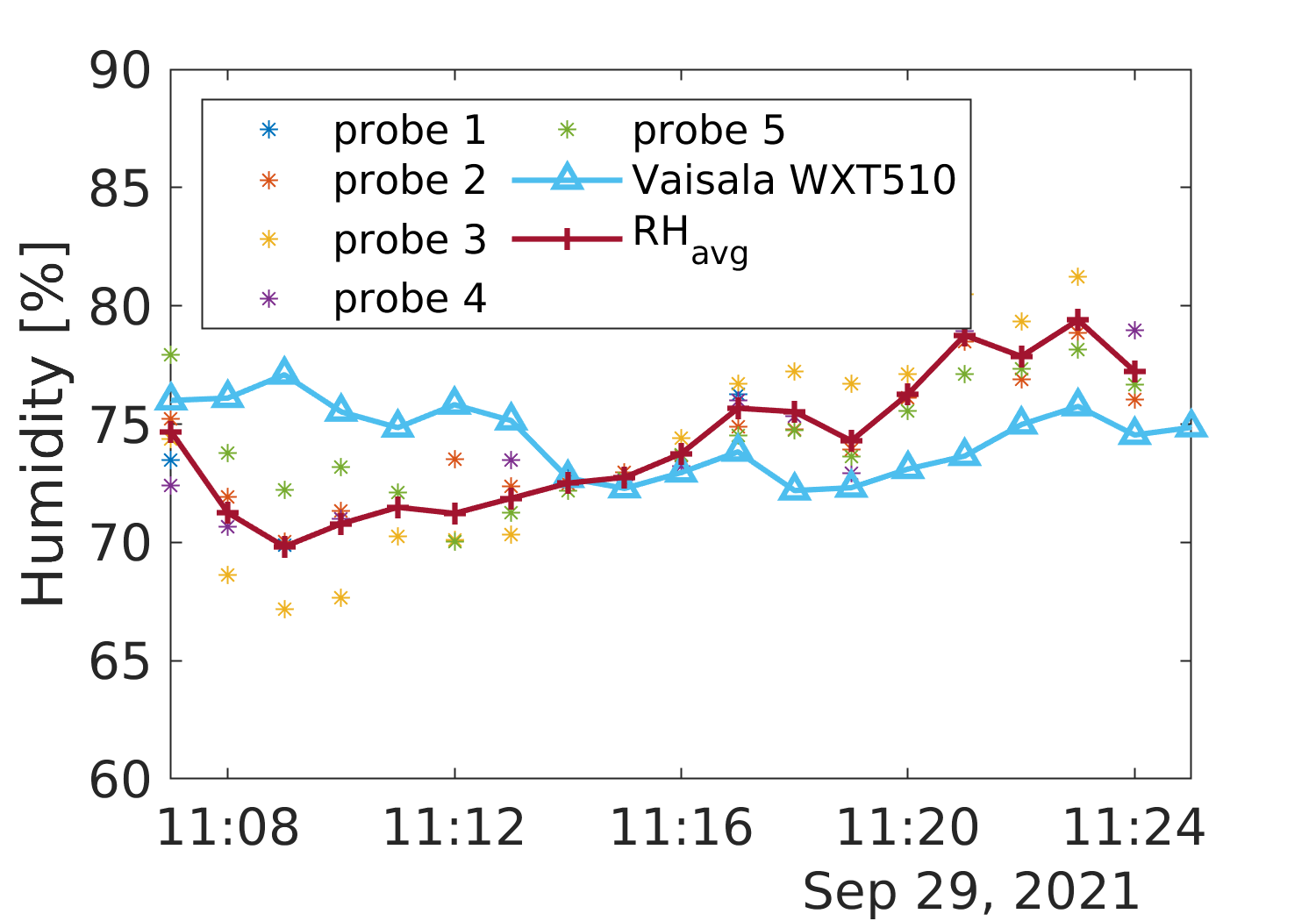}
		\caption{}
	\end{subfigure}
	\hspace{0.01\textwidth}
	\begin{subfigure}{0.45\textwidth}
		\includegraphics[width=\linewidth]{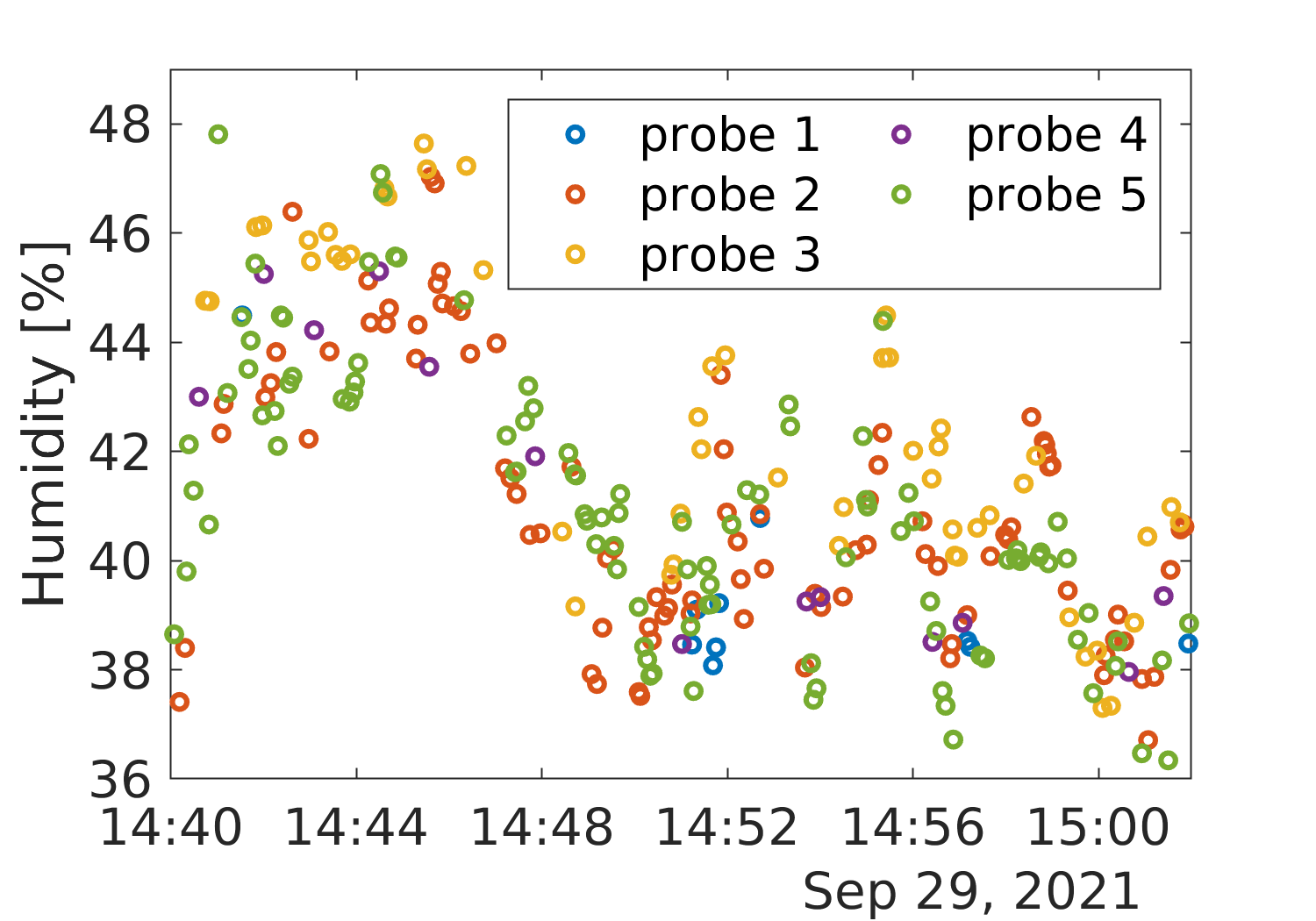}
		\caption{}
	\end{subfigure}
	\vspace{0.01\textwidth}
	\begin{subfigure}{0.45\textwidth}
		\includegraphics[width=\linewidth]{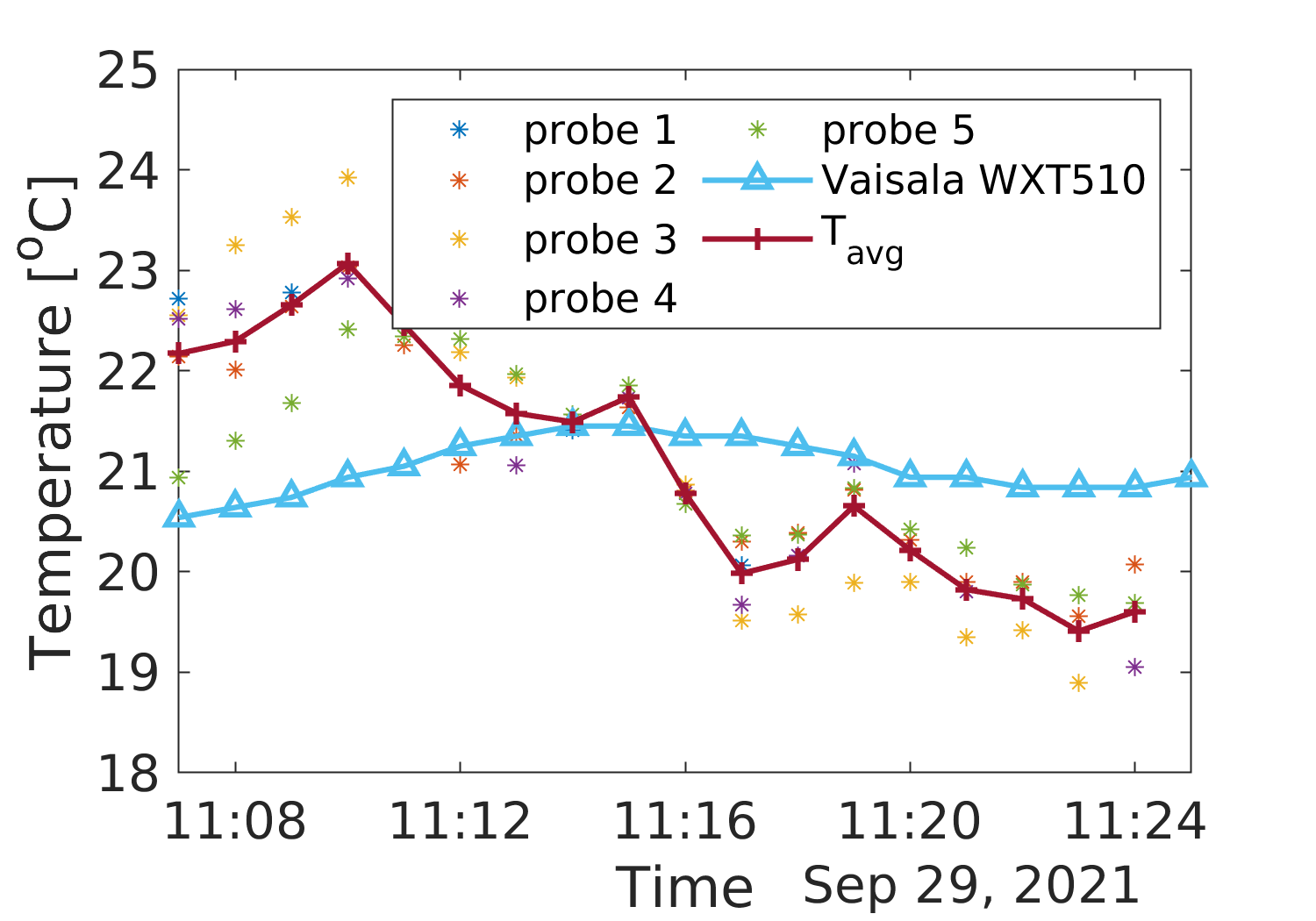}
		\caption{}
	\end{subfigure}
	\hspace{0.01\textwidth}
	\begin{subfigure}{0.45\textwidth}
		\includegraphics[width=\linewidth]{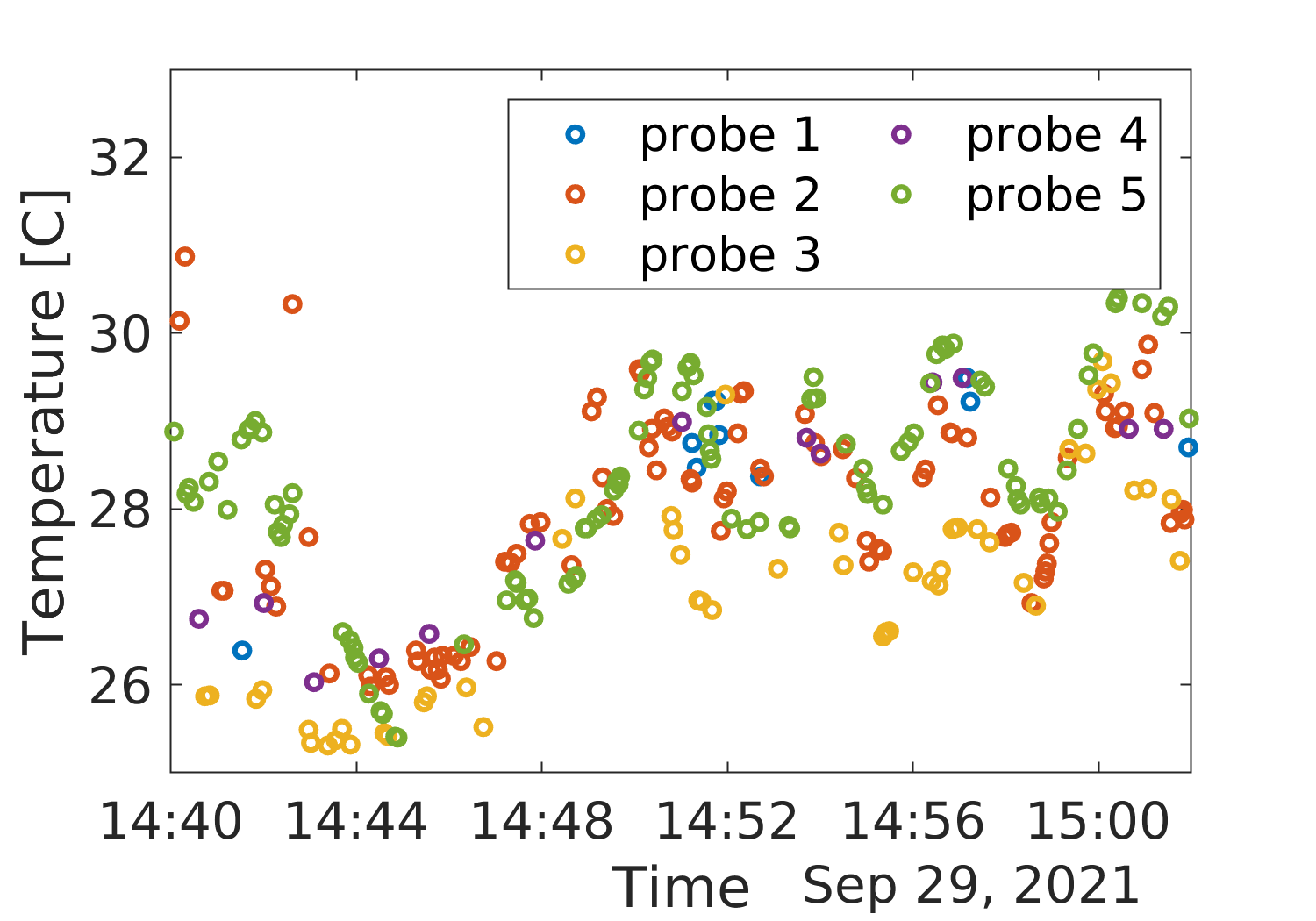}
		\caption{}
	\end{subfigure}
	\caption{{Pressure, humidity, and temperature readings during the INRIM campus experiment on September 29, 2021. The first column (panels a, c, and e) displays minutely averaged sensor readings from a pre-launch test near the Vaisala WXT510 station.  The solid red line represents the average values  for each minute computed over the readings of the 5 COMPLETE probes not shielded by the solar radiation as the Vaisala station which is shielded by a solar helical protection. The second column (panels b, d, and f) shows measurements of the radiosondes from 14:40 to 15:02. During this phase, radiosondes were launched in tethered mode as shown in Figure \ref{fig:inr2_setup}.}}
	\label{fig:inr2_pht}
\end{figure}
{Figure \ref{fig:inr2_pht} displays pressure, humidity, and temperature measurements, including pre-launch checks and tethered cluster launching. Sensor readings were validated against the INRIM campus Vaisala WXT510 weather station, situated two meters above the ground and providing one-minute interval data. Pre-launch check results are summarized in Table \ref{tbl:inr2_pht_comp}. Pressure readings were highly accurate as we observed during dual-soundings, while humidity and temperature sensors without compensation for solar irradiance performed well with an average RMSD (Root Mean Square Deviation) of 3.57\% and 1.21°C, respectively, closely aligning with manufacturer specifications. In addition, the sensor uncertainties could be further reduced by performing extensive comparative tests over a wide range of atmospheric conditions.}

\begin{table}[ht!]
	\centering
	\caption{RMSD of the pressure, humidity, and temperature readings of five radiosondes relative to measurements of the Vaisala WXT510 automatic weather station at the INRIM campus on September 29, 2021. Average values  of the RMSD are provided in the last two rows, together with the values declared by the manufacturer \cite{docs:pht_datasheet}.}
	\small
	\begin{tabular}{p{1.6cm} p{1.3cm} p{1.3cm} p{1.8cm}} 
		\hline
		Quantity/ \newline probe & Pressure [hPa] & Humidity [\% RH] & Temperature [$^o$C] \\ [0.5ex] 
		\hline\hline
		probe 1 & 0.08 & 3.45 & 1.15 \\
		probe 2 & 0.06 & 3.06 & 1.11 \\
		probe 3 & 0.06 & 5.24 & 1.69 \\
		probe 4 & 0.06 & 3.38 & 1.29 \\
		probe 5 & 0.06 & 2.70 & 0.82 \\
		\hline
		Average & 0.065 & 3.57 & 1.21 \\ \hline
		Datasheet & 1.0 & 3.0 & 0.5 - 1.5 \\
		[1ex]
		\hline
	\end{tabular}
	\label{tbl:inr2_pht_comp}
\end{table}

\subsection{Validation of the position with the stereo vision analysis}\label{sec:pos_validation}

\begin{figure}[bht!]
	\centering
	\begin{subfigure}{0.8\textwidth}
		\includegraphics[width=\linewidth]{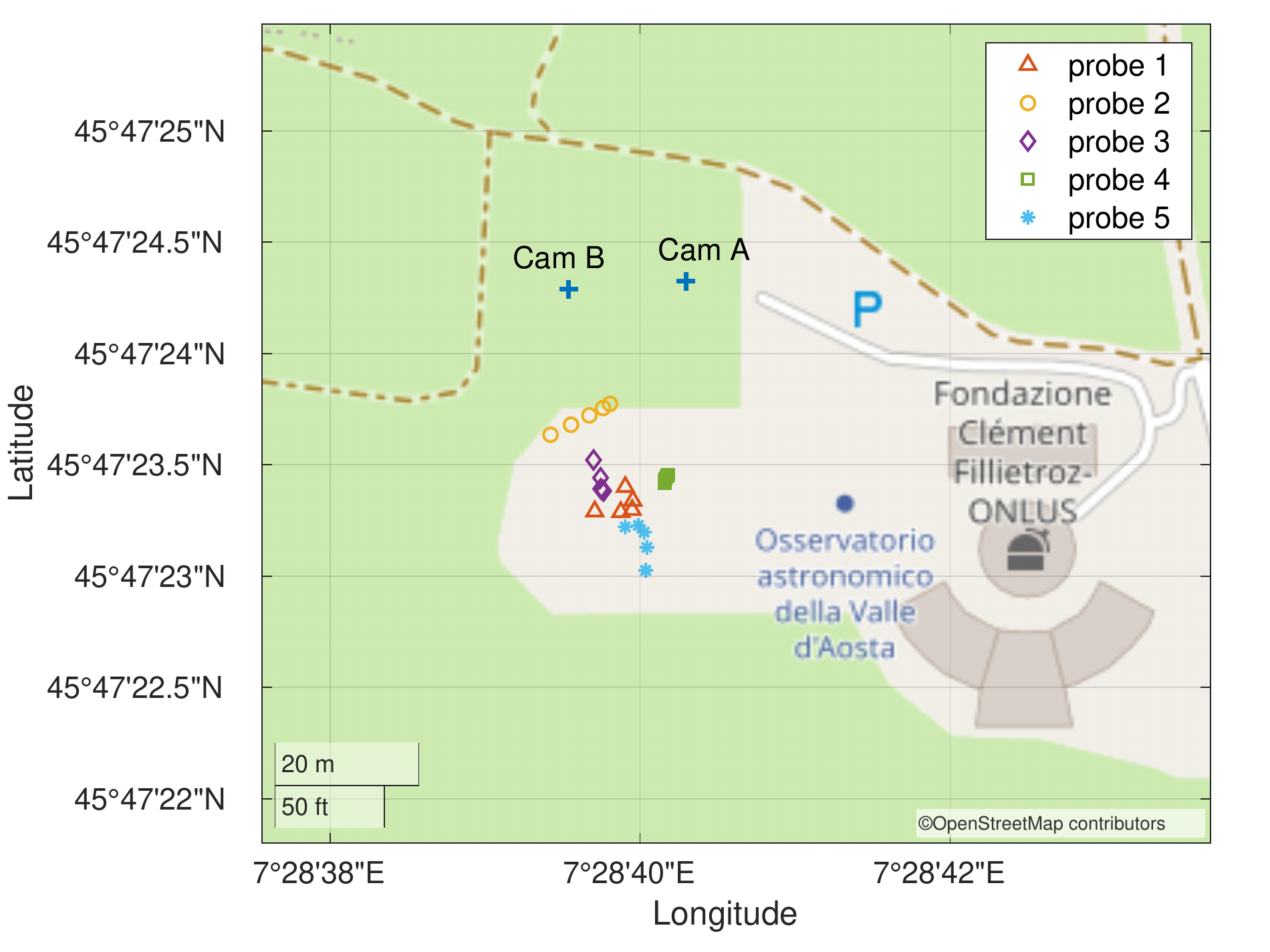}
	\end{subfigure}
	\caption{{Stereo vision experiment setup in Saint-Barthelemy, Italy, on February 10, 2022, with tethered radiosondes. Multiple positions were recorded between 16:15 and 16:17 using two Sony HDV cameras, 16 meters apart from each other(Cam A and Cam B).}}
	\label{fig:ao1_setup_cam}
\end{figure}
The position and trajectory information of each radiosonde are crucial to acquire Lagrangian statistics on various flow quantities. Several tests were performed during the development of the radiosonde system, to validate the well-functioning of the positioning sensors (GNSS and IMU). The results of the first validation tests were presented in a previous work, \cite{miryam_sensors2021}, where 2D position data of a radiosonde were compared with those of a phone positioning dataset. The phone position dataset showed a higher precision and accuracy, thanks to A-GNSS (Assisted GNSS), and can be thus considered a good reference. Furthermore, two comparative launches with the Vaisala RS41-SG probe were used to assess the 3D position dataset of the radiosonde as well as other sensor measurements, as described in Section \ref{results_vertical_launch}.

{In a recent in-field experiment campaign at OAVdA, St. Barthelemy, on February 10, 2022, we evaluated relative distance tracking using a multi-radiosonde setup. In this experiment, we measured and validated the relative changes in positioning information of the radiosondes compared to the distances tracked by a stereo vision camera. The experimental setup, illustrated in Figure \ref{fig:ao1_setup_cam}, consisted of five radiosondes filled with helium and tethered to the ground, two ground stations ready to receive packets from all radiosondes, both connected to PCs for real-time data storage and analysis, and two Sony HDV cameras to capture movements of the marked radiosondes during the experiment.}

\begin{figure}[bht!]
	\centering
	\begin{subfigure}{0.45\textwidth}
		\includegraphics[width=\linewidth]{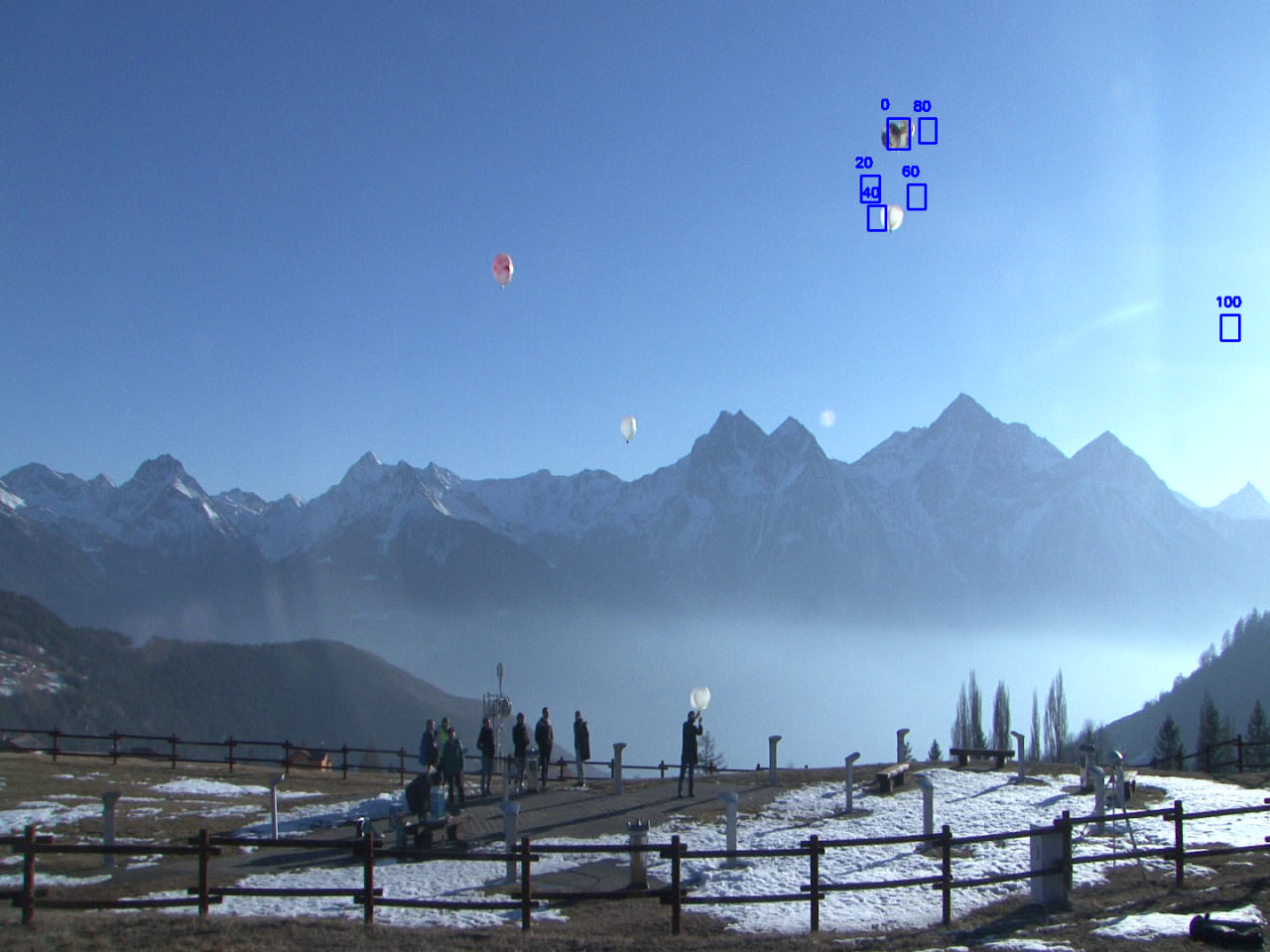}
		\caption{}
	\end{subfigure}
	\hspace{0.01\textwidth}
	\begin{subfigure}{0.45\textwidth}
		\includegraphics[width=\linewidth]{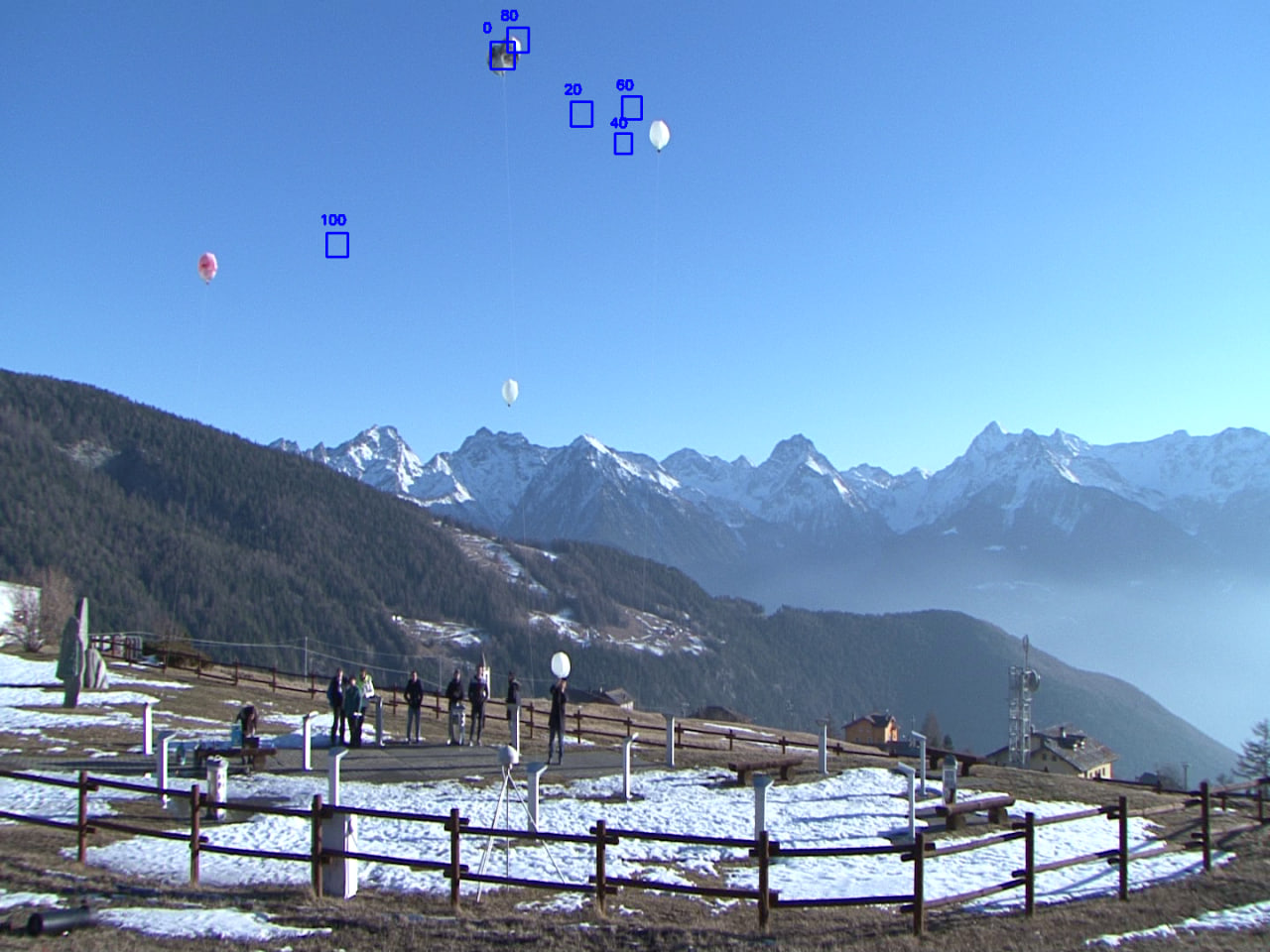}
		\caption{}
	\end{subfigure}
	\vspace{0.01\textheight}
	\begin{subfigure}{0.7\textwidth}
		\includegraphics[width=\linewidth]{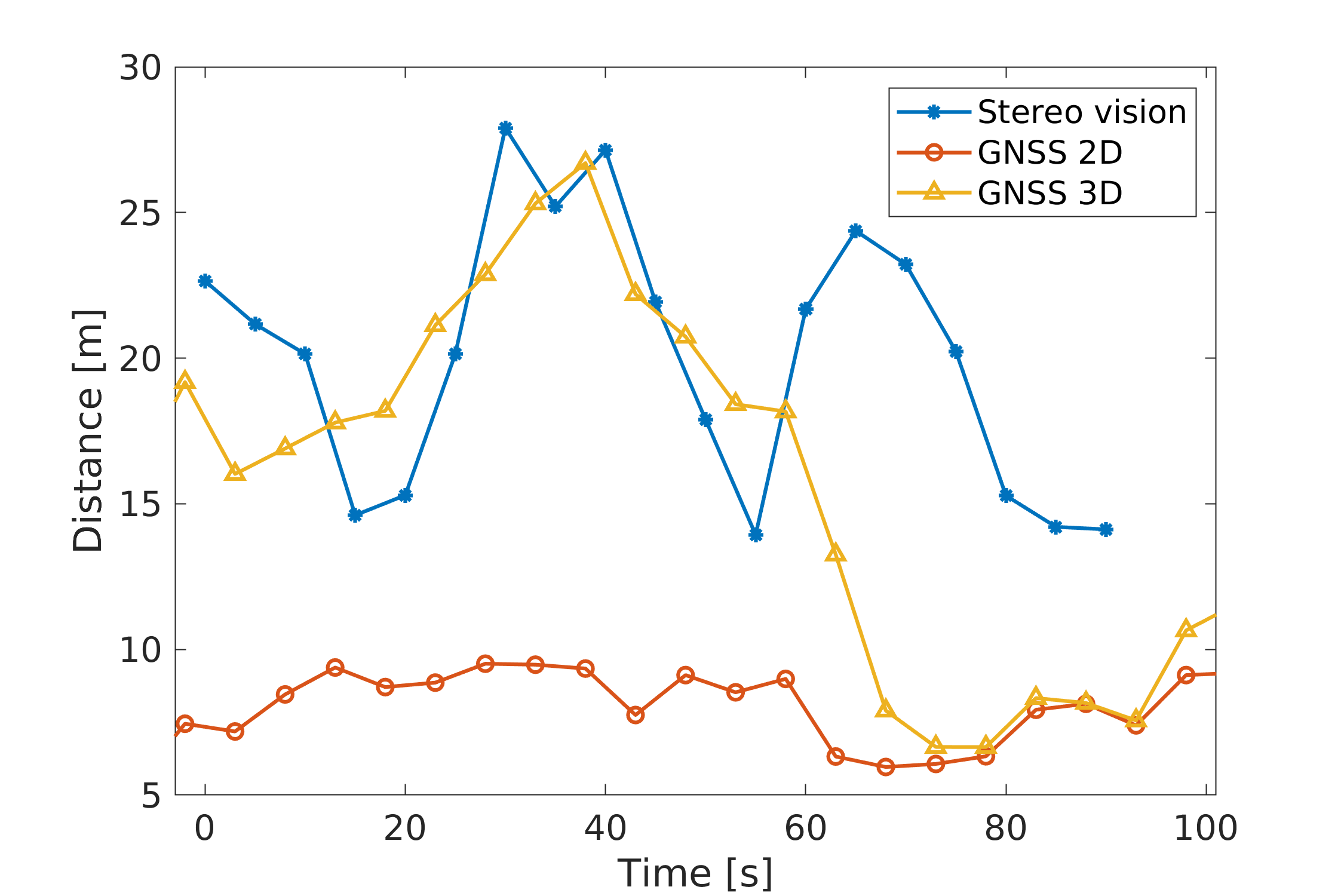}
		\caption{}
	\end{subfigure}
	\caption{{The relative distances between radiosondes, determined by stereo vision and GNSS coordinates during the OAVdA experiment (Figure \ref{fig:ao1_setup_cam})}. (a) Initial frame from camera A. (b) Initial frame from camera B. The blue rectangles in these frames indicate the subsequent positions of the radiosonde with a black balloon, with the corresponding time (in seconds relative to the initial frame) displayed in the upper left corner of each blue rectangle. {(c) Relative distance over time with reference to the initial frame at 16:25:37.}}
	\label{fig:ao1_stereo_comparison}
\end{figure}

The relative movement of two radiosondes, marked in red and black colors, was chosen as a quantity of interest for validation. Figure \ref{fig:ao1_stereo_comparison} shows the relative distances between the marked radiosondes, which were produced from two datasets: GNSS (longitude, latitude, and altitude) coordinates and video frames recorded by two Sony HDV cameras. \deleted{In this study, the performance of the GNSS sensors was compared with the distance as determined by means of the stereo vision analysis of two Sony HDV cameras.} The pinhole, linear camera model with basic camera calibration \cite{onln:simplestereo}, was employed for distance computation using stereo vision. {The tracking of marked radiosondes in the subsequent video frames was conducted using the CSRT tracker (Discriminative Correlation Filter with Channel and Spatial Reliability)\cite{Lukezic_2017_CVPR} from the OpenCV object tracking API \cite{onln:opencvtracking}. The CSRT tracker was selected due to its relatively superior performance with respect to other OpenCV tracking algorithms, as demonstrated by its approximate 90\% accuracy and an approximately 85\% success rate of detection, along with its resistance to interference from overlapping objects.} {Regarding the camera coordinate frame used in the stereo vision technique, the following accuracy values were estimated: (i) horizontal direction perpendicular to the camera axis, $\Delta$x = $\pm$ 0.5 m within a range of 20 m (2.5\%); (ii) vertical direction $\Delta$y = $\pm$  0.8 m within a range of 20 m (4\%); and horizontal depth direction $\Delta$z = $\pm$  4 meters within a range of 100 m (4\%).}

The relative distances obtained from the GNSS dataset and through stereo vision in Figure \ref{fig:ao1_stereo_comparison}c showed a good alignment for major part of the selected time window. In fact, the mean absolute difference for the first 60 seconds of the time window was 2.6 meters, while it was 5.2 meters for the entire time window. The trajectory of the radiosondes rapidly changed at the end of the time window due to the presence of a strong wind. Both balloons were close to the ground and visible at the edges of the frames of both camera frames. {A comparison of 2D and 3D distances from the GNSS dataset confirmed this observation.} While the stereo vision dataset was taken as a reference, we believe, on the basis of a visual analysis of the experiment scene, that the GNSS sensor performed better. {Additionally, it's worth noting that the differences in GNSS distances, in relation to those obtained through stereo vision, fall within the accuracy range provided by the GNSS sensor manufacturer ($\pm$ 4-8 meters). The manufacturer specifies a horizontal accuracy of $\pm$ 4 meters in the Super-E power-saving mode, but vertical accuracy is not typically provided, though it is generally known to be about 1.7–2 times that of horizontal accuracy (see page 6 in \cite{docs:ubx_rec_desc}).}

In this investigation, we primarily used the GNSS dataset to evaluate the positioning and trajectory tracking capabilities of the proposed system. {However, the IMU dataset could be utilized to enhance and optimize the GNSS sensor's positioning data. This can be achieved by using a combination of sensor fusion algorithms, such as Madgwick \cite{madgwick2011} Kalman \cite{Kalman1960ANA}. However, to maintain clarity and simplicity, we defer the analysis and the discussion of position tracking for future studies.}

\subsection{Free launching of multiple radiosondes}\label{sec:cluster_launch}
\begin{figure}[ht!]
	\centering
	\begin{subfigure}{0.71\textwidth}
		\centering
		\includegraphics[width=\textwidth]{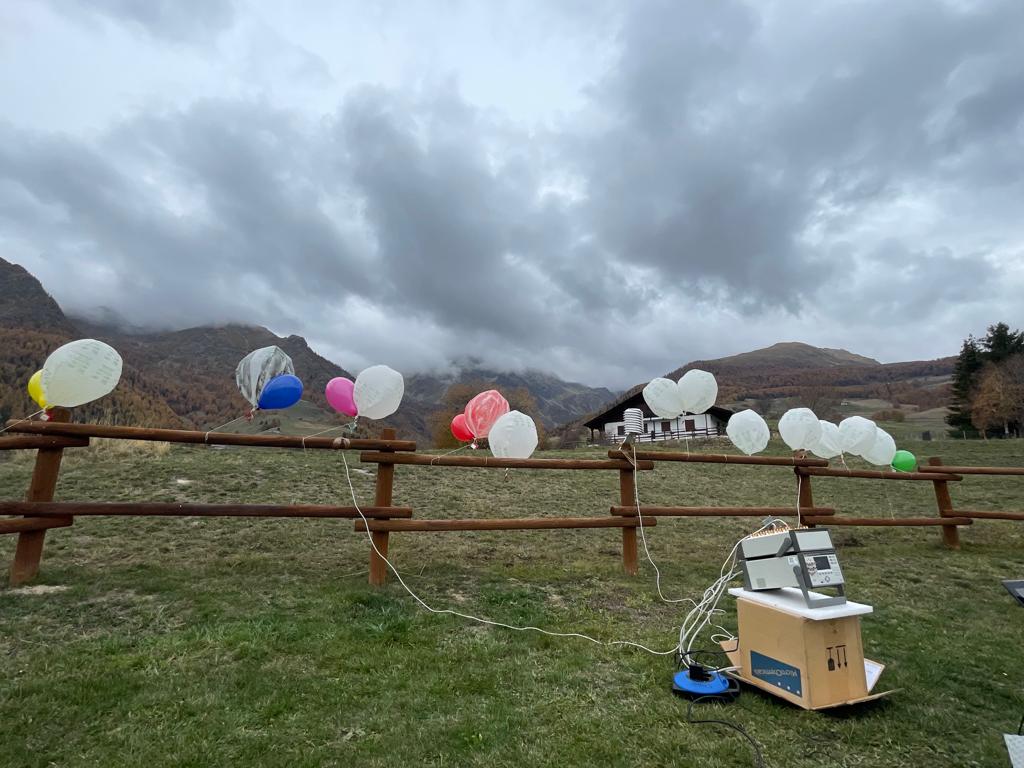}
		\caption{Pre-launch check}
	\end{subfigure}
	\vspace{0.01\textwidth}
	\begin{subfigure}{0.29\textwidth}
		\centering
		\includegraphics[width=\textwidth]{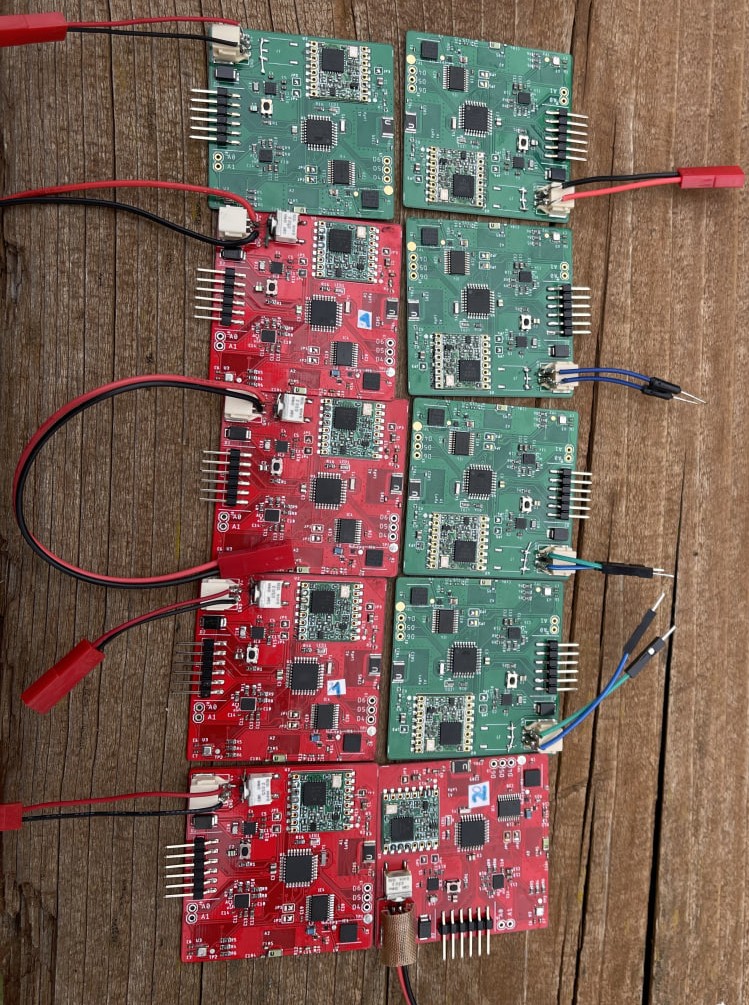}
		\caption{radioprobe boards}
	\end{subfigure}
	\hspace{0.01\textwidth}
	\begin{subfigure}{0.29\textwidth}
		\centering
		\includegraphics[width=\textwidth]{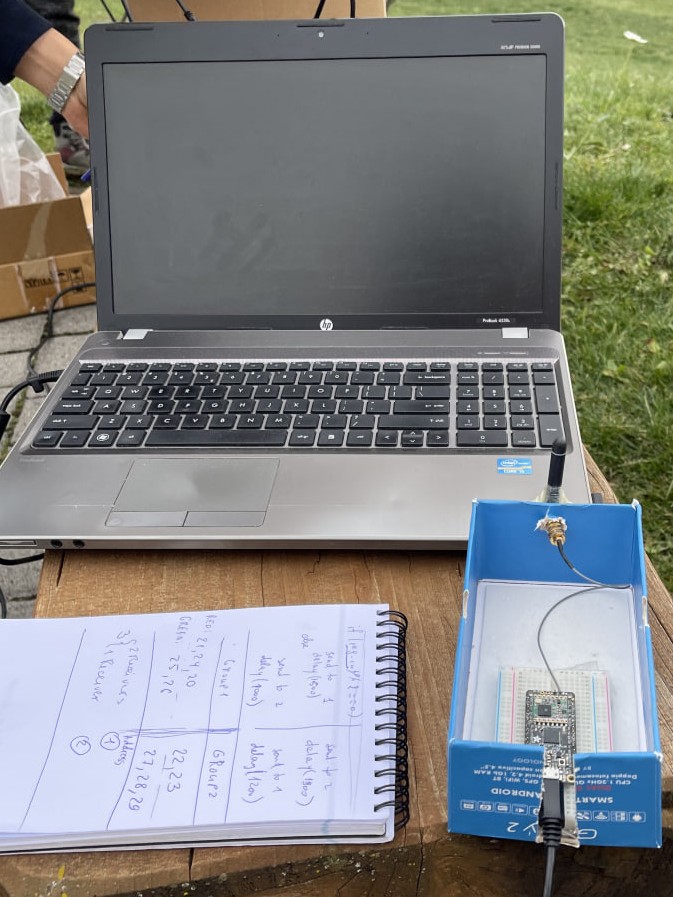}
		\caption{receiver station}
	\end{subfigure}
	\hspace{0.01\textwidth}
	\begin{subfigure}{0.32\textwidth}
		\centering
		\includegraphics[width=\textwidth]{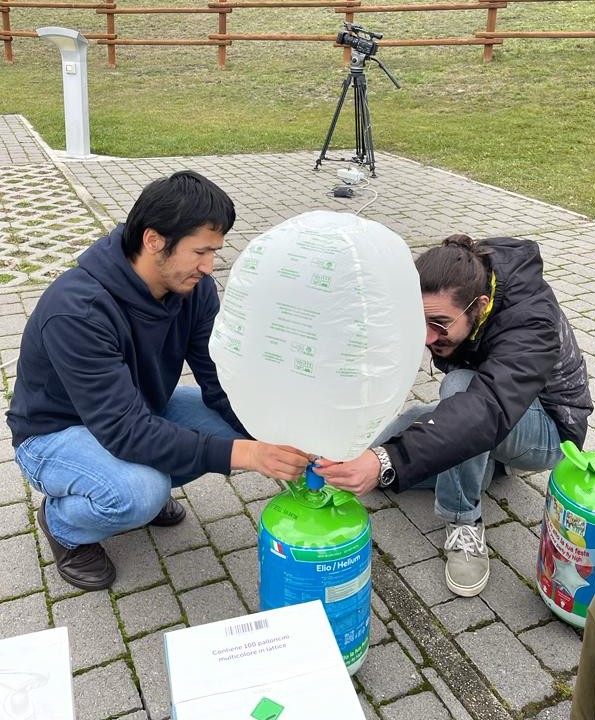}
		\caption{balloon inflation}
	\end{subfigure}
	\caption{Experiment setup at OAVdA \deleted{(L’Osservatorio Astronomico della Regione Autonoma Valle d’Aosta)}, St. Barthelemy, Aosta, Italy, on November 3, 2022. (a) The radiosondes during pre-launch calibration with INRIM reference instrumentation. (b) Prepared radioprobe electronic boards. (c) A ground station connected to the laptop PC. (d) Balloon preparation for radiosonde assembly.
	}
	\label{fig:ao2_experiment_setup}
\end{figure}

\begin{figure}[ht!]
	\centering
	\begin{subfigure}{0.7\textwidth}
		\centering
		\includegraphics[width=\textwidth]{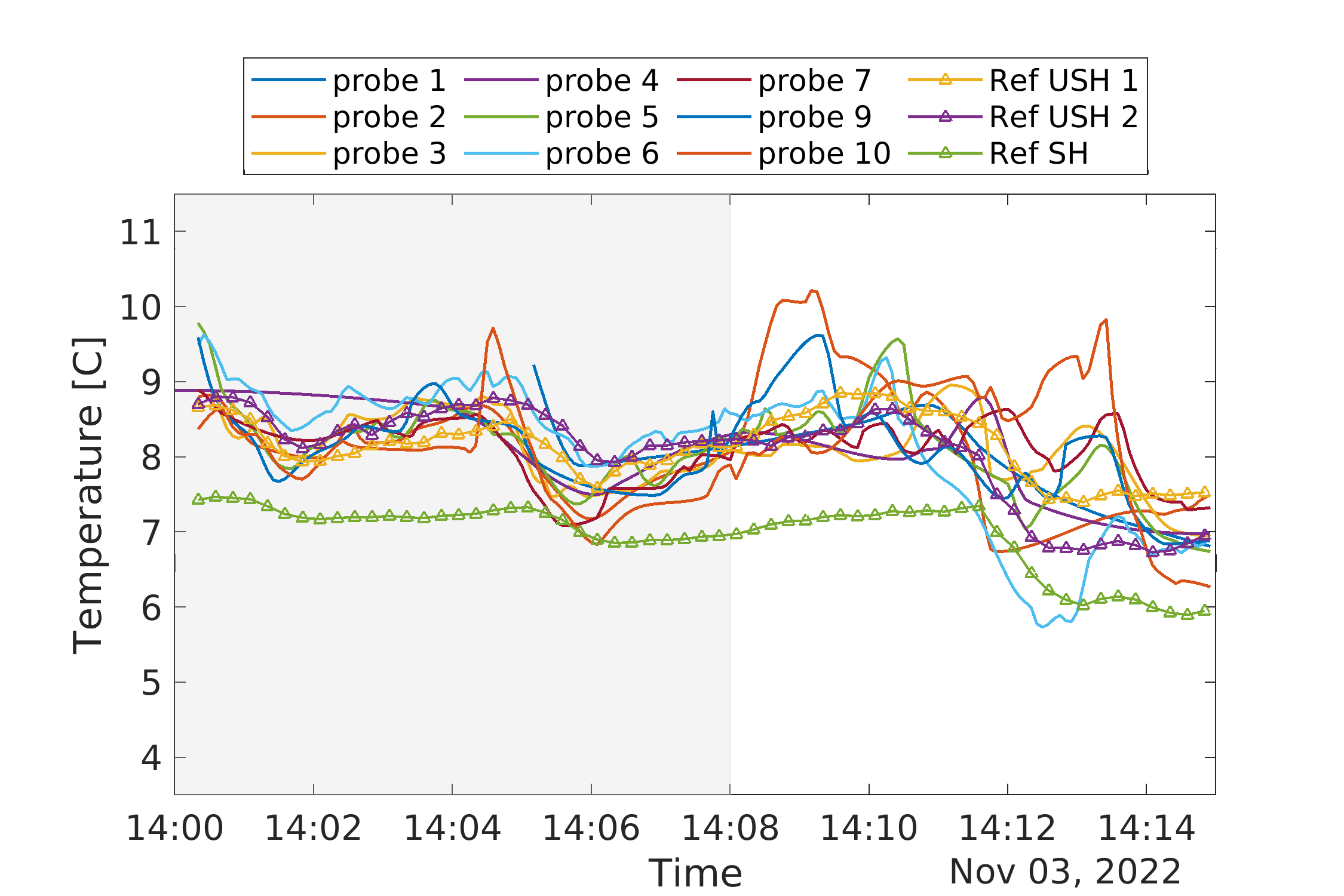}
	\end{subfigure}
	\caption{\textit{Pre-launch} calibration during the OAVdA experiment at Saint-Barthelemy, Aosta, Italy, on November 3, 2022. Temperature measurements were compared with INRIM reference instrumentation using readings from a Fluke DAQ 1586A multimeter. All the radiosondes were fixed to the fence during the first phase (highlighted in light gray) as in Figure \ref{fig:ao2_experiment_setup}a, while the radiosondes were picked up for the free launching in the second phase. See Table \ref{tbl:ao2_pht_comp} for  the standard deviations and mean differences in the temperature measurements.}
	\label{fig:ao2_cal_pht}
\end{figure}

The conducted preliminary tests and in-field experiments allowed us to carry out the recent experiment with a cluster of freely floating radiosondes at OAVdA, St. Barthelemy, Aosta, Italy, on November 3, 2022. To the best of our knowledge, this is one of the first observation experiments to have been conducted using a cluster of radiosondes to track fluctuations of the quantities within clouds and the atmospheric flow field.

{The preparation and instrumentation of the experimental setup are detailed in Figure \ref{fig:ao2_experiment_setup}. Prior to launching, all radiosondes were affixed to wooden fences, as illustrated in Figure \ref{fig:ao2_experiment_setup}a, for pre-launch checks and calibration.} The pre-launch checks were divided into two parts: (i) The radiosondes remained attached to the fences from 13:58 to 14:08; (ii) they were picked up one by one and kept together from 14:08 to 14:15. We considered the two above ranges for temperature calibration relative to the INRIM reference instrumentation, the Fluke DAQ 1586A multi-meter (see Figure \ref{fig:ao2_experiment_setup}a). Three PT100 platinum resistance thermometers were connected to multi-meter: sensors 1 and 2, without a solar shield; sensor 3, equipped with a helical passive solar shield, was positioned in between two unshielded sensors. Figure \ref{fig:ao2_cal_pht} shows the measurements conducted during the pre-launch checks. The probe measurements were compared with the reference sensor readings with (Ref SH) and without solar shields (Ref USH1 or USH2). {The two phases of sensor readings during the pre-launch checks are represented by light gray and white backgrounds.} During the initial phase, sensor readings of probes and the reference sensors (Ref USH1 and USH2) were in good agreement. {In the second phase (after 14:08), some spikes were observed, primarily attributed to manual handling of the radiosondes during their preparation for free launch.}

\begin{table}[ht!]
	\centering
	\caption{RMSD ($\sigma_{USH 1}$, $\sigma_{USH 2}$, $\sigma_{SH}$) of the temperature measurements from the radiosondes with respect to the reference temperature sensors of the Fluke DAQ 1586A multimeter. Three sensors were connected to the multimeter: sensors \textit{USH 1} and \textit{USH 2} without solar shields, and the third sensor (\textit{SH}), with a solar shield. The temperature bias offsets ($\mu_{USH}$) are provided in the last column with respect to unshielded reference sensors, which were effectively compensated from the temperature readings. The experienced radiation offset was also computed as the difference between $\sigma_{SH}$ and $\sigma_{USH 1}$ (and $\sigma_{USH 2}$) and it was found to basically stay constant in the 1.15–1.40 $^{\circ}$C region with an average value of 1.28 $^{\circ}$C. All the reported quantities are given in Celcius degrees, $^{\circ}$C. The dataset for probe 8 is not available, for the reasons explained in the text.}
	\begin{tabular}{l c c c c} 
		\hline
		Probes & $\sigma_{USH 1}$ & $\sigma_{USH 2}$ & $\sigma_{SH}$ & $\mu_{USH}$ \\ [0.2ex] 
		\hline\hline
		Probe 1 & 0.10 & 0.09 & 1.33 & 0.36 \\
		Probe 2 & 0.30 & 0.40 & 1.73 & 0.37 \\
		Probe 3 & 0.20 & 0.33 & 1.48 & 0.63 \\
		Probe 4 & 0.21 & 0.17 & 1.40 & 0.21 \\
		Probe 5 & 0.14 & 0.21 & 1.48 & 0.31 \\ 
		Probe 6 & 0.51 & 0.18 & 1.71 & 1.55 \\
		Probe 7 & 0.17 & 0.43 & 1.51 & 1.41 \\
		\textit{Probe 8} & - & - & - & - \\
		Probe 9 & 0.21 & 0.33 & 1.56 & 1.77 \\
		Probe 10 & 0.53 & 0.86 & 1.90 & 2.64 \\[0.2ex] 
		\hline \rule{0pt}{3ex}
		Average & 0.26 & 0.33 & 1.57 & 1.03 \\[1ex] 
		\hline
	\end{tabular}
	\label{tbl:ao2_pht_comp}
\end{table}

Table \ref{tbl:ao2_pht_comp} presents the RMSD \deleted{(Root Mean Square Deviation)}of the temperature measurements from the radiosondes, in comparison to INRIM reference sensors. Depending on the sensor compensation method and environmental conditions, sensors may introduce an internal bias offset. During the pre-launch calibration, the offset was identified as the mean difference between the probe measurement and reference sensor measurement (as shown in column 5 of Table \ref{tbl:ao2_pht_comp}). This procedure was performed on the reference sensors without a solar shield in order to distinguish between the \textit{radiation offset} and the \textit{bias offset}.

RMSD deviations of the sensor readings were computed after compensating for the bias offsets as displayed in columns 2-4 of Table \ref{tbl:ao2_pht_comp}. The RMSD values and their averages, in comparison to each reference sensor, aligned well with the values declared by the manufacturer, falling within the range of 0.5–1.5 $^{\circ}$C\cite{docs:pht_datasheet}. The deviation values were higher than the first reference sensor (with a solar shield) due to the radiation effect naturally experienced by radiosondes. Figure \ref{fig:ao2_cal_pht} also shows the radiation effect, with a roughly constant offset observed between the shielded (Ref SH) and unshielded (Ref USH) reference sensors. {The offset induced by the radiation effect was computed as the difference between $\sigma_{SH}$ and $\sigma_{USH 1}$ (also $\sigma_{USH 2}$), and was predominantly within the range 1.15–1.40 $^{\circ}$C, with an average value of 1.28 $^{\circ}$C.}

\begin{figure}[ht!]
	\centering
	\begin{subfigure}{0.7\textwidth}
		\centering
		\includegraphics[width=\textwidth]{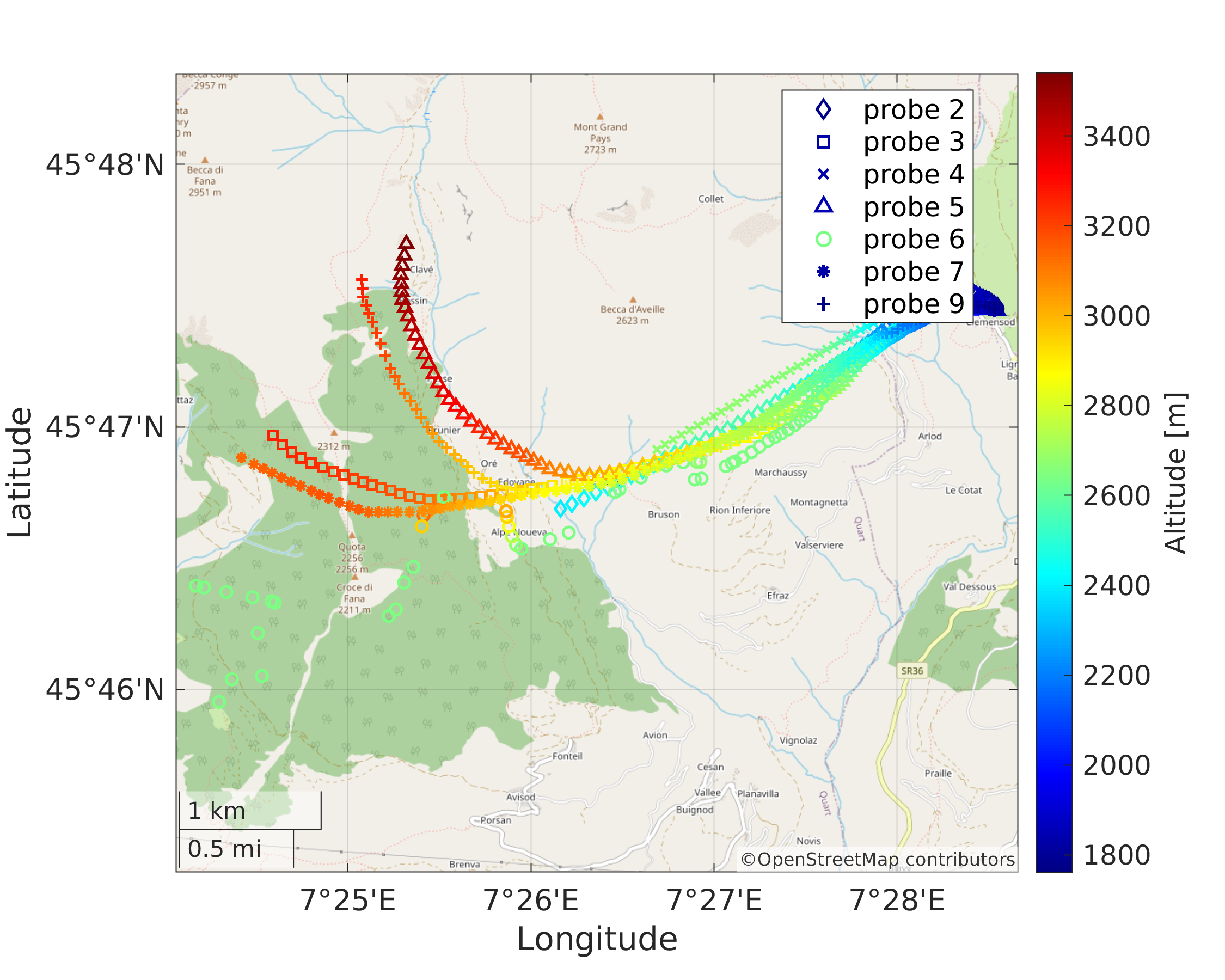}
	\end{subfigure}
	\caption{\textit{Free launching} of a cluster of radiosondes during the OAVdA experiment, St. Barthelemy, Aosta, Italy, on November 3, 2022. The trajectory of the radiosondes during the first 25 minutes of the free-launch period, from 14:15 to 14:40. The marker colors indicate the altitude reached by the radiosondes along the trajectory, starting from an altitude of 1700 meters.}
	\label{fig:ao2_lla_map}
\end{figure}	

After the pre-launch checks and the calibration phase, the radiosondes were simultaneously launched to free-float from the same initial position. Figure \ref{fig:ao2_lla_map} highlights the trajectory of the radiosondes during the first 25 min of the free-launch period.
{These radiosondes ascended from an initial altitude of 1700 meters to a maximum altitude of 3950 meters, covering a distance of up to 8300 meters. It can be noted that Figures \ref{fig:ao2_cal_pht} and \ref{fig:ao2_lla_map} do not include all radiosonde datasets.} This is due to radiosonde dataset was either not sufficient to consider it for this analysis or the radiosonde being powered off due to mechanical stress during release. For example, probe 1 transmitted readings during the pre-launch checks and for a few seconds after launching. Probe 10 transmitted measurements without issues, but it could not acquire a proper GNSS signal from satellites (no GNSS fix). Probe 8, on the other hand, did not transmit any readings during the launch, even though it had passed the initial firmware and sensor checks between 8:47 to 9:17. However, we believe that the disconnection due to mechanical stress and oscillations during movement could be eliminated by introducing a lightweight, robust case for the electronic components.

\begin{figure}[ht!]
	\centering
	\begin{subfigure}{0.45\textwidth}
		\centering
		\includegraphics[width=\textwidth]{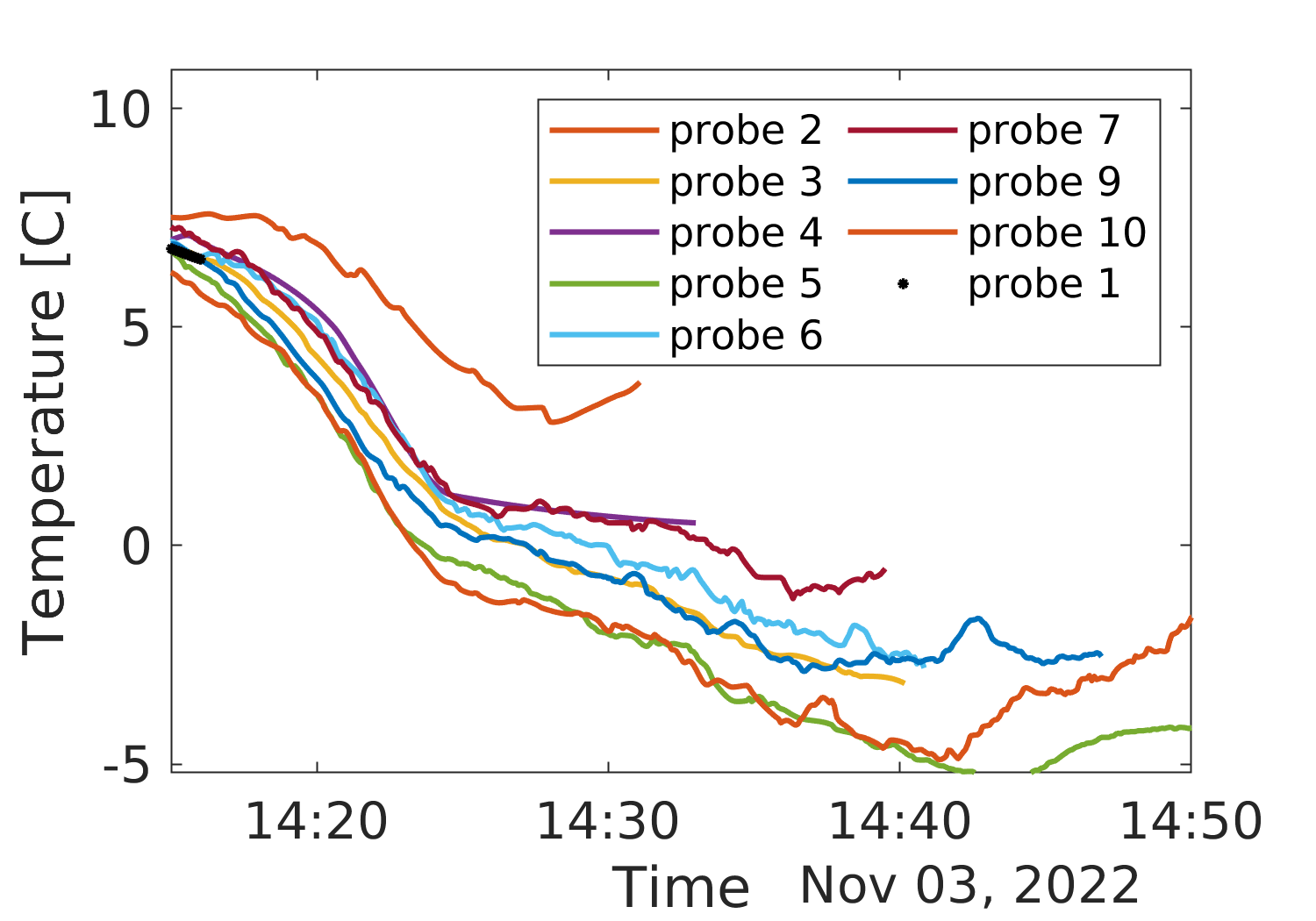}
		\caption{}
	\end{subfigure}
	\hspace{0.01\textwidth}
	\begin{subfigure}{0.45\textwidth}
		\centering
		\includegraphics[width=\textwidth]{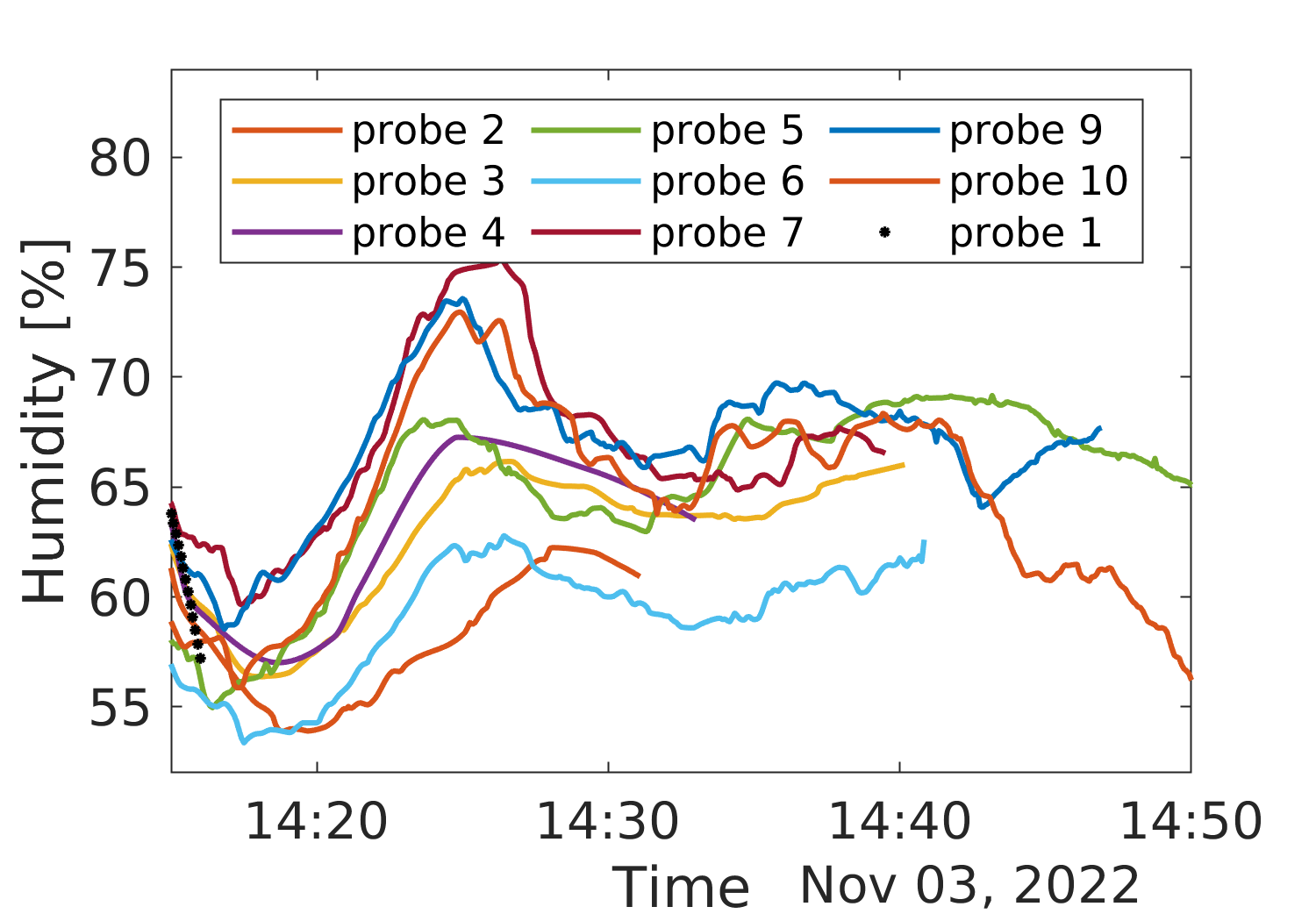}
		\caption{}
	\end{subfigure}
	\begin{subfigure}{0.45\textwidth}
		\centering
		\includegraphics[width=\textwidth]{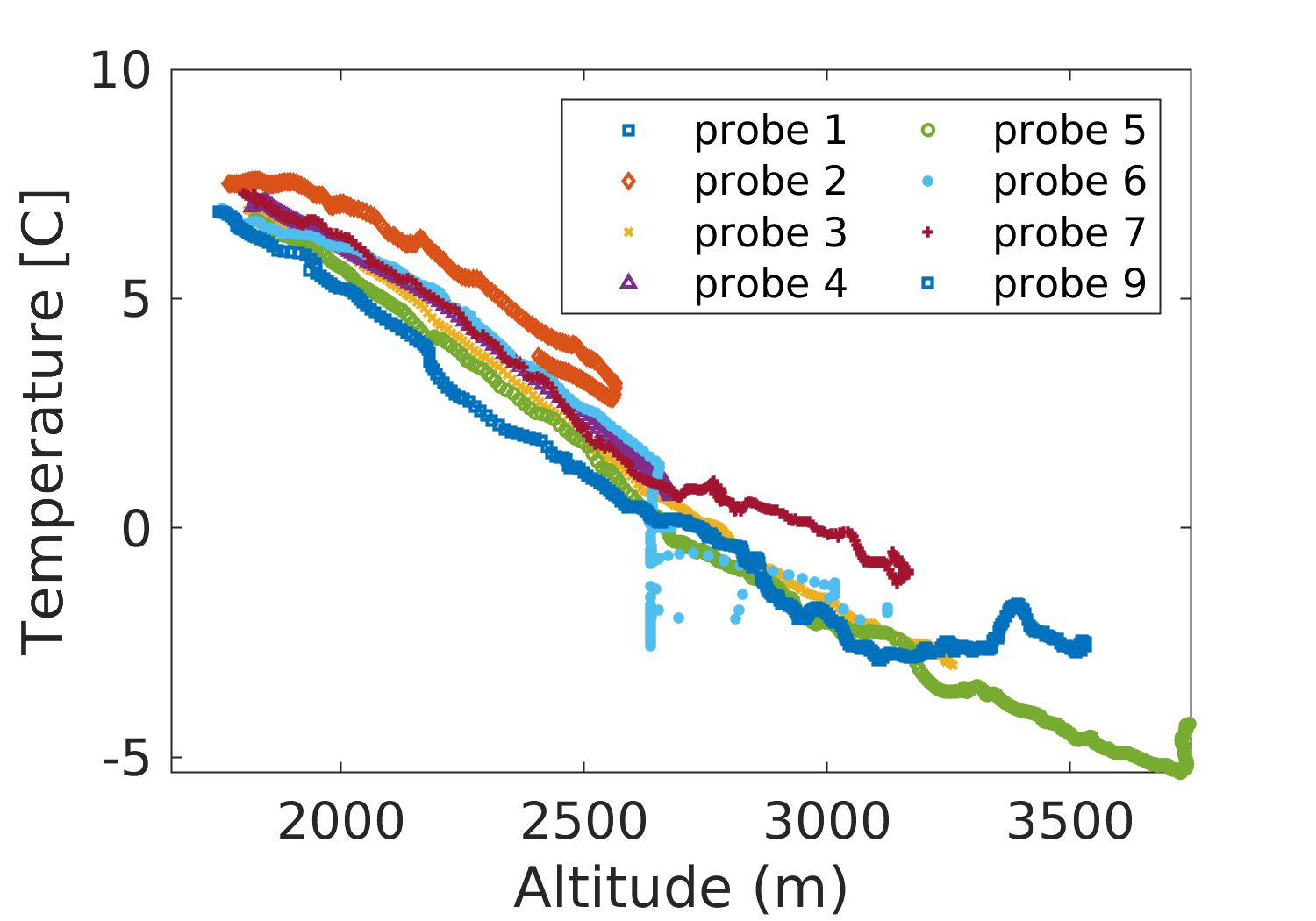}
		\caption{}
	\end{subfigure}
	\hspace{0.01\textwidth}
	\begin{subfigure}{0.45\textwidth}
		\centering
		\includegraphics[width=\textwidth]{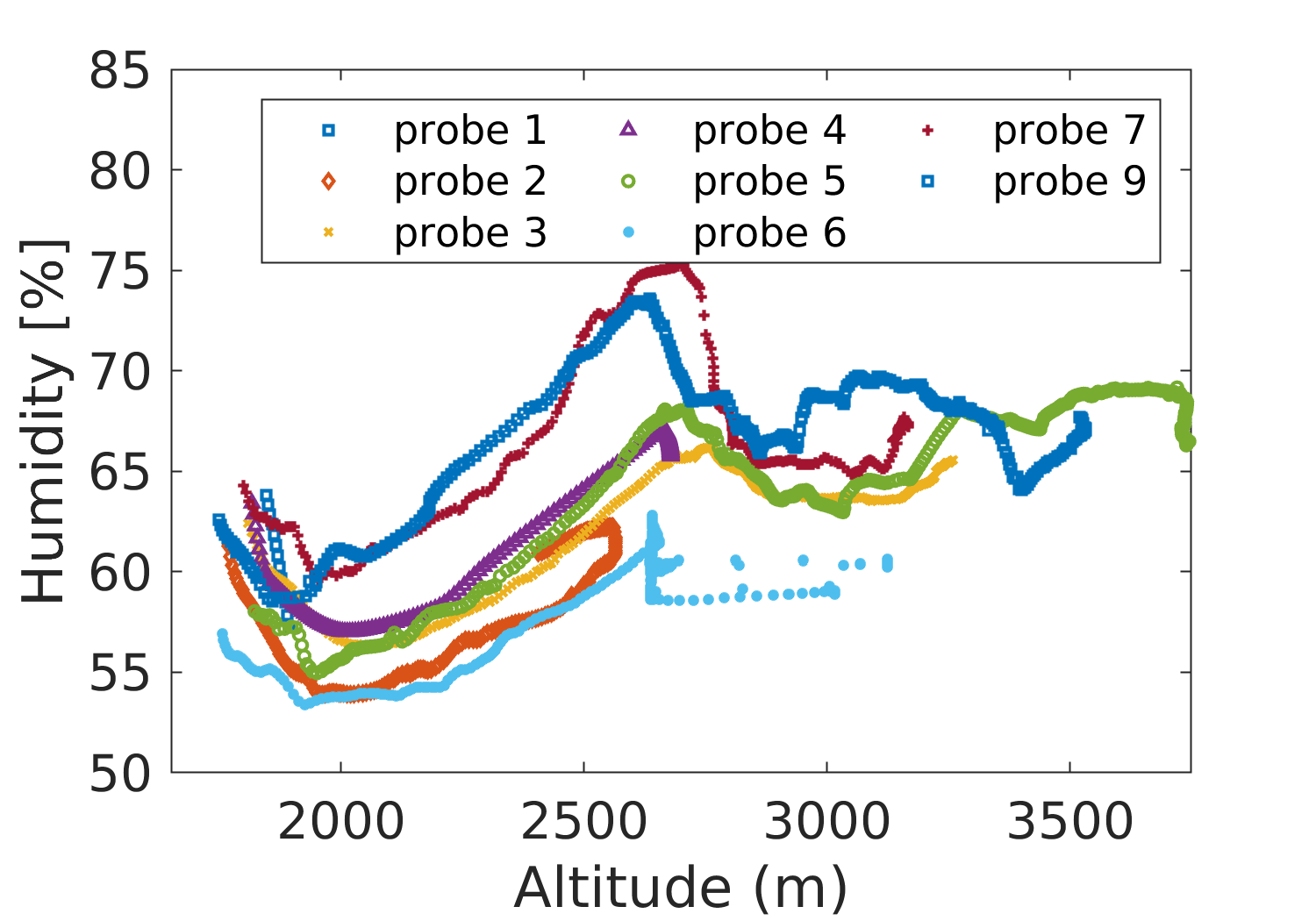}
		\caption{}
	\end{subfigure}
	\begin{subfigure}{0.45\textwidth}
		\centering
		\includegraphics[width=\textwidth]{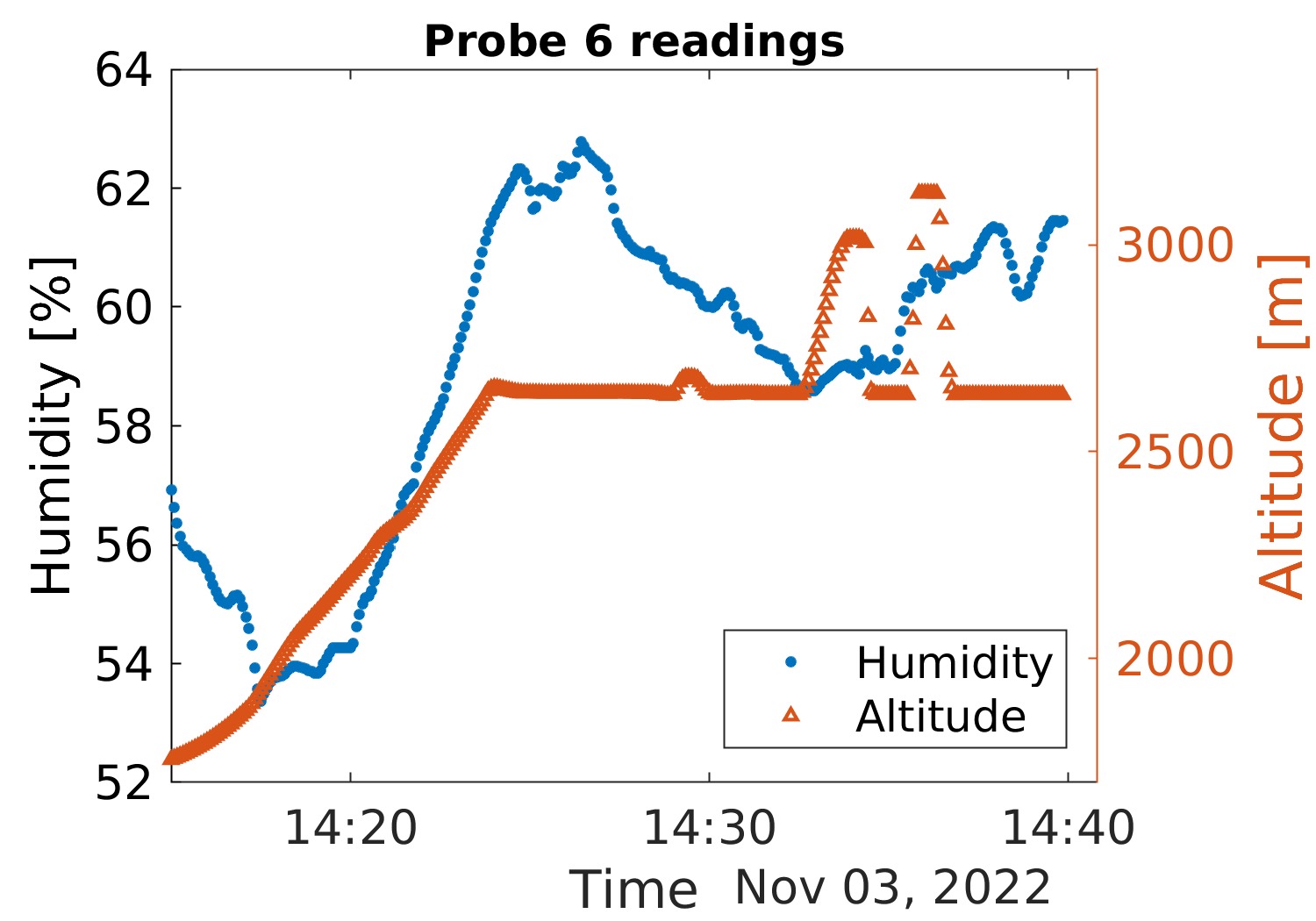}
		\caption{}
	\end{subfigure}
	\hspace{0.01\textwidth}
	\begin{subfigure}{0.45\textwidth}
		\centering
		\includegraphics[width=\textwidth]{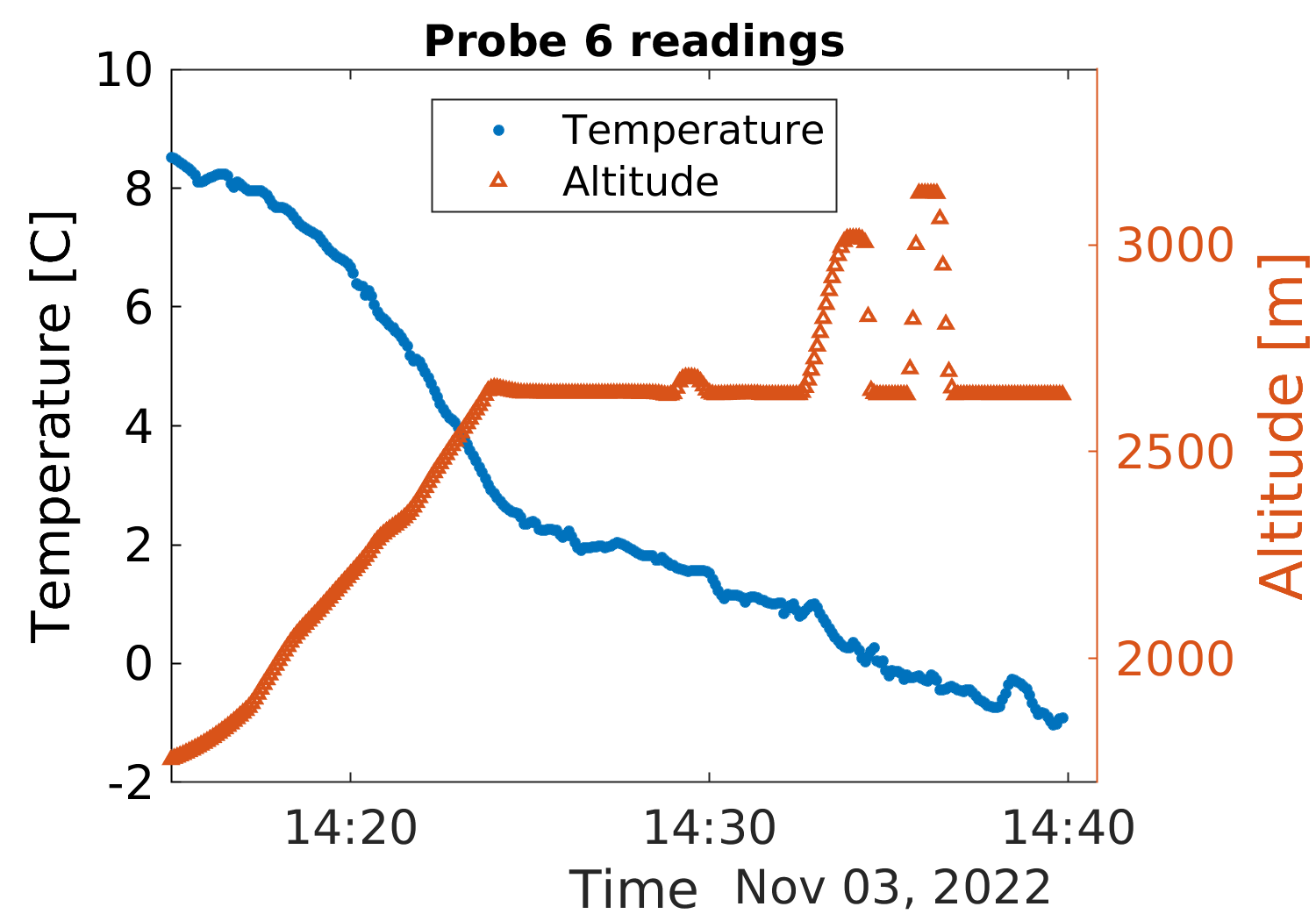}
		\caption{}
	\end{subfigure}
	\caption{{Temperature and humidity measurements during the experiment in St. Barthelemy on November 3, 2022. Panels a and b show temperature and humidity measurements during the launch from 14:15 to 14:50. Panels c and d display measurements along the altitude range of 1740 to 3800 meters. Panels e and f present temperature and humidity readings of Probe 6 with corresponding altitudes.}}
	\label{fig:ao2_raw_pht}
\end{figure}

{Figure \ref{fig:ao2_raw_pht} shows temperature and humidity measurements over a 35-minute period, from 14:15 to 14:50. Some radiosondes ascended to altitudes as high as 3800 meters, starting from 1700 meters.} However, sondes do not always ascend and they tend to reach an equilibrium altitude and to float horizontally across the isopycnic layer. As can be seen in panels c and d in Figure \ref{fig:ao2_raw_pht}, we were able to observe this tendency for probes 2 (orange), 6 (light blue), and 4 (violet). {Probe 2 initially ascended to 2600 meters  due to the updraft along the mountain slope. Then the probe descend a bit after having passed the mountain  top and stay aloft horizontally by returning  to its  equilibrium altitude (see vertical velocity plots in panel c of Figure \ref{fig:ao2_temp_spectra}).} Probes 4 and 6 reached about 2700 meters in altitude and then stayed at that altitude, {until we lost the communication with them.} However, they were able to stay at the equilibrium altitude for a few minutes (10-15 min. for probe 6 and 3-4 min. for probe 4). This aspect of horizontal floating is highlighted for the probe 6 in panels e and f of Figure \ref{fig:ao2_raw_pht}, {plotting the probe's altitude readings alongside the corresponding humidity and temperature measurements.}. Height of the hills below probes was much lower than the equilibrium altitude, about 1700-2000 m.
\begin{figure}[ht!]
	\centering
	\caption*{\footnotesize\bf Probe 5. LATITUDE: 45$^\circ$47'N - 45$^\circ$50'N; LONGITUDE:  7$^\circ$27'E - 7$^\circ$28'E}
	\begin{subfigure}{0.39\textwidth}
		\centering
		\includegraphics[width=\textwidth]{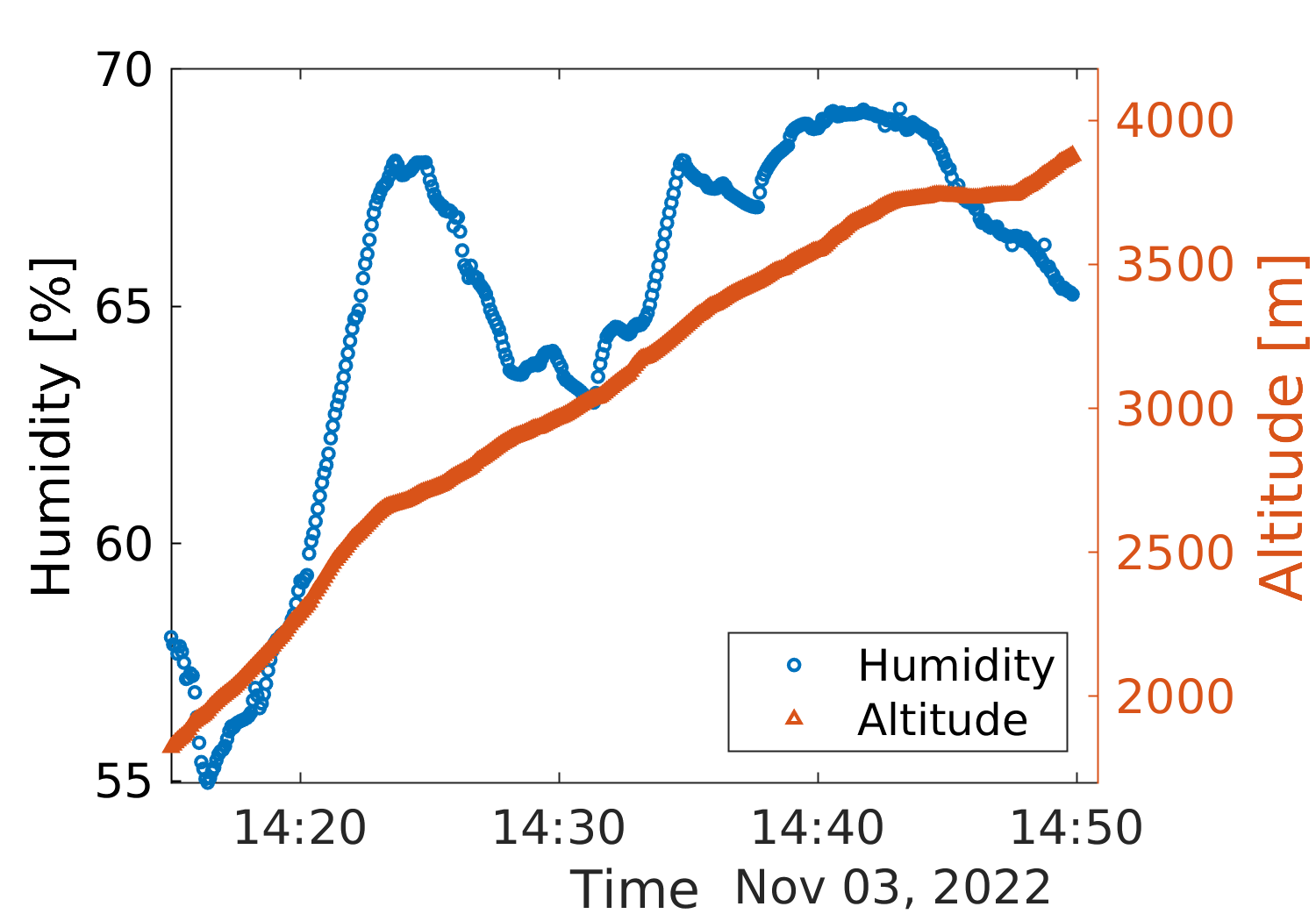}
		\caption{}
	\end{subfigure}
	\hspace{0.01\textwidth}
	\begin{subfigure}{0.39\textwidth}
		\centering
		\includegraphics[width=\textwidth]{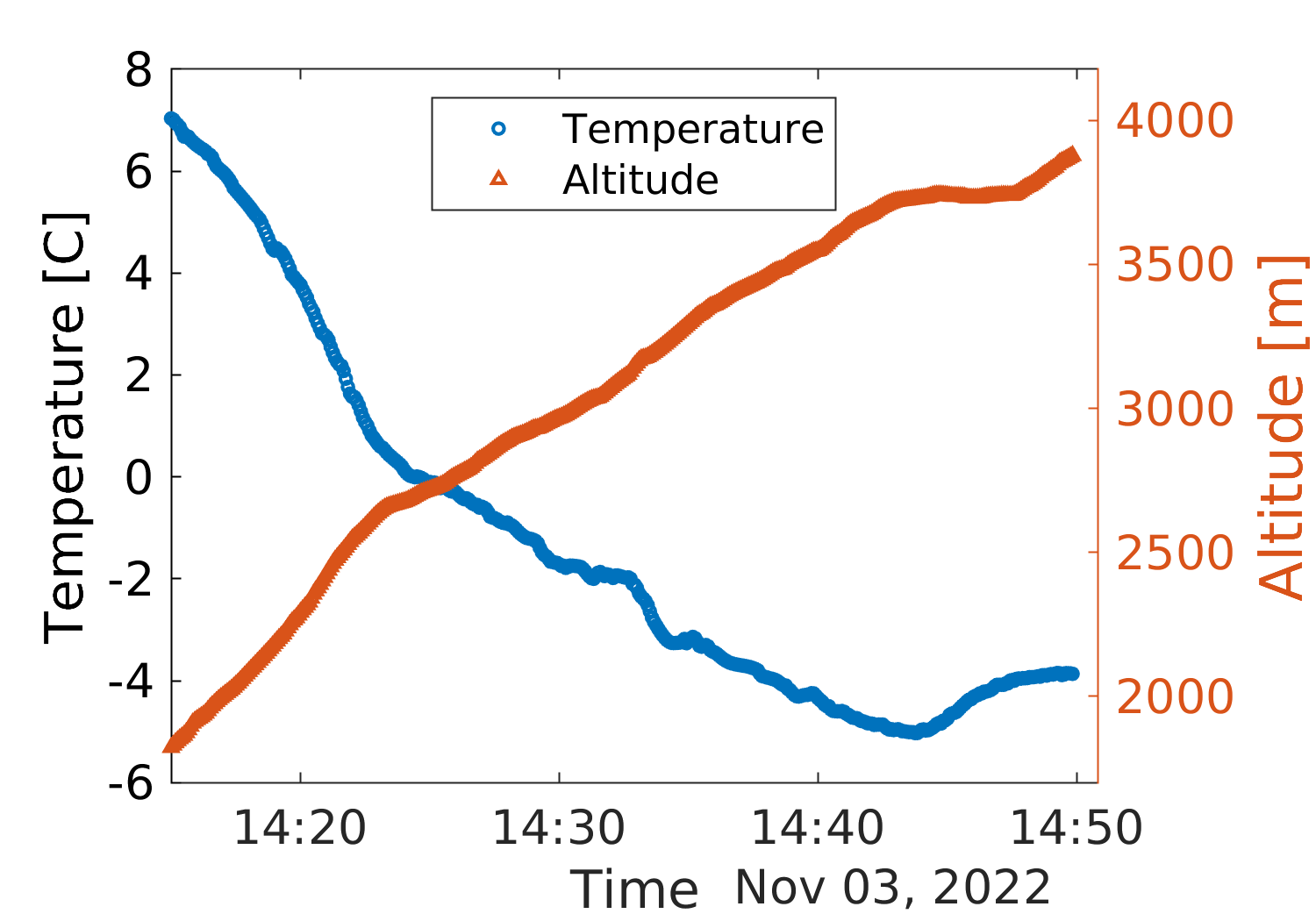}
		\caption{}
	\end{subfigure}
	\caption*{\footnotesize\bf Probe 7. LATITUDE: 45$^\circ$46'N - 45$^\circ$47'N; LONGITUDE:  7$^\circ$24'E - 7$^\circ$28'E}
	\begin{subfigure}{0.39\textwidth}
		\centering
		\includegraphics[width=\textwidth]{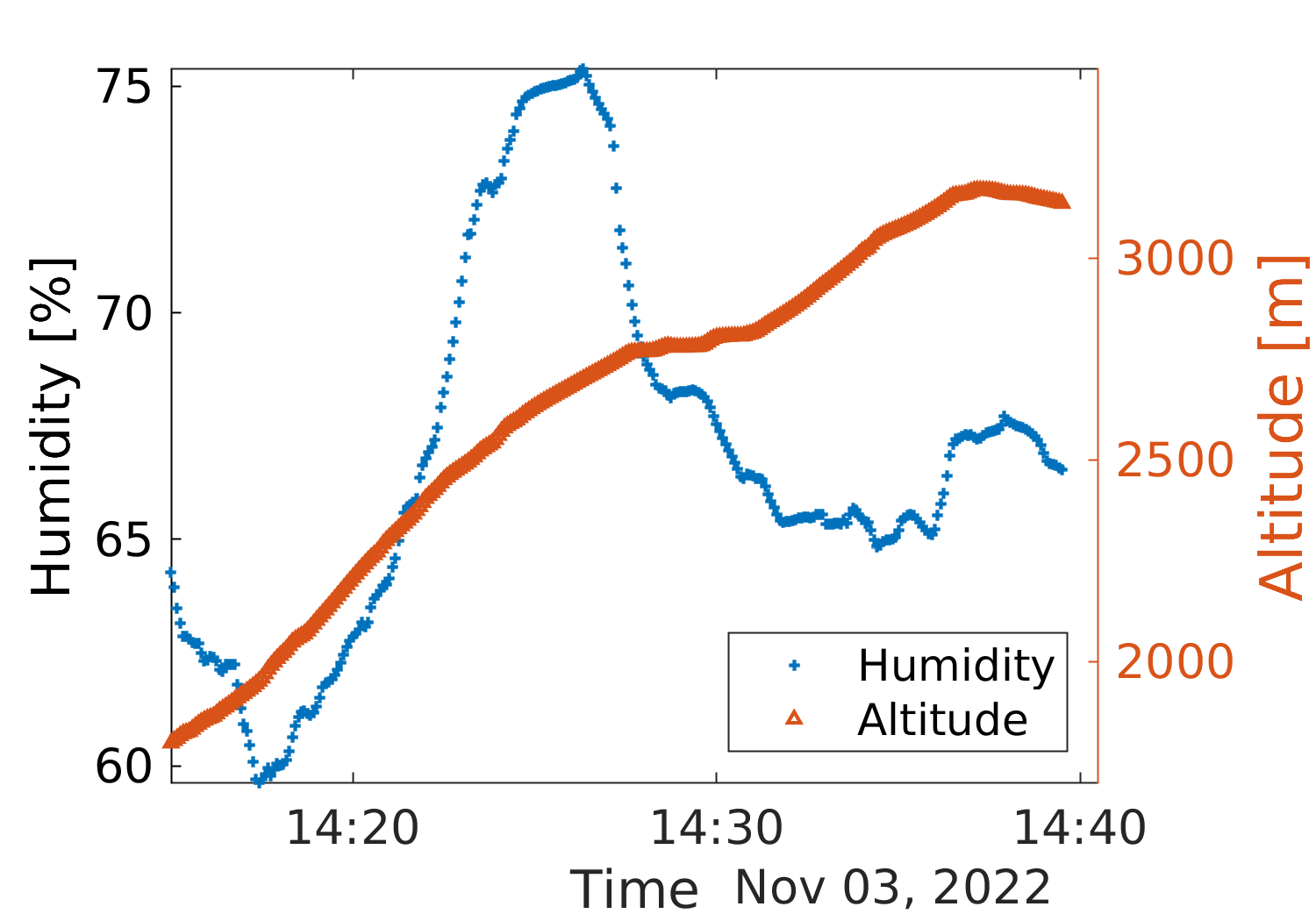}
		\caption{}
	\end{subfigure}
	\hspace{0.01\textwidth}
	\begin{subfigure}{0.39\textwidth}
		\centering
		\includegraphics[width=\textwidth]{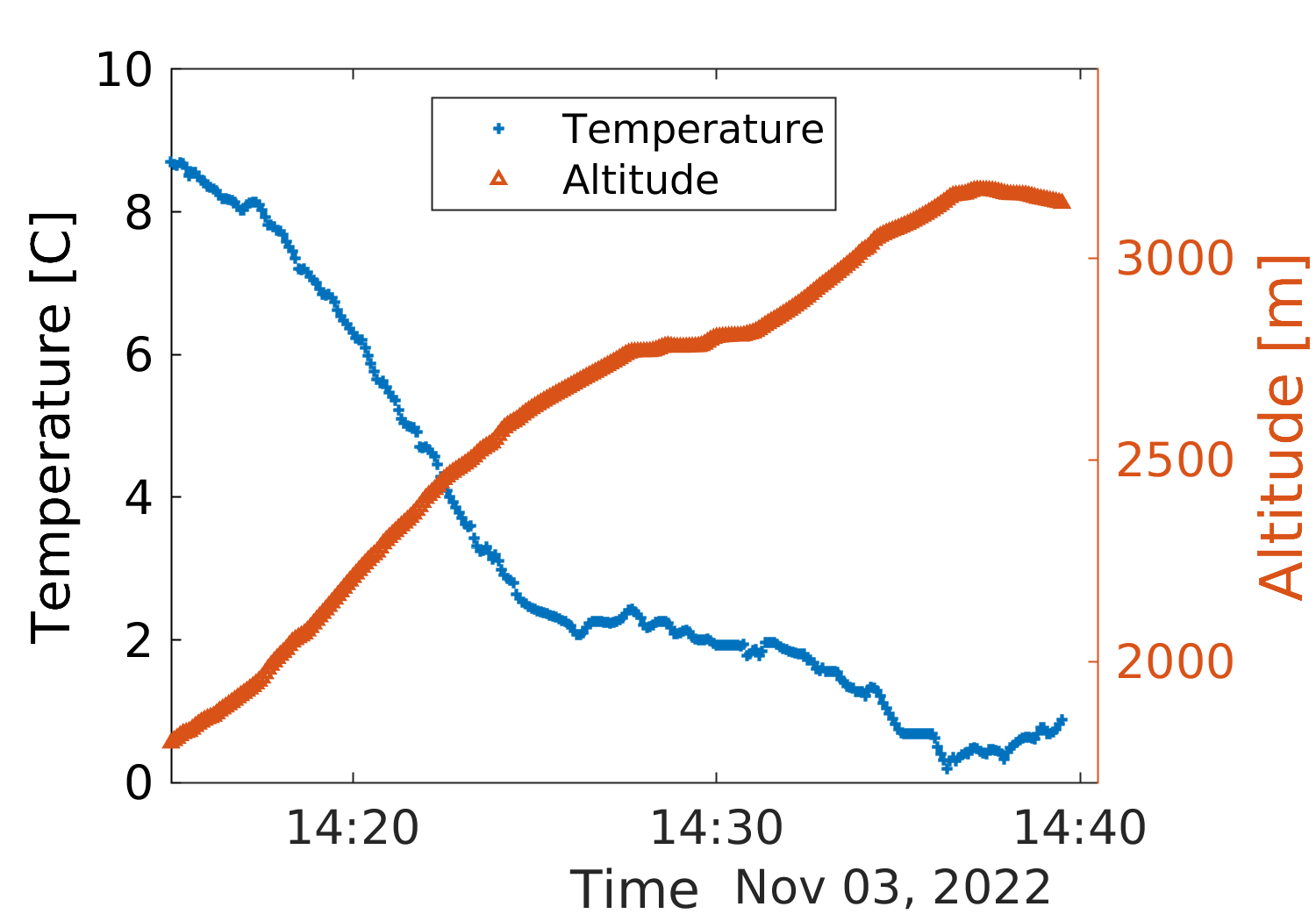}
		\caption{}
	\end{subfigure}
	\caption*{\footnotesize\bf Probe 9. LATITUDE: 45$^\circ$47'N - 45$^\circ$48'N; LONGITUDE:  7$^\circ$24'E - 7$^\circ$28'E}
	\begin{subfigure}{0.39\textwidth}
		\centering
		\includegraphics[width=\textwidth]{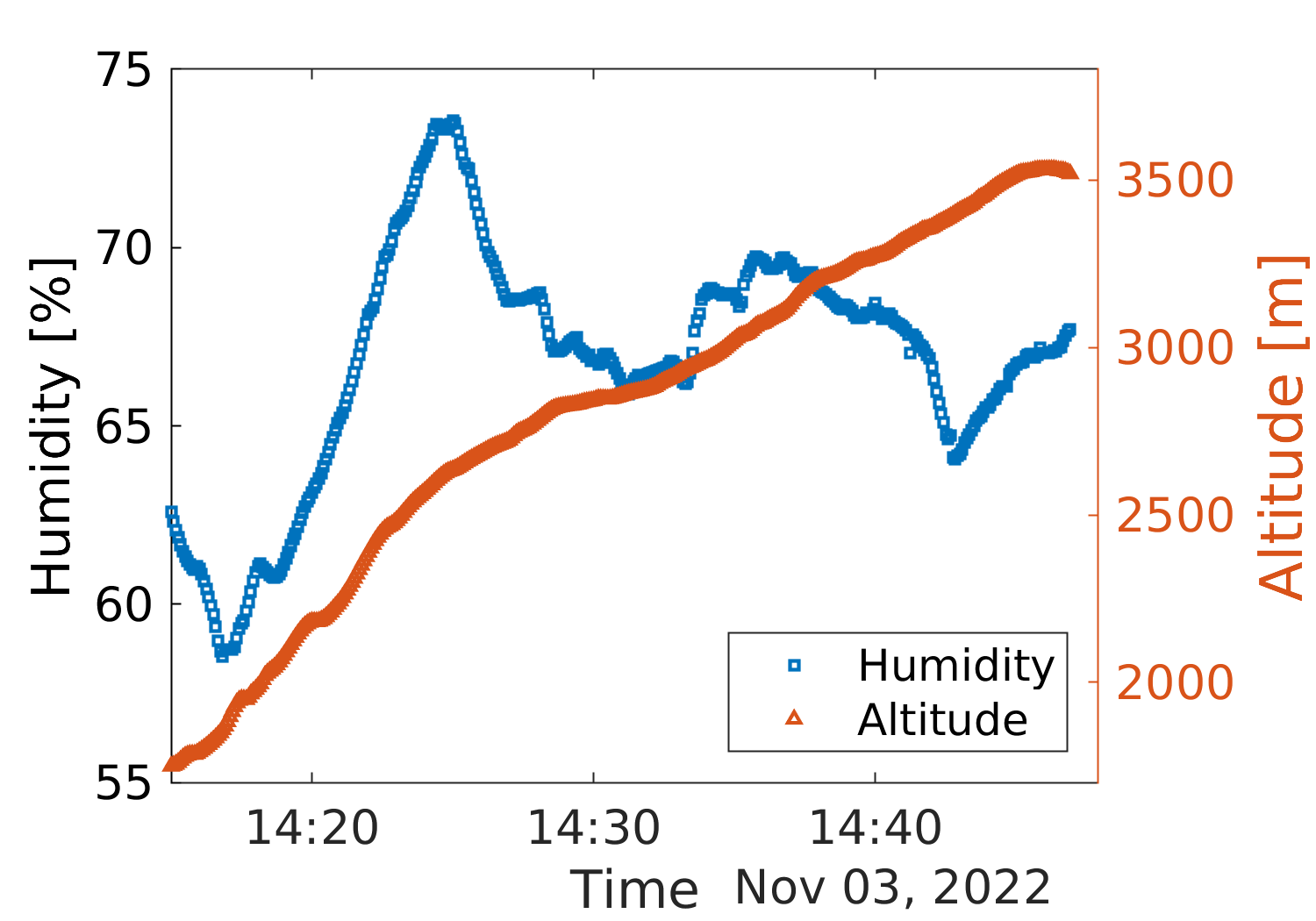}
		\caption{}
	\end{subfigure}
	\hspace{0.01\textwidth}
	\begin{subfigure}{0.39\textwidth}
		\centering
		\includegraphics[width=\textwidth]{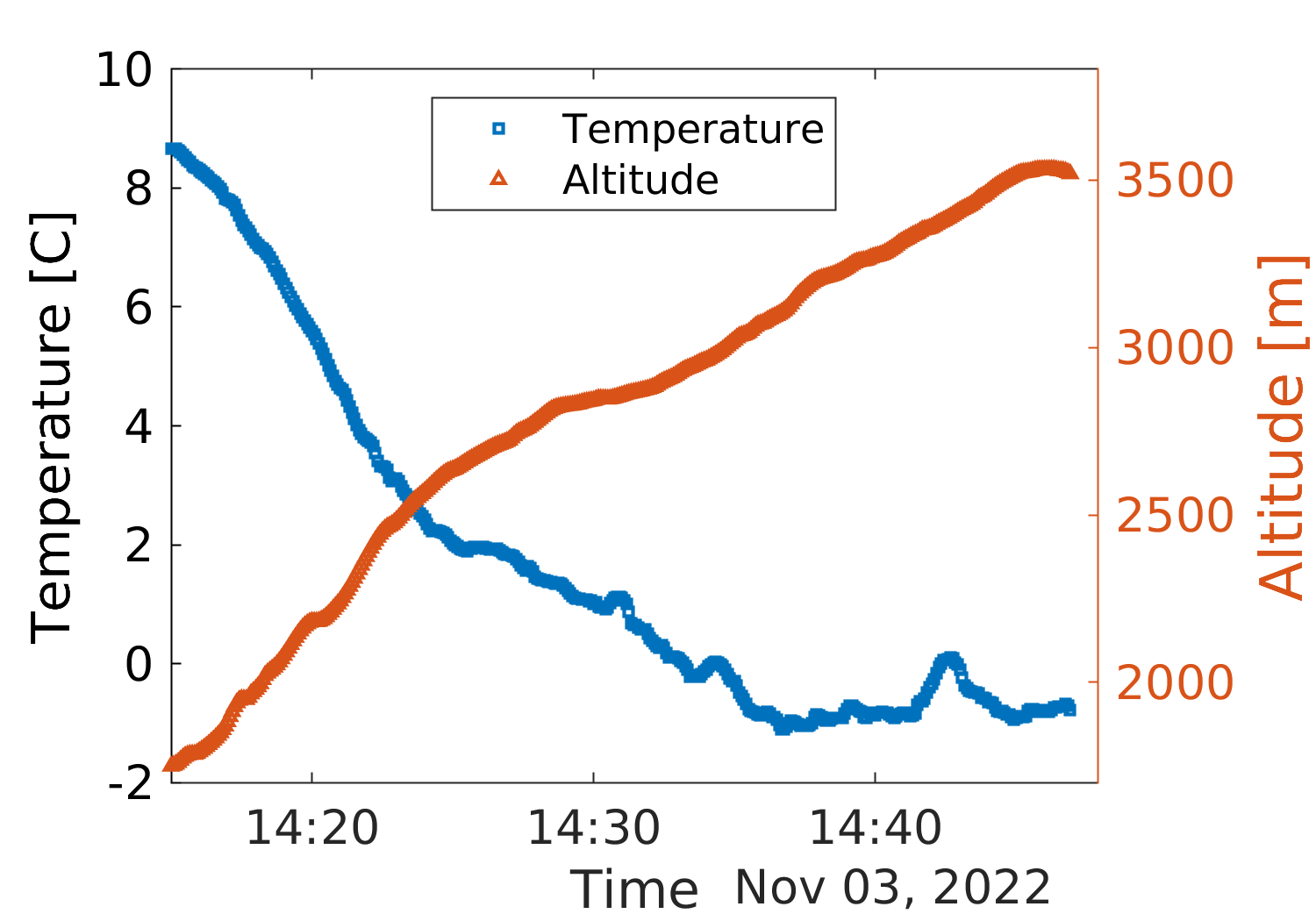}
		\caption{}
	\end{subfigure}
	\caption{{Humidity and temperature measurements from the OAVdA experiment on November 3, 2022, are presented for selected probes. Panels a, c, and e display humidity measurements and the corresponding altitudes for probes 5, 7, and 9. Temperature readings for these probes are depicted in panels b, d, and f.}}
	\label{fig:ao2_raw_ht_alt_time}
\end{figure}

\subsubsection{Tracking temperature, humidity, and wind speed fluctuations}\label{sec:cluster_fluctuations}	

\begin{figure}[bht!]
	\centering
	\begin{subfigure}{0.45\textwidth}
		\includegraphics[width=\linewidth]{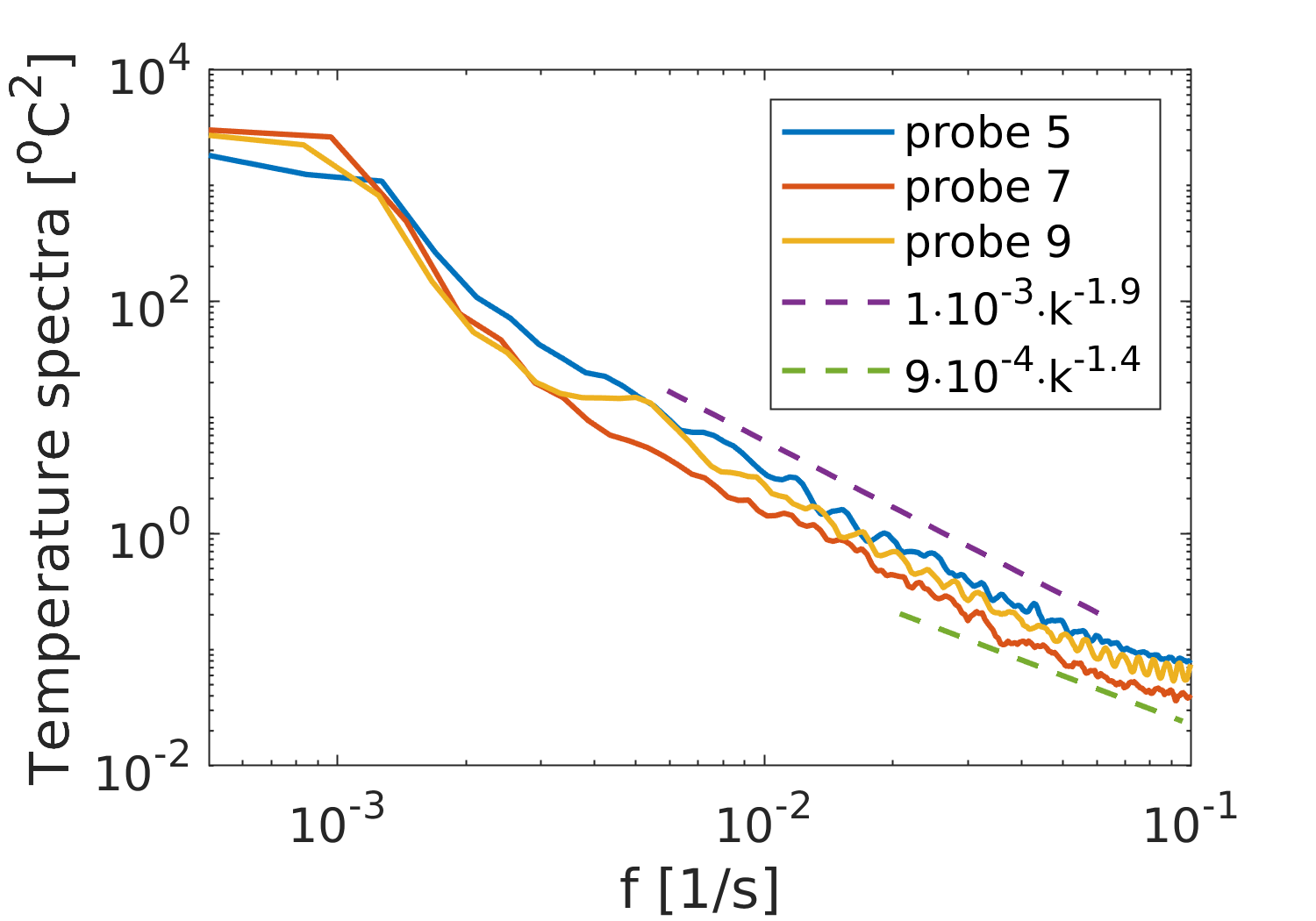}
		\caption{}
	\end{subfigure}
	\hspace{0.01\textwidth}
	\begin{subfigure}{0.45\textwidth}
		\includegraphics[width=\linewidth]{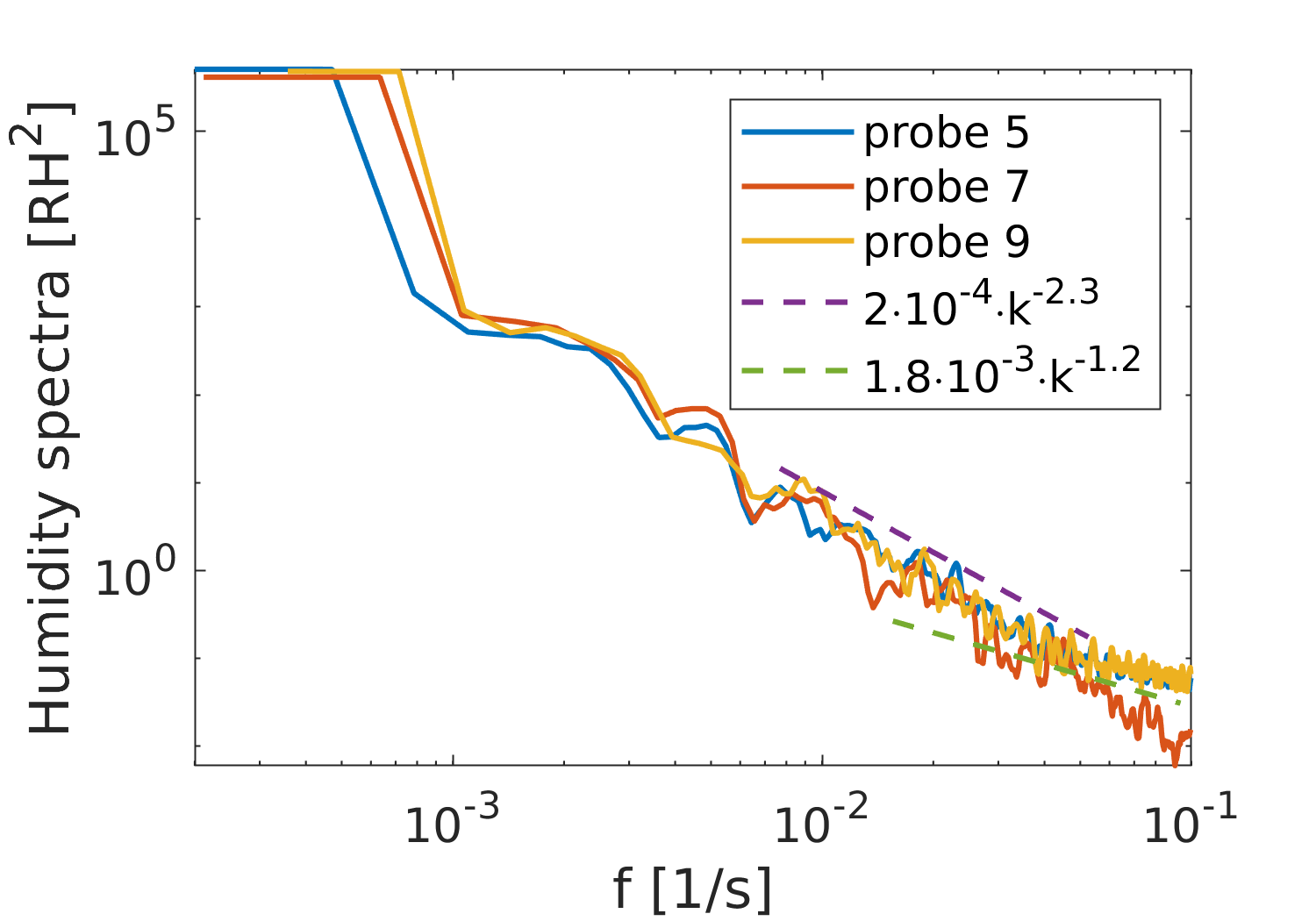}
		\caption{}
	\end{subfigure}
	\begin{subfigure}{0.45\textwidth}
		\includegraphics[width=\linewidth]{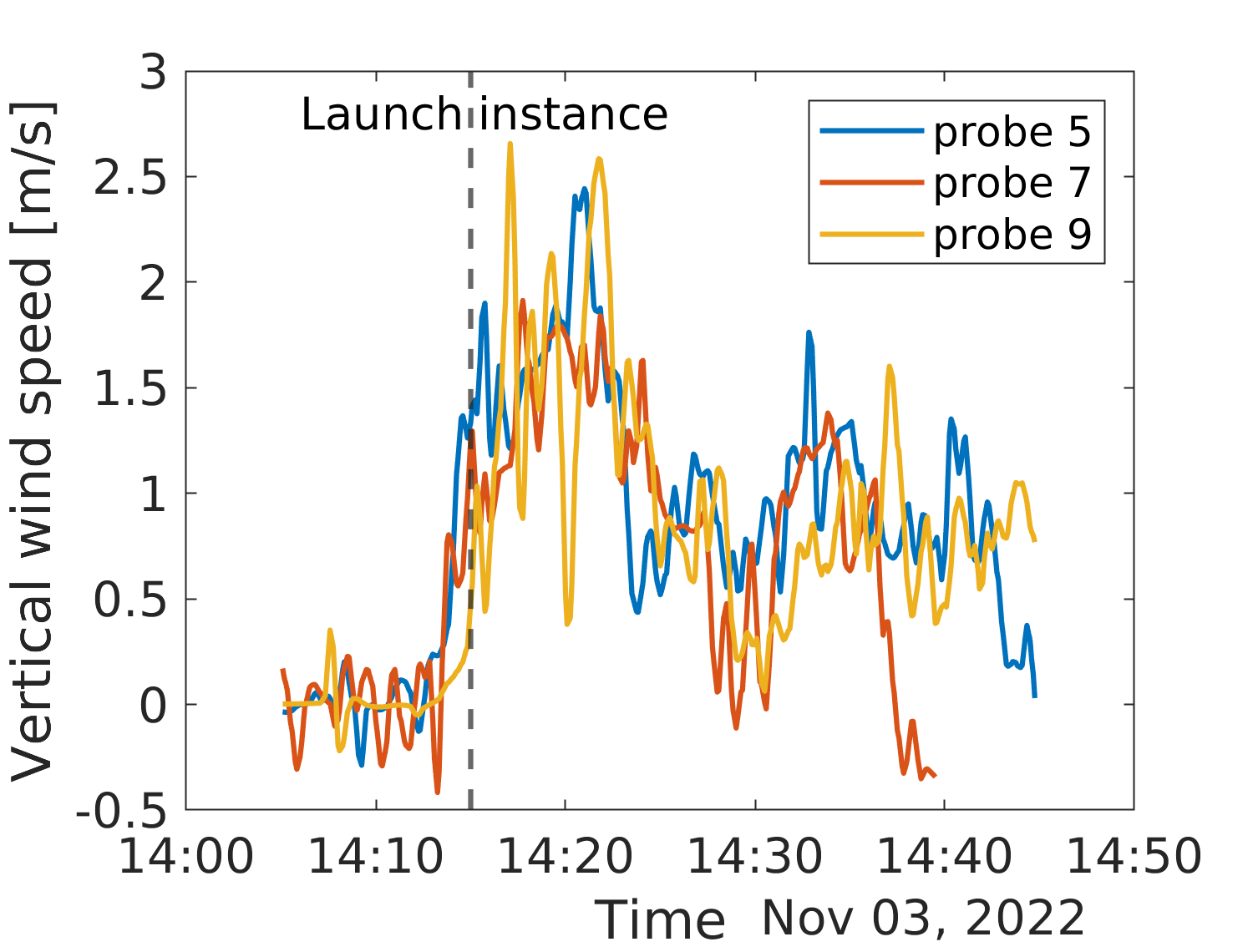}
		\caption{}
	\end{subfigure}
	\hspace{0.01\textwidth}
	\begin{subfigure}{0.45\textwidth}
		\includegraphics[width=\linewidth]{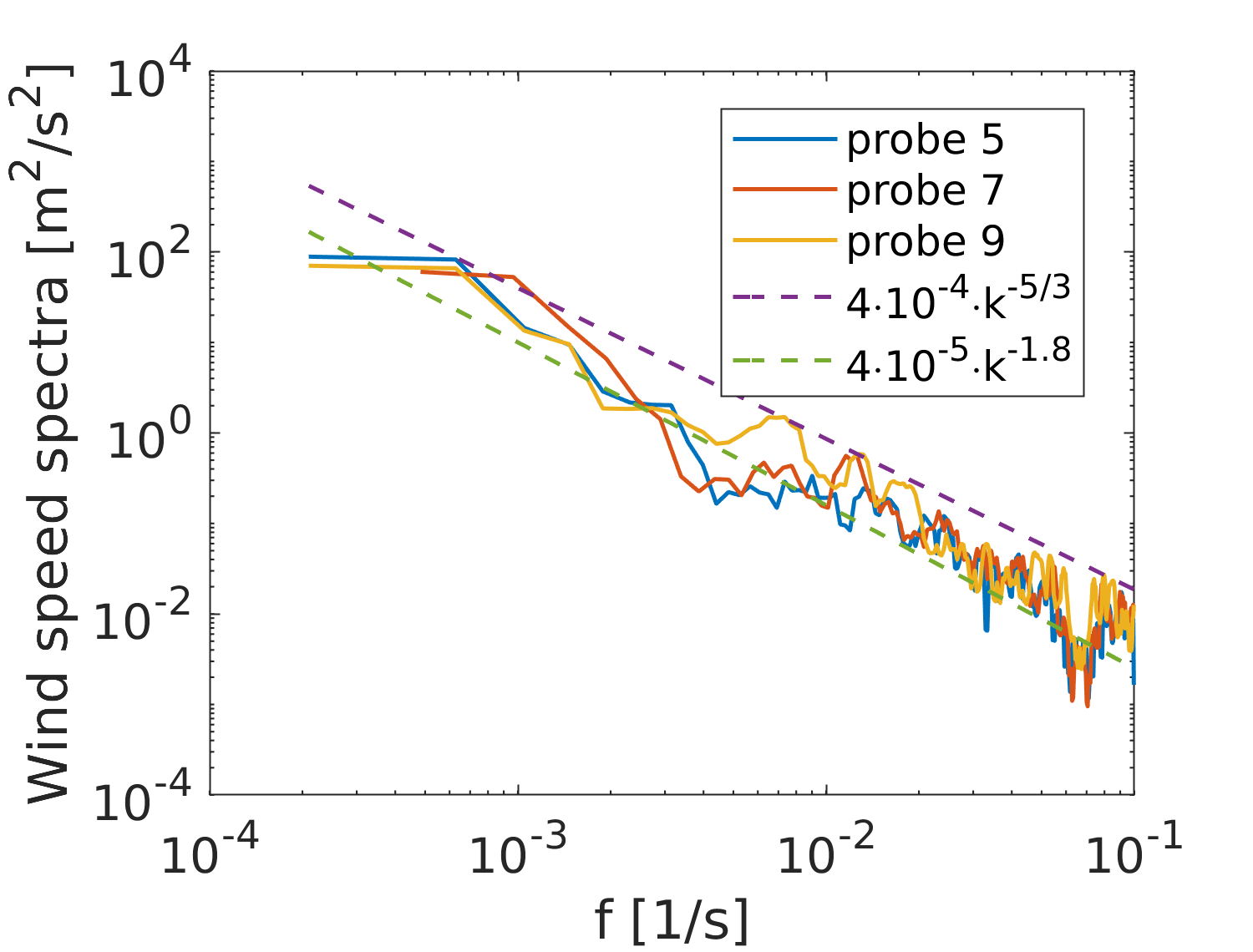}
		\caption{}
	\end{subfigure}
	\caption{{Spectral analysis of temperature, humidity, and vertical velocity fluctuations during the free flight of the radiosondes (OAVdA, Nov 3, 2022). Panels a and b show power spectra of temperature and humidity readings from three radiosondes. Panel c presents vertical wind speed, and panel d displays power spectra of wind speed fluctuations. Two trend lines (violet and green) are included for reference. The dataset was resampled at 5-second intervals within a frequency range based on the Nyquist frequency, $f_s/2 = 0.1 s^{-1}$.}}
	\label{fig:ao2_temp_spectra}
\end{figure}

As stated in the previous sections, one of the objectives of the present work has been to to track the fluctuations of the physical quantities along the Lagrangian trajectories and to perform relative measurements within a cluster of radiosondes. In this way, the measurement technique enables us to obtain a broader understanding of the turbulent intermittency, dispersion, and diffusion that occurs inside isopycnic layers of atmospheric flows. {Figure \ref{fig:ao2_temp_spectra} presents the results of the spectral analysis for temperature, humidity and vertical velocity fluctuations analyzed for a subset of radiosondes (probes 5, 7 and 9). Humidity and temperature readings for these radiosondes, along with their reached altitudes are illustrated in Figure \ref{fig:ao2_raw_ht_alt_time}. The dataset covers a 30-minute period, specifically from 14:15 to 14:45, with a 5-second sampling rate.} The dataset was then transformed from a time domain to a spectral domain with FFT, as described for the wind speed analysis in Section \ref{results_vertical_launch}.

\begin{figure}[bht!]
	\centering
	\begin{subfigure}{0.3\textwidth}
		\includegraphics[width=\linewidth]{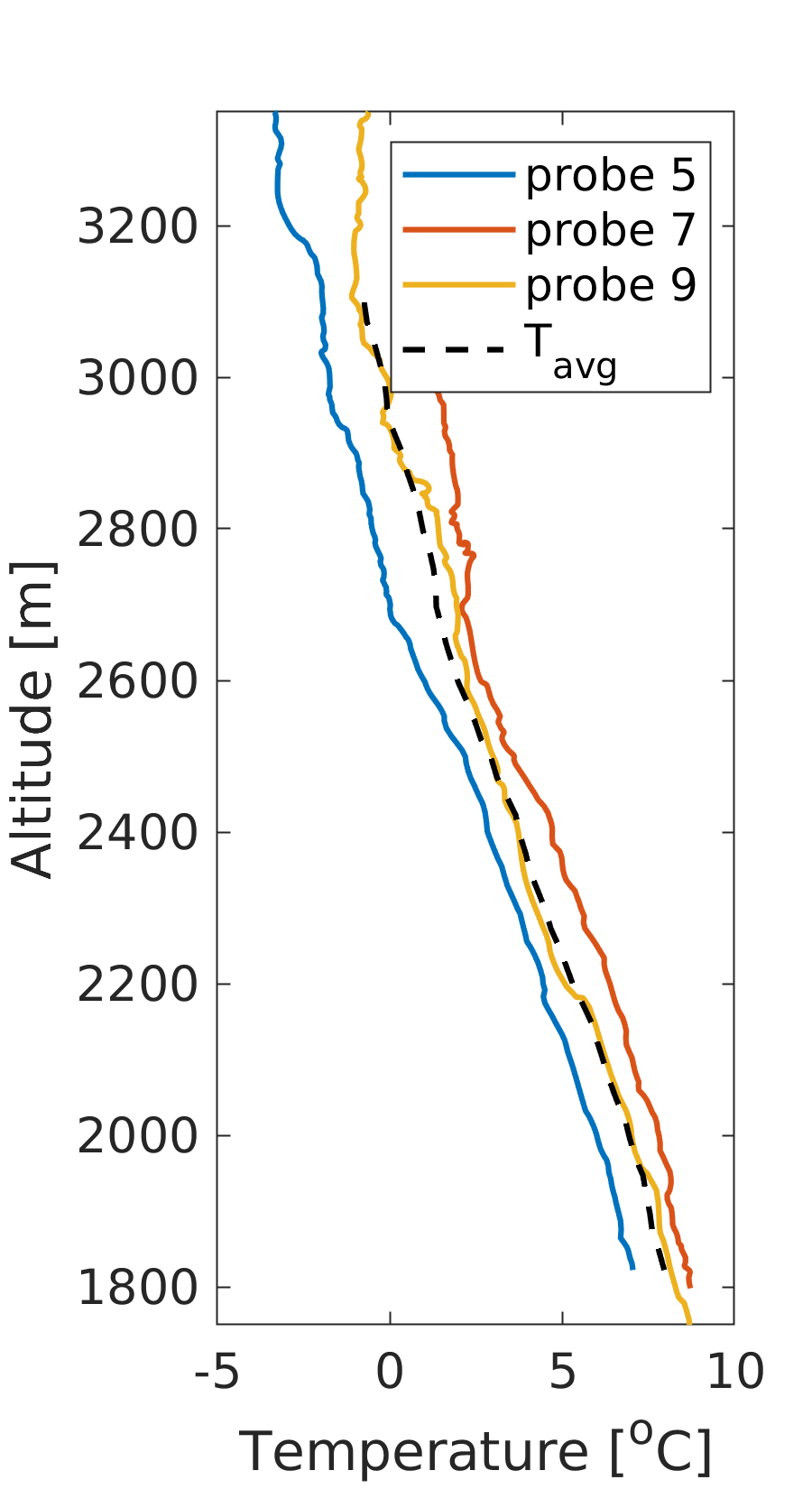}
		\caption{}
	\end{subfigure}
	\hspace{0.01\textwidth}
	\begin{subfigure}{0.3\textwidth}
		\includegraphics[width=\linewidth]{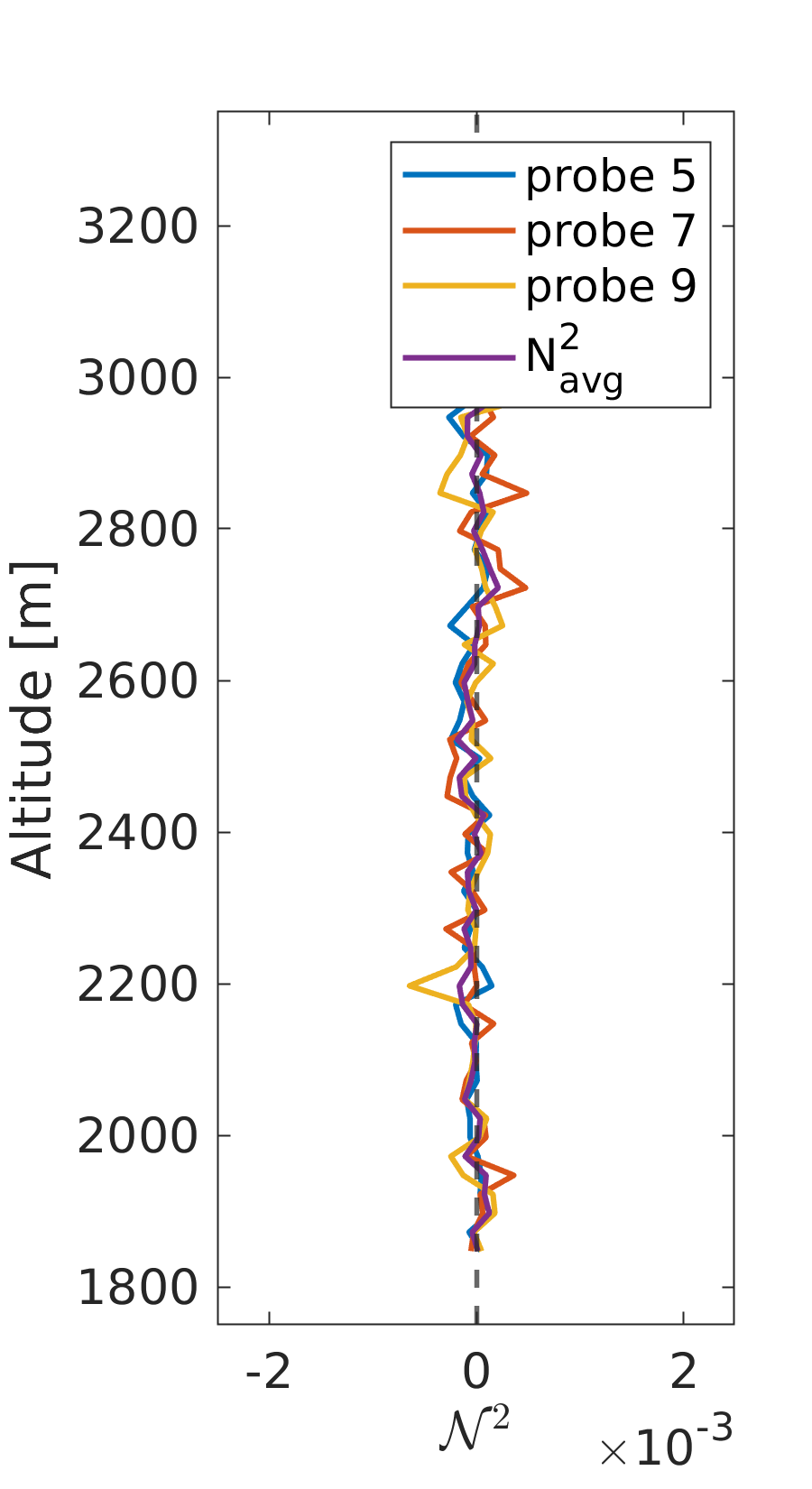}
		\caption{}
	\end{subfigure}
	\hspace{0.01\textwidth}
	\begin{subfigure}{0.3\textwidth}
		\includegraphics[width=\linewidth]{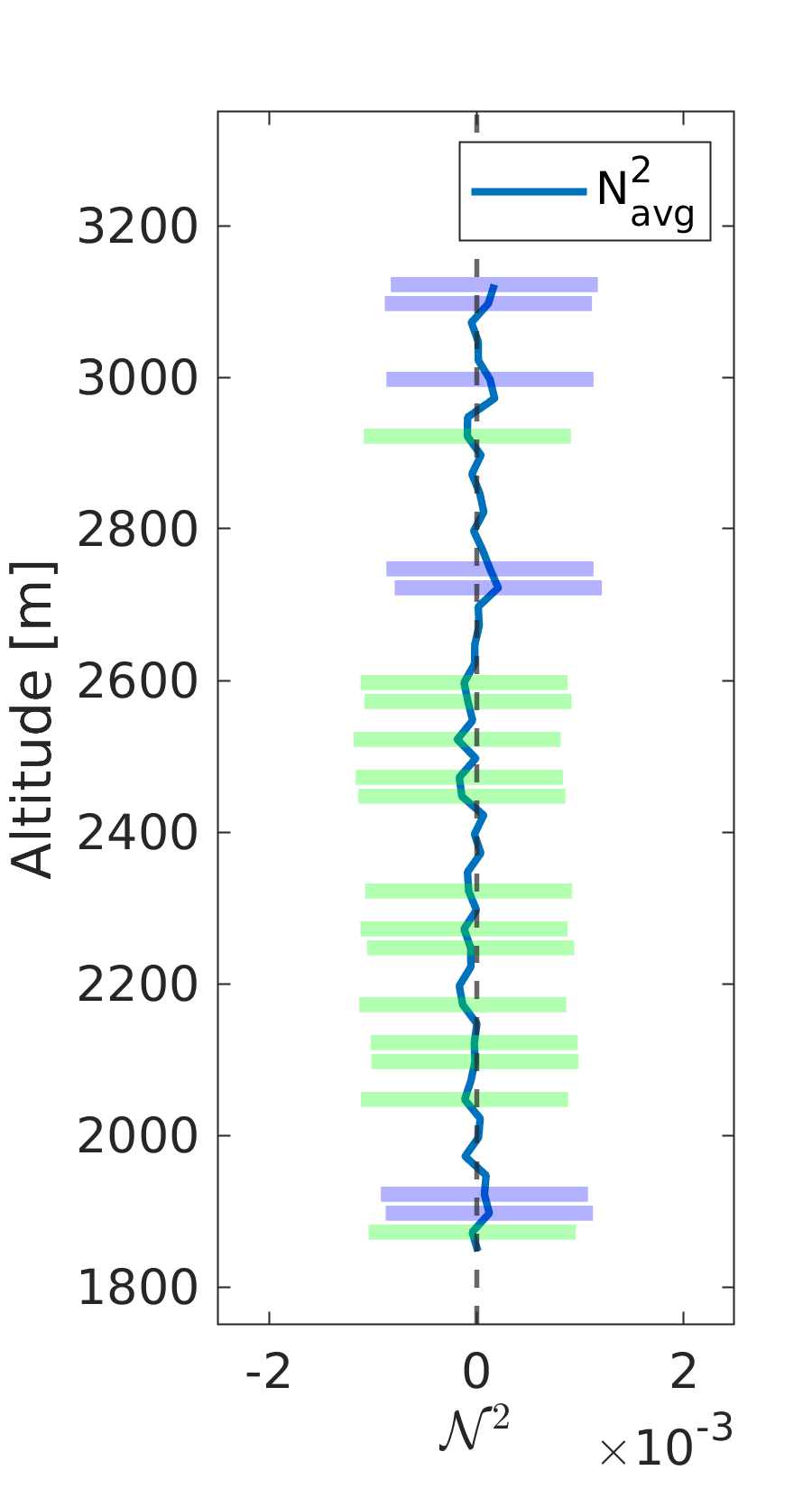}
		\caption{}
	\end{subfigure}
	\caption{Vertical profile of the temperature measurements and computed Brunt-Vaisala frequency during the free-floating experiment in OAVdA, Nov 3, 2022. (a) Temperature along the altitude. (b) Vertical profile of the Brunt-Vaisala frequency, $\mathcal{N}^2=g \frac{\delta T}{T_0} \frac{1}{\Delta z}$, where $T_0$ = 281 K and g = 9.81 m/s$^2$. {(c) Averaged BV profile computed from BV profiles of probes 5, 7 and 9. The violet color highlights the altitude ranges where all the three profiles had a positive (stable) temperature gradient, while the green color  indicates where all three profiles had a negative (unstable) temperature gradient.}}
	\label{fig:ao2_bv_profile}
\end{figure}

In atmospheric dynamics and geophysics terms, the Brunt–Vaisala frequency serves as a measure of the stability or instability (in this case, the parameter is complex and represents an instability growth rate) of a stratified layer of air at a specific altitude within the atmosphere. {These stratified layers can vary at different altitudes, resulting in alternating stable and unstable stratification in the lower atmosphere.} Figure \ref{fig:ao2_bv_profile} shows the vertical profiles of the temperature and Brunt-Vaisala (BV) frequency for the radiosonde dataset. The BV values were computed using the $\mathcal{N}^2=g \frac{\delta T}{T_0} \frac{1}{\Delta z}$ relation \cite{wilson2013effect}, where $T_0$ = 281 K, $\delta T, \delta z$ were derived from the temperature and altitude readings and g = 9.81 m/s$^2$. To derive the BV profiles shown in panel (b), we averaged the temperature readings within each 25-meter altitude interval. The frequency range lies between 0.002 and 0.007 for the positive values of $\mathcal{N}^2$, which corresponds to a local stable stratification. {Panel c illustrates the computed statistical average of the BV profile within specific altitude intervals. This average was derived from BV profiles collected by probes 5, 7, and 9. Notably, the figure highlights altitude regions where all three probes consistently experienced stable conditions (marked in violet) or unstable conditions (marked in green). In the unmarked regions, different probes displayed different stability conditions.}
\deleted{Negative $\mathcal{N}^2$ appears at latitudes within an altitude range of 2750-3200 m, and in a very localized altitude layer at about 2200 m.} 

\subsubsection{Distributions of relative neighbor distance, temperature, humidity, wind speed based on the readings of probes 2 to 7 and probe  9.}\label{sec:cluster_relative}	
\begin{figure}[ht!]
\centering
\begin{subfigure}{0.45\textwidth}
	\centering
	\includegraphics[width=\textwidth]{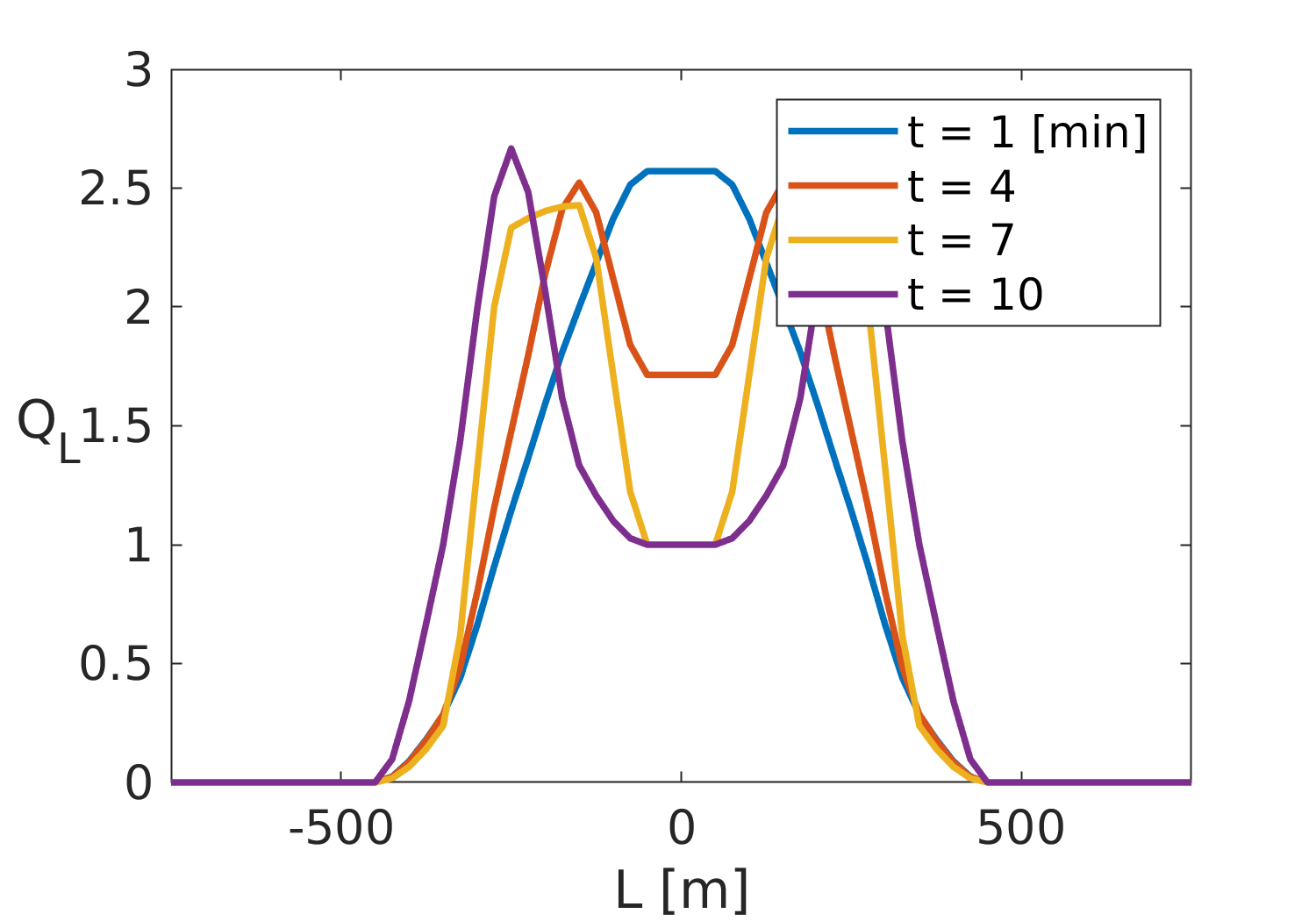}
	\caption{}
\end{subfigure}
\begin{subfigure}{0.45\textwidth}
	\centering
	\includegraphics[width=\textwidth]{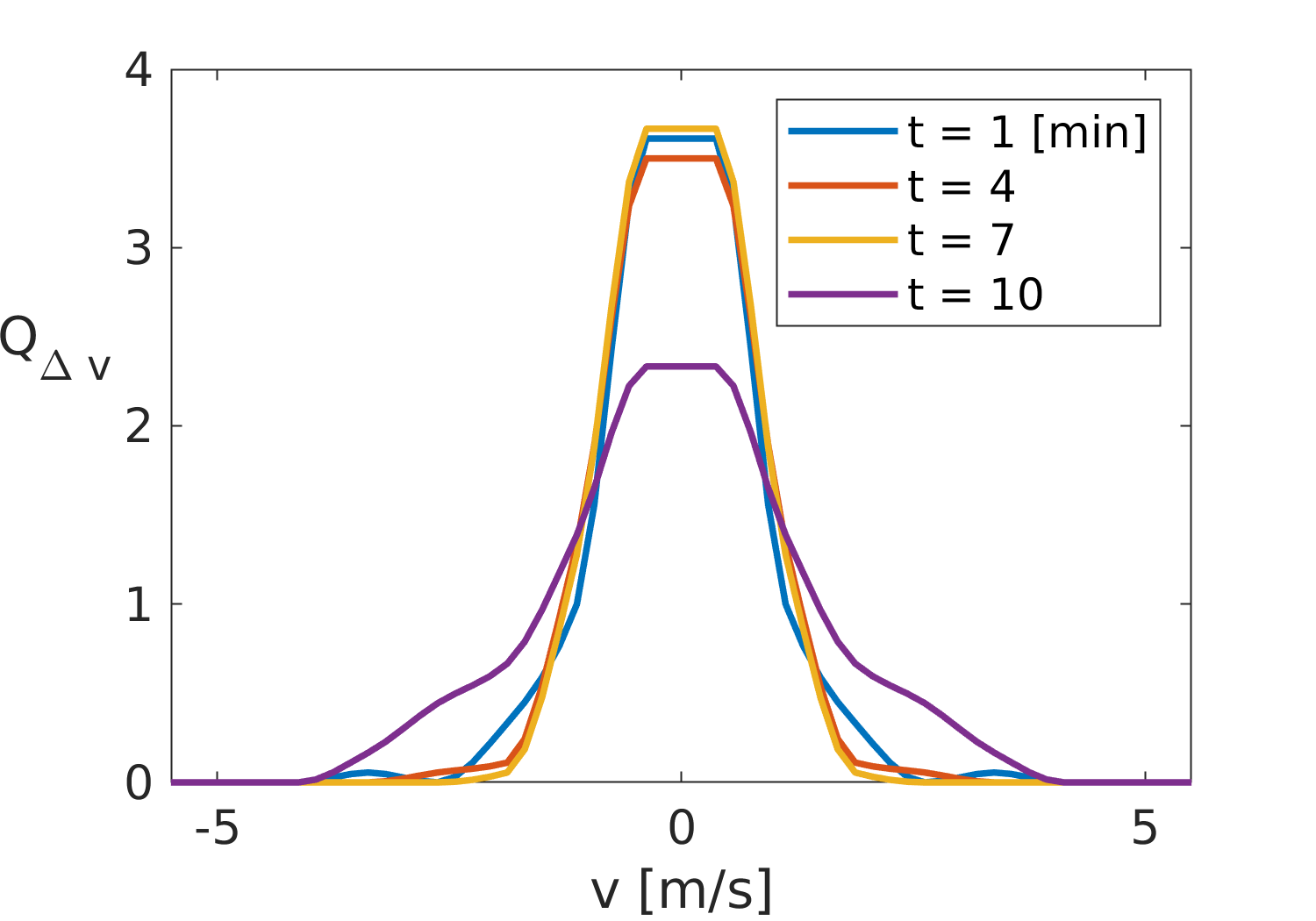}
	\caption{}
\end{subfigure}
\begin{subfigure}{0.45\textwidth}
	\centering
	\includegraphics[width=\textwidth]{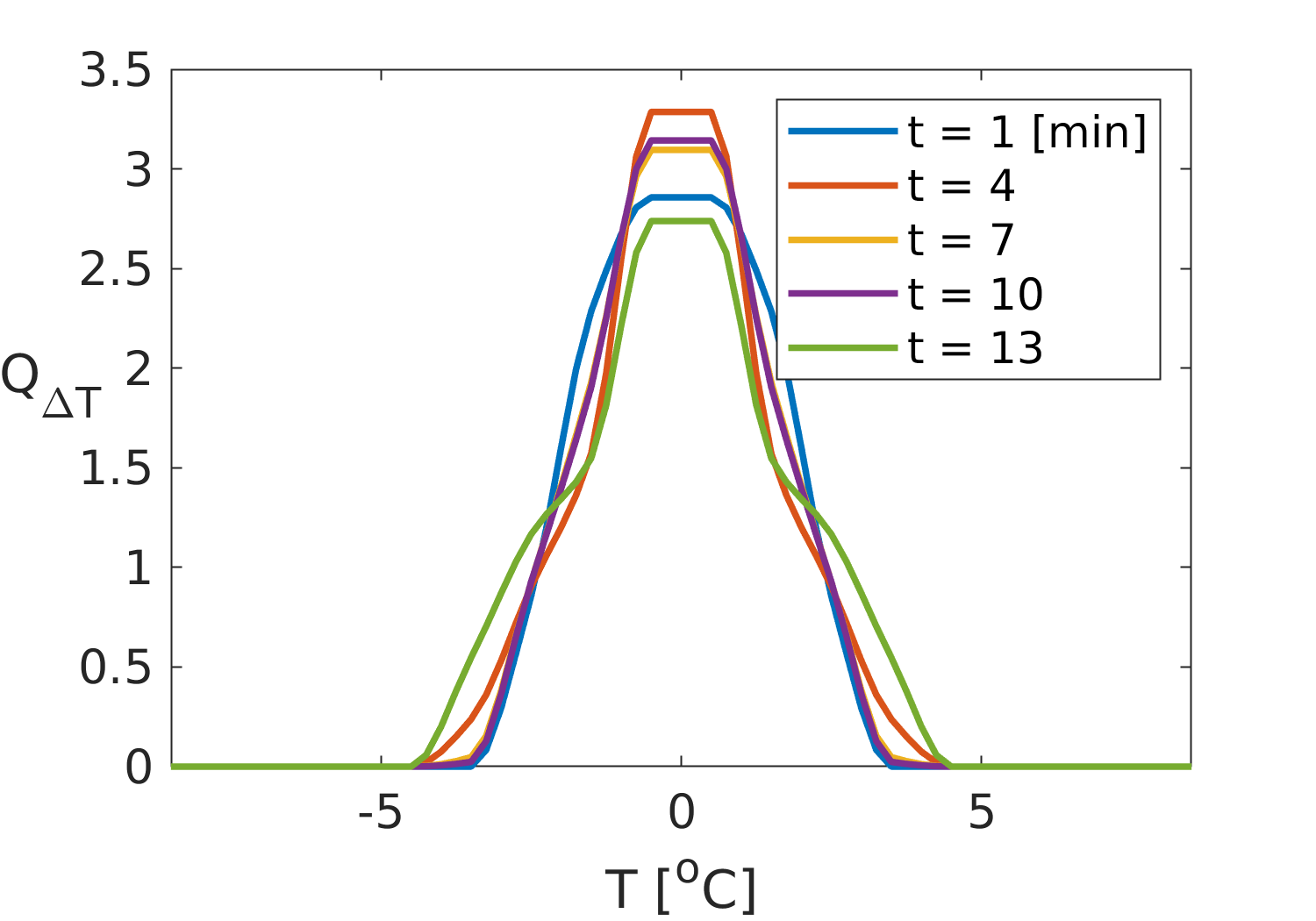}
	\caption{}
\end{subfigure}
\hspace{0.01\textwidth}
\begin{subfigure}{0.45\textwidth}
	\centering
	\includegraphics[width=\textwidth]{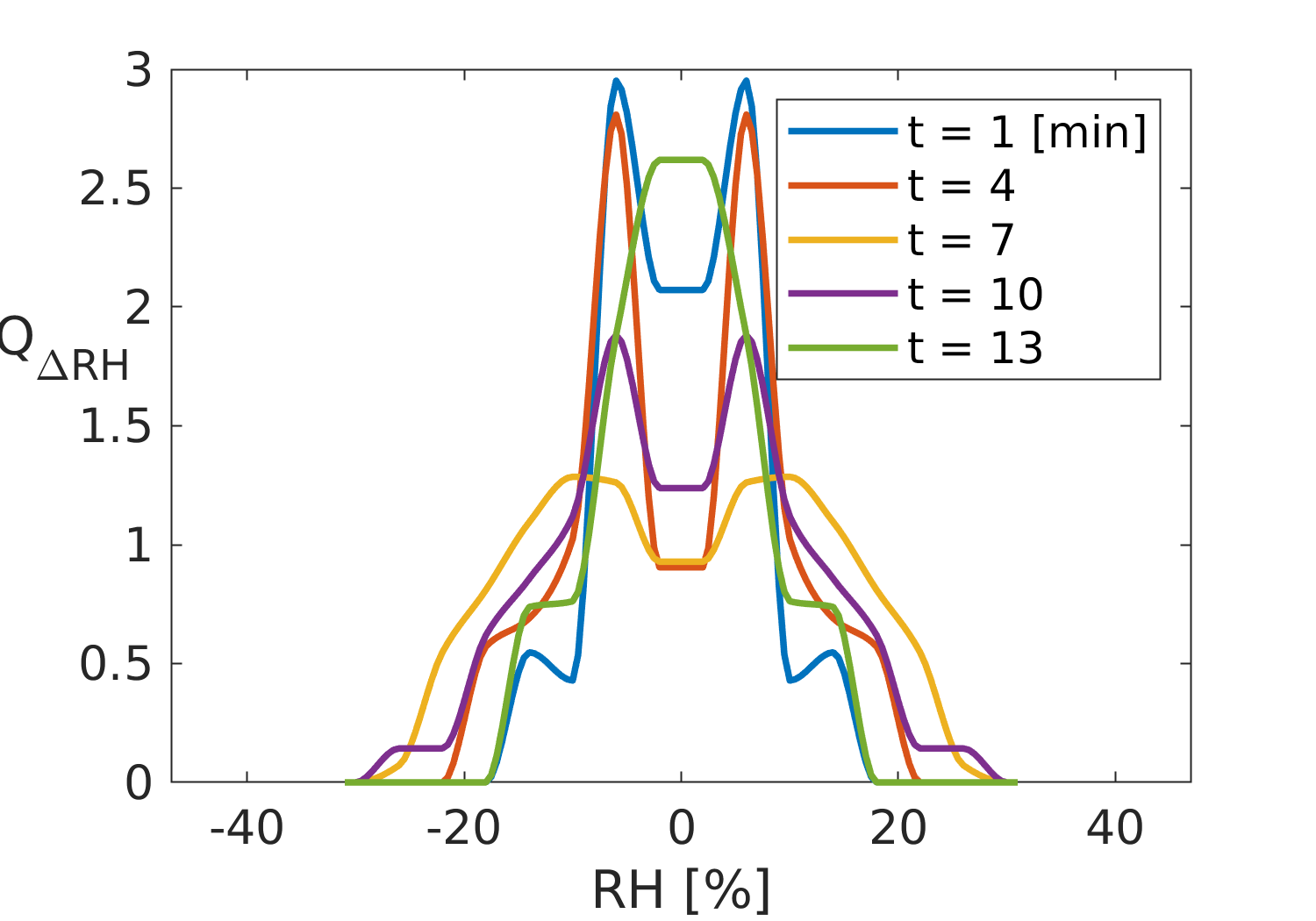}
	\caption{}
\end{subfigure}
\caption{Relative measurements of the temperature, humidity, distance, and wind speed during the free-floating experiment in OAVdA, November 3, 2022. Distributions of the these quantities were computed using the distance-neighbor graph function (eq. \ref{eq:q}) between the radiosondes inside a cluster: (a) relative distance, (b) relative wind speed, (c) relative temperature, and (d) relative humidity. The analysis was initiated at 14:18.}
\label{fig:ao2_Qgraphs}
\end{figure}

\citet{richardson1926atmospheric}, in his classical framework for turbulent dispersion, suggested analyzing the relative motion of a large number of "marked particles" in terms of the probability distribution function (PDF) of the relative distances between particles. {In practice, "marked particles" can be replaced by radiosondes by simply ensuring that the radiosonde rises to a pre-configured altitude and floats through the air without disturbing the carrier flow.} Recent studies have placed emphasis on the PDF of the pair separation. However, we would like to return to the original definition of Richardson's PDF via the distance-neighbor graph function. Here, $Q$ is the \textit{distance-neighbor graph}, where its element, $Q_{n, n+1}$, is the average number of neighbor {radiosondes per distance interval of constant size $h$. Let us write, Richardson's relation, for a general N number of radiosondes:} 
\begin{equation}
Q_{n, n+1} = \frac{1}{N}(P^1_{n, n+1} + P^2_{n, n+1} + P^3_{n, n+1} + ... + P^N_{n, n+1}),
\label{eq:q}
\end{equation}
{where the integer n enumerates the distance intervals. $P^k_{n, n+1}$ represents the number of neighbors in each distance range (n, n + 1) for each k-th sonde, where k runs from 1 to N. To obtain $P$, for each sonde, we compute its separation distance from all the other sondes (the other N-1 sondes). Then, we compute the histogram of the obtained distances over all neighborhood intervals. For example, if $h$ = 10 m, then $P^\textbf{1}_{0, 1}$ represents the number of neighbor sondes for the sonde \textbf{1} in the distance range from 0 to 10 m, while $P^1_{1, 2}$ denotes the number of neighbors found between 10 and 20 meters, always for the sonde \textbf{1} and so on. Then, $Q_{0, 1}$ is calculated as the arithmetic average of the $P^k_{0, 1}$ values for all sondes. The result provides the average count of neighboring radiosondes within the range of 0 to 10 meters.}

\deleted{The distance-neighbor graph function can easily be adopted for quantities other than a relative distance, such as relative temperature and relative humidity measurements within a cluster of particles. In practice, "marked particles" can be replaced by balloons by simply ensuring that the balloon rises to a pre-configured altitude and floats through the air without disturbing the carrier flow. It is expected that a combination of knowledge, gained from in-field experiments and numerical simulations, will enable us to better understand the relative dispersion and diffusion in the atmosphere. Indeed, simulation results can provide preliminary insights for the setup of the in-field measurements, such as selecting the initial launching point and the initial distance interval size, $h$, for distance-neighbor graph analysis.}

Figure \ref{fig:ao2_Qgraphs} shows relative measurements of the radiosondes as obtained by means of the Q (distance-neighbor graph) function. We extracted a 20-minute dataset for the temperature and humidity readings for this analysis, starting at 14:18 and a 12-minute dataset for the position and velocity readings. First, a $Q$ graph was computed for the relative distance between the radiosondes ($Q_L$ in panel a), where the distance interval  size, $h$, was 100 m, in terms of 3D distance. The computation of $Q$ was performed every 10 seconds and then averaged over each minute interval. We can see the values of $Q$ in the graph for the 1st, 4th, 7th, and 10th minutes over the length range between -400 and 400 meters (n = 0,1,2,3,4). It can be noted that the Q graph becomes wider over time. Initially, there were approximately 2.5 neighbor radiosondes in the first distance interval (100 meter proximity range), but as time passed, the number dropped to 1. The opposite trend can be seen for the other distance intervals on the right and on the left side of the graph.

By generalizing the definition of the "distance interval" to the definition of one among the  multiple "relative measurement" performed by the cluster, we can extend  the computation of $Q$ to temperature ($Q_{\Delta T}$), humidity ($Q_{\Delta RH}$) and wind speed ($Q_{\Delta v}$) relative differences. {We replaced the definition of the distance interval with the absolute value of the difference of readings ($\Delta T$, $\Delta RH$, $\Delta v$) among  all the radiosondes. The size, $h$, of the absolute difference range (distance interval) were 1 $^o$C, 2 \% and 0.75 m/s, respectively, for the relative temperature, humidity, and wind speed measurements.} One of the reasons for choosing the distance-neighbor graph function is that this is a truly direct quantification of the turbulent  dispersion of the physical quantities, see Figure \ref{fig:ao2_Qgraphs}. Furthermore, since all the panels share the same structure as that of eq. \ref{eq:q}, it can be easy to verify possible high values of the cross-correlation  among different field quantities. 

\section{Conclusions and future works}\label{sec:conclusion}
This work describes a new balloon-borne radiosonde network system and offers considerations on a new measurement technique involving a cluster of radiosondes. We have presented the results of field-tests and in-field experiments that helped us validate and bring the measurement system closer to realization. We have demonstrated that the proposed measurement system is able to track the Lagrangian fluctuations of physical quantities, such as position, velocity, pressure, humidity, temperature, acceleration, and magnetic field. {Post-processing analysis of spatial and temporal measurements using distance-neighbor graphs can provide a Lagrangian quantification of turbulent dispersion}. In the future, we would like to combine the results from numerical simulations and in-field measurements in a more comprehensive analysis of {warm} clouds {and atmospheric boundary layer, by considering} cloud microphysics, turbulent fluctuations{, and related diffusion processes}.

The data transmission and acquisition modules of the system are currently under optimization. The present data transfer rate is one packet every three-to-four seconds, which is adequate for the current prototype. However, the new prototype has been proposed with the optimum computational characteristics, weight, and size. A new ground station is also  being developed that will enable users to receive data at higher rates and concurrently via multiple radio channels. For this reason, we are developing custom-built gateways based on a LoRa peer-to-peer architecture. In the future, the radioprobe board sensors will also be protected from radiation and precipitation sources through the use of a lightweight shield.

\deleted{The developed post-processing module of the system can provide an extensive analysis of the relative measurements via distance-neighbor graph functions and, possibly, space-time Lagrangian correlations. Furthermore, the proposed system can provide magnetic field strength over space and time, which could be used in lightning detection applications in the future.}

The radiosonde cluster presented in this study was tested {for measuring various}\deleted{in the context of a few} physical atmospheric quantities , with a particular focus on capturing their temporal and spatial fluctuations. Its use could be extended to cover a wide range of applications, to observe the topology of atmospheric and marine boundary layers, which cannot be conducted with a single-sonde system. Furthermore, the proposed system can provide fluctuations of the magnetic field strength over space and time, which could be used in lightning detection applications in the future. Additionally, the measurement system can be employed for environmental monitoring over urban and industrial areas.

It is worth noting that the integration of low-cost and lightweight MEMS sensors into this system enables a reduction in the unit cost of the radiosonde. This, in turn, opens up the possibility of conducting cost-effective radiosonde measurements in diverse atmospheric experiments, including vertical profiling.

\section*{Acknowledgments}
This project has received funding from the Marie-Sklodowska Curie Actions (MSCA ITN ETN COMPLETE) project under the European Union’s Horizon 2020 research and innovation program, grant agreement 675675, \href{http://www.complete-h2020network.eu}{\textit{COMPLETE ITN-ETN NETWORK}}. The project has also received funding from the Links Foundation within \textit{"PoC Instrument – IV cut off"} initiative, for the development of the second prototype and conducting the recent in-field tests with the cluster of the radiosondes. We thank for the kind hospitality in performing dual sounding in-field tests with \href{http://www.arpa.piemonte.it/chi-siamo}{\textit{ARPA-Piemonte}} at LIMZ (Levaldigi Airport, Dr. Luca Tommassone) and for the kind hospitality in the alpine context at \href{https://www.oavda.it/la-fondazione}{\textit{L’Osservatorio Astronomico della Regione Autonoma Valle d’Aosta}, St. Barthelemy (Dr. Jean Marc Christille)}. We thank also the UK Meteorological Office (Dr. Jeremy Price, Dr. Simon Osborne, Dr. Paul Barrett) and NCAS, Chilbolton Observatory Facility (Dr. Chris Walden and Dr. Darcy Ladd) for the hospitality during the Wessex Convection Campaign 2023 and for data sharing. 


\bibliographystyle{elsarticle-num-names-mod} 
\bibliography{references}

\end{document}